\documentclass[a4paper,preprintnumbers,showpacs,onecolumn,superscriptaddress,nofootinbib,amsmath,amssymb,notitlepage]{revtex4-1}

\usepackage{enumitem}
\usepackage{times}
\usepackage{dcolumn}
\usepackage{mathrsfs}
\usepackage{comment}
\usepackage{hyperref}
\usepackage{amsmath,accents}
\usepackage{mathtools}
\usepackage{bbm}
\usepackage{bm}
\usepackage{amsfonts}
\usepackage{amssymb}
\usepackage{mathrsfs}
\usepackage[scr=rsfs,cal=boondox]{mathalfa}
\usepackage{wasysym}
\usepackage{graphicx}
\usepackage[]{subfigure}
\usepackage{filecontents}
\usepackage[dvipsnames]{xcolor}
\usepackage{bbold}
\usepackage[scr=rsfs,cal=boondox]{mathalfa}
\usepackage{stackengine}
\stackMath
\newcommand\tsup[2][2]{%
 \def\useanchorwidth{T}%
  \ifnum#1>1%
    \stackon[-1.3ex]{\tsup[\numexpr#1-1\relax]{#2}}{\mathchar"307E}%
  \else%
    \stackon[-1ex]{#2}{\mathchar"307E}%
  \fi%
}

\hypersetup{
    bookmarks=true,         
    pdftoolbar=true,        
    pdfmenubar=true,        
    pdffitwindow=true,     
    pdfstartview={FitH},     
    pdftitle={My title},     
    pdfauthor={author},      
    pdfsubject={Subject},    
    pdfcreator={Creator},    
    pdfproducer={Producer},  
    pdfkeywords={keyword1} 
    pdfnewwindow=true,       
    colorlinks=true ,        
    linkcolor=Blue  ,        
    citecolor=Blue,      
    filecolor=Blue,       
    urlcolor=Blue            
}

\newcommand{\arccosh}{\operatorname{arccosh}}

\newcommand{\ed}{\mathrm{d}}

\newcommand{\tet}{\tilde{\eta}}
\newcommand{\tch}{\tilde{\chi}}
\newcommand{\bF}{\operatorname{\mathbf{F}}}
\newcommand{\bPi}{\operatorname{\mathbf{\Pi}}}
\newcommand{\bE}{\operatorname{\mathbf{E}}}
\newcommand{\bK}{\operatorname{\mathbf{K}}}
\newcommand{\bJ}{\operatorname{\mathbf{{J}}}}
\newcommand{\sn}{\operatorname{\mathbf{sn}}}
\newcommand{\cn}{\operatorname{\mathbf{cn}}}
\newcommand{\dn}{\operatorname{\mathbf{dn}}}
\newcommand{\m}{\mathcal{k}}
\newcommand{\s}{\mathrm{s}}
\newcommand{\ob}{\mathrm{o}}
\newcommand{\Q}{\mathscr{Q}}
\newcommand{\ph}{\mathrm{ph}}
\newcommand{\po}{\mathrm{po}}
\newcommand{\DFK}{\mathrm{DFK}}
\newcommand{\DSK}{\mathrm{DSK}}
\newcommand{\COFK}{\mathrm{COFK}}

\newcommand{\PO}{\mathrm{PO}}
\newcommand{\oalpha}[1]{\accentset{\circ}{\alpha}}
\newcommand{\obf}[1]{\accentset{\circ}{\mathbf{f}}}
\newcommand{\boR}[1]{\accentset{\circ}{\mathbf{R}}}
\newcommand{\obF}[1]{\accentset{\circ}{\mathbf{F}}}
\newcommand{\obPi}[1]{\accentset{\circ}{\mathbf{\Pi}}}

\begin{document}

\title{Light propagation around a Kerr-like black hole immersed in an inhomogeneous anisotropic plasma in Rastall gravity: Analytical solutions to the equations of motion}

\author{Mohsen Fathi}
\email{mohsen.fathi@usach.cl}
\affiliation{Departamento de F\'{i}sica, Universidad de Santiago de Chile,
Avenida V\'{i}ctor Jara 3493,  Estaci\'{o}n Central, 9170124, Santiago, Chile}

\author{Marco Olivares}
\email{marco.olivaresr@mail.udp.cl}
\affiliation{Facultad de Ingenier\'{i}a y Ciencias, Universidad Diego Portales, Avenida Ej\'{e}rcito Libertador 441, Santiago, Chile}

\author{J.R. Villanueva}
\email{jose.villanueva@uv.cl}
\affiliation{Instituto de F\'{i}sica y Astronom\'{i}a, Universidad de Valpara\'{i}so,
Avenida Gran Breta\~{n}a 1111, Valpara\'{i}so, Chile}

\author{Norman Cruz}
\email{norman.cruz@usach.cl}
\affiliation{Departamento de F\'{i}sica, Universidad de Santiago de Chile,
Avenida V\'{i}ctor Jara 3493,  Estaci\'{o}n Central, 9170124, Santiago, Chile}


\begin{abstract}

In this paper, we explore the behavior of light ray trajectories in the exterior geometry of a rotating black hole within the Rastall theory of gravity, which is surrounded by an inhomogeneous anisotropic electronic cold plasma. By specifying the plasma's frequency profile, we derive fully analytical solutions for the temporal evolution of spacetime coordinates using elliptic integrals and Jacobi elliptic functions. These solutions illustrate various possible orbits. Throughout the study, we compare the results with those in the vacuum case, emphasizing the influence of plasma. Additionally, we utilize the analytical solutions to establish the lens equation for the considered spacetime. The investigation also addresses the significance of spherical photon orbits on critical trajectories, by presenting several examples.

\bigskip

{\noindent{\textit{keywords}}: Photon motion, Rastall gravity, plasmic medium, gravitational lensing
}\\

\end{abstract}

\maketitle


\section{Introduction and Motivation}\label{sec:intro}


While the general theory of relativity remains highly regarded in astrophysics and cosmology for its elegance and beauty, recent observations like Type Ia supernovae, cosmic microwave background radiation anisotropies, flat galactic rotation curves, and weak lensing have shifted scientific perspectives. These observations suggest that only a small portion of the observable universe consists of ordinary baryonic matter, with the remainder attributed to dark matter and dark energy \cite{riess_type_2004,komatsu_seven-year_2011}. To date, scientists have devised numerous theoretical models to explain the phenomena of dark matter and dark energy. (for example, see Refs. \cite{turner_coherent_1983,press_single_1990,dubinski_structure_1991,sin_late-time_1994,navarro_structure_1996,navarro_universal_1997,hu_fuzzy_2000,peebles_fluid_2000,peebles_cosmological_2003,kiselev2003quintessential,viel_constraining_2005,amendola_dark_2006,copeland_dynamics_2006,de_la_macorra_dark_2006,rahaman_perfect_2010,rahaman_modeling_2011,li_dark_2011,schive_cosmic_2014,marsh_axion_2016,hui_ultralight_2017,wang_holographic_2017}). Nevertheless, the enigmatic nature of dark matter and dark energy persists as unresolved quandaries in modern cosmology. Consequently, some scientists speculate that by adjusting the general theory of relativity and reshaping our understanding of gravitational behavior, it might be plausible to elucidate the anomalous phenomena currently attributed to the dark side of the universe. Among the various modified gravity models proposed, the Rastall theory of gravity stands as one such alternative \cite{Rastall19723357,rastall_theory_1976}, which follows a broader energy-momentum conservation law, directly connected to the Mach principle. Since its introduction, this theory has found various applications in both gravitation and cosmology (for example, see Refs. \cite{batista_rastall_2012,fabris_rastall_2012,fabris_rastalls_2015,heydarzade_black_2017-1,heydarzade_black_2017,darabi_einstein_2018,lobo_thermodynamics_2018,guo_shadow_2021,guo_observable_2022}). Additionally, the Rastall theory produces interesting outcomes, particularly in its black hole solutions that incorporate additional terms to accommodate the universe's dark components. The incorporation of dark matter/energy effects into black hole spacetimes has been a longstanding focus of scientific inquiry. Kottler, for instance, investigated the impact of the cosmological constant on the Schwarzschild black hole spacetime \cite{kottler_uber_1918}, later extended to its Kerr-like solution by Carter \cite{Carter:1973rla}. Kiselev was among the first to reconcile quintessential dark energy with the Schwarzschild black hole spacetime \cite{kiselev_quintessence_2003}, a concept later extended to Kerr and Kerr-Newman-anti de Sitter black holes \cite{toshmatov_rotating_2017}. A similar solution has been proposed for Rastall gravity in Ref. \cite{heydarzade_black_2017}, involving a Reissner-Nordstr\"{o}m black hole spacetime embedded within perfect fluid matter. This solution was further generalized in Refs. \cite{kumar_rotating_2018,xu_kerrnewman-ads_2018} to describe a Kerr-Newman-like black hole within Rastall gravity, examining certain characteristics of the dark fluid within this spacetime context. However, in this study, we focus on a similar black hole with zero net charge. This leads to a Kerr-like black hole within a cosmological fluid background, allowing us to investigate the behavior of light rays within its exterior spacetime in this work.

In fact, the study of light behavior has held significant importance in relativistic astrophysics. From Eddington's groundbreaking confirmation of light deflection during the 1919 solar eclipse \cite{Eddington:1920}, validating Einstein's predictions \cite{Einstein:1911}, to recent revelations from the Event Horizon Telescope (EHT) capturing the shadows of M87* \cite{Akiyama:2019} and Sgr A* \cite{Akiyama:2022}, the role of light as our primary source of information from celestial objects has become increasingly evident. Black holes, in particular, remain enigmatic as they reside beyond an event horizon where light becomes trapped, rendering their interiors inscrutable. Thus, our only means of studying light behavior around black holes is beyond this event horizon. This study focuses on the limits to which light can approach a Kerr-like Rastall black hole, employing fully analytical methods to analyze photon behavior within these specific regions. 

Indeed, the study of geodesics in black hole spacetimes began soon after the advent of general relativity. However, it was Hagihara's 1930 derivation of particle trajectories in Schwarzschild spacetime that marked a reliable analytical breakthrough. Efforts since then have consistently pursued gravitational lensing caused by black holes, yielding significant formulations, including methods to constrain the black hole shadow \cite{Bardeen:1972a,Bardeen:1973a,Bardeen:1973b,Chandrasekhar:2002}. Analytical treatment of particle geodesics aids in strict theoretical predictions and their comparison with observational data, attracting interest among scientists. Such analysis also helped revive the study of gravitational lensing \cite{Virbhadra:2000}.
However, non-static black hole spacetimes involve complex equations requiring advanced mathematical methods, leading to numerous investigations. In the realm of elliptic integrals' application in calculating light ray geodesics (relevant to this work), various criteria have been discussed and resolved in Refs. \cite{kraniotis_precise_2004,beckwith_extreme_2005,kraniotis_frame_2005,hackmann_complete_2008,hackmann_geodesic_2008,bisnovatyi-kogan_strong_2008,kagramanova_analytic_2010,hackmann_analytical_2010,hackmann_complete_2010,enolski_inversion_2011,kraniotis_precise_2011,enolski_inversion_2012,gibbons_application_2012,munoz_orbits_2014,kraniotis_gravitational_2014,de_falco_approximate_2016,soroushfar_detailed_2016,barlow_asymptotically_2017,uniyal_null_2018,villanueva_null_2018,chatterjee_analytic_2019,hsiao_equatorial_2020,gralla_null_2020,hendi_simulation_2020,fathi_classical_2020,kraniotis_gravitational_2021,fathi_study_2022,battista_geodesic_2022,fathi_spherical_2023}. Most recently, these methods have been employed to conduct analytical ray-tracing of the shadow of rotating black holes in cosmological backgrounds, as demonstrated in Refs. \cite{omwoyo_black_2023, Wang:2023}. Furthermore, Synge's discussions in Ref.~\cite{Synge:1960} enabled mathematical methods for determining and constraining light propagation in theoretical (non-)static black hole spacetimes within plasmic or dark fluid environments \cite{bisnovatyi-kogan_gravitational_2009,bisnovatyi-kogan_gravitational_2010,tsupko_gravitational_2013,morozova_gravitational_2013,bisnovatyi-kogan_gravitational_2015,perlick_influence_2015,atamurotov_optical_2015,abdujabbarov_shadow_2016,bisnovatyi-kogan_gravitational_2017,perlick_light_2017,schulze-koops_sachs_2017,abdujabbarov_shadow_2017,liu_effects_2017,haroon_shadow_2019,kimpson_spatial_2019,babar_optical_2020,junior_spinning_2020,Badia:2021,fathi_analytical_2021}. However, the analytical treatment of light ray trajectories in non-vacuum black hole environments demands greater attention, offering the potential to identify additional orbits often overlooked in determining the photon spheres and shadows of black holes. Hence, building upon the introductory context outlined earlier, we utilize these mathematical methods to compute light ray paths within a non-vacuum, non-static spacetime governed by Rastall gravity. {We will study the light ray trajectories outside a rotating black hole, whose exterior spacetime is derived from the Rastall theory of gravity.} Accordingly, we focus on the possible photon orbits around a Kerr-like black hole immersed in an inhomogeneous and anisotropic plasmic medium whose structural properties remain constant throughout.

To obviate this objective, we have organize this study as follows: In Sect. \ref{sec:BHsol} we provide a brief introduction to the Rastall theory of gravity, outlining its static and Kerr-like black hole solutions, and emphasizing the parameter $\zeta$ in the cosmic fluid. We explore three distinct cases for this parameter, discussing the resulting causal structure of the black hole. We argue that $\zeta=3$ serves as the foundational choice, forming the basis for subsequent sections. In Sect. \ref{sec:HJEOM}, we delve into the Hamiltonian dynamics of light propagation within a refractive medium, particularly focusing on a non-magnetized plasma. We derive and present the general equations of motion governing light propagation within a stationary spacetime filled with such plasma. Section \ref{sec:eqofmot} constitutes the core of our study, analyzing the evolution of spacetime coordinates, beginning with polar motion. We classify feasible photon orbits using the radial effective potential and provide analytical solutions to the equations of motion for all cases. We express these solutions, when necessary, in terms of elliptic integrals and Jacobian elliptic functions. Additionally, this section addresses spherical photon orbits and their variations concerning the photons' impact parameter. Further insights into spherical orbits are discussed in Sect. \ref{sec:spherical}, followed by a comprehensive summary of our findings in Sect. \ref{sec:conclusions}. Throughout this work, we adopt a geometrized unit system ($G=c=1$) and adhere to the sign convention $(-, +, +, +)$.

\section{Rastall gravity and its black hole solution}\label{sec:BHsol}

The field equations of the Rastall theory of gravity is given by \cite{Rastall19723357}
\begin{equation}
R_{\mu\nu}-\frac{1}{2}g_{\mu\nu}R=\kappa\left(T_{\mu\nu}-\lambda g_{\mu\nu} R\right),
    \label{eq:field0}
\end{equation}
in which $R_{\mu\nu}$, $R$, and $T_{\mu\nu}$ represent the Ricci tensor, Ricci scalar, and energy-momentum tensor, respectively. The gravitational coupling constant in Rastall gravity, denoted as $\kappa$, contributes to our discussions via a new parameter, $\psi=\kappa\lambda$, recognized as the Rastall coupling parameter \cite{KUMAR2021100881}. This parameter holds significant importance in distinguishing between non-vacuum general relativistic solutions and those derived in Rastall gravity. Rastall gravity is founded on the Rastall hypothesis, which, instead of the conventional energy-momentum conservation law, ${T^{\mu\nu}}_{;\mu}=0$, adheres to the modified conservation law ${T^{\nu}}_{\mu;\nu}=\lambda R_{,\mu}$. Notably, this theory reinstates the common conservation law in flat spacetime. The parameter $\lambda$ delineates the degree to which energy-momentum conservation is upheld. As a result, the theory aligns with the Mach principle, postulating that the inertia of a local mass correlates with the total mass/energy distribution in the universe \cite{majernik2006rastalls}.

In fact, the quest for exact solutions in Rastall gravity remains a relatively recent trend. However, certain aspects of spherically symmetric black hole solutions have surfaced in recent discussions. Notably, Ref. \cite{heydarzade_black_2017-1} explored aspects of spherically symmetric black hole solutions within the context of Rastall gravity, introducing a solution grounded in an anisotropic fluid background, initially reported in Ref. \cite{heydarzade_black_2017}. This solution is formulated based on the line element
\begin{equation}
\ed s^2 = -f(r) \ed t^2+\frac{\ed r^2}{f(r)} + r^2\left(\ed\theta^2+\sin^2\theta\ed\phi^2\right),
    \label{eq:ds_0}
\end{equation}
in the usual Schwarzschild coordinates, where the lapse function is given by
\begin{equation}
f(r) = 1-\frac{2M}{r}-N_s r^{-\frac{1+3w_s-6\psi(1+w_s)}{1-3\psi(1+w_s)}},
    \label{eq:f(r)_0}
\end{equation}
where $N_s$ represents the matter density for a perfect fluid characterized by the state parameter $w_s$, enveloping a static black hole with the mass $M$.
Note that, the lapse function \eqref{eq:f(r)_0} is valid for $\psi\neq\frac{1}{3}(1+w_s)^{-1}$.
This fluid is characterized by an anisotropic energy-momentum tensor, as described in Ref. \cite{Visser_2020}. In Refs. \cite{kumar_rotating_2018,xu_kerrnewman-ads_2018}, the above static spacetime was extended to a Kerr-Newman-like spacetime, by means of a modified Newman-Janis algorithm, proposed by Azreg-A\"{I}nou \cite{Azreg:2014,azreg-ainou_static_2014}. In the Boyer–Lindquist coordinates, this stationary black hole spacetime is characterized by the line element \cite{xu_kerrnewman-ads_2018}
\begin{multline}\label{metr}
\mathrm{d}s^2 = -\left(\frac{\Delta-a^2\sin^2\theta}{\rho^2}\right)\ed t^2 + \frac{\rho^2}{\Delta}\ed r^2+\rho^2 \ed\theta^2 
-2a \sin^2\theta\left(\frac{r^2+a^2-\Delta }{\rho^2}\right) \ed t \ed\phi\\
+\sin^2\theta\left[\frac{(r^2+a^2)^2-\Delta a^2\sin^2\theta}{\rho^2}\right]\ed\phi^2,
\end{multline}
where for the case of a black hole with zero electric charge, we have
\begin{subequations}\label{eq:rho,Delta}
	\begin{align}
	& \Delta(r) = r^2+a^2-2 M r-N_s r^\zeta,\label{eq:Delta}\\
	& \rho^2(r,\theta) = {r^2 + a^2 \cos^2 \theta},\label{eq:rho}\\
	\end{align}
\end{subequations}
with
\begin{equation}
    \zeta = {1-3w_s\over 1-3\psi (1+w_s)}.
    \label{eq:zeta}
\end{equation}
This way, the line element \eqref{metr} characterizes the exterior spacetime of a Kerr-like Rastall black hole (KRBH) with mass $M$ and spin parameter $a$, existing within a fluid defined by the parameter $N_s$. In instances where $N_s=0$, the KRBH converges to a Kerr black hole. However, for $N_s\neq0$ and $-1<w_s<-1/3$, the configuration yields a Kerr black hole encompassed by a quintessential fluid when $\psi=0$. When $N_s=a=0$, the spacetime reverts to that of a Schwarzschild black hole. It is noteworthy that akin to Kerr black holes, the KRBH spacetime exhibits two Killing vectors $\bm{\xi}_t$ and $\bm{\xi}_\phi$, aligning with symmetries respecting the $t$ and $\phi$-coordinates.

We proceed by examining the crucial hypersurfaces that delineate the causal boundaries within the KRBH spacetime.

\subsection{The causal structure of the KRBH}\label{subsec:horizons}

Based on the specific spacetime metric, it is evident that the formation of horizons around the KRBH diverges from that in general relativistic spacetimes. This divergence arises due to contributions from the structure parameter $N_s$ and the Rastall coupling $\psi$. Equation \eqref{eq:Delta} indicates the potential existence of up to three horizons within the spacetime. However, the black hole's ability to manifest these horizons hinges strictly upon the value of $\zeta$ within the function $\Delta$, determined by both $w_s$ and $\psi$. In the subsequent subsection, we explore various specific cases, discussing the availability of horizons and furnishing analytical solutions for the radii of the characteristic hypersurfaces within the spacetime.

\subsubsection{The case of $\zeta =1$}\label{subsub:zeta1}

\begin{figure}[!h]
\centering
 \includegraphics[width=6cm]{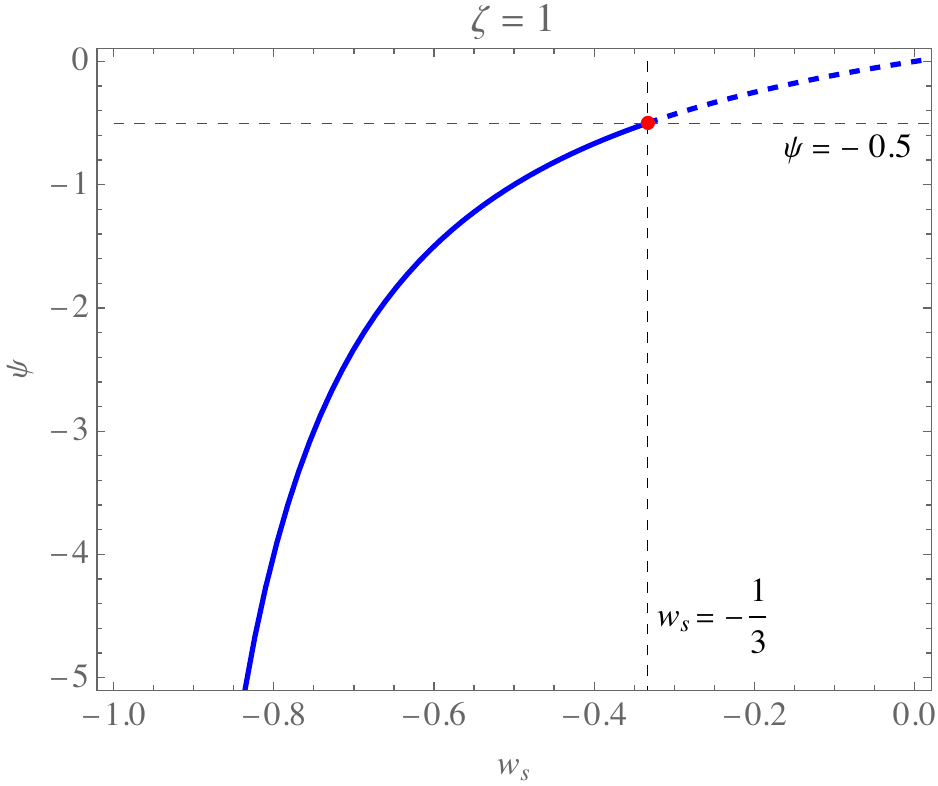}~(a)\qquad\qquad
 \includegraphics[width=6cm]{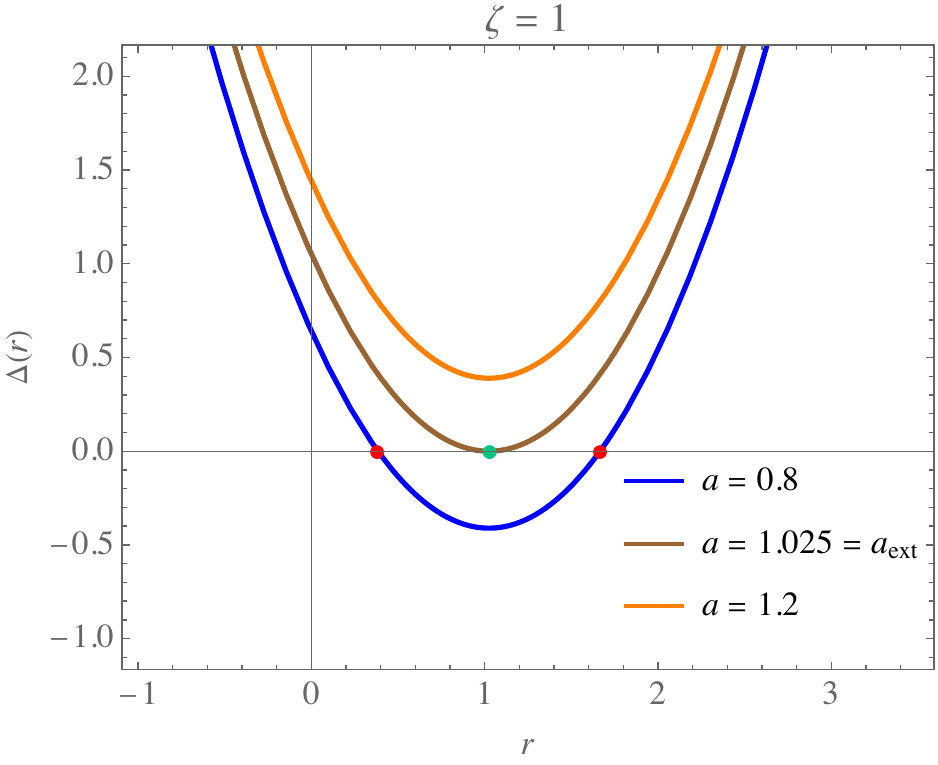}~(b)
	\caption{The diagrams depict (a) the interrelation between $\psi$ and $w_s$, and (b) the radial profile of $\Delta(r)$ for three distinct spin parameters, specifically for $\zeta = 1$, with $N_s = 0.05$. Throughout this paper, including these diagrams and all subsequent ones, the unit of length along the axes is standardized to $M$.}
	\label{fig:zeta1}
\end{figure}

The polynomial equation $\Delta=0$ identifies the black hole horizons, which, when $\zeta=1$, leads to a quadratic equation. Solving this equation yields the analytical solutions
\begin{equation}
r_\mp={2M+N_s\over 2}\mp\sqrt{\left( {2M+N_s\over 2}\right) ^2-a^2},
\label{eq:rHzeta1}
\end{equation}
which corresponds, respectively, to the Cauchy and event horizons of the black hole, each present distinctly when the condition $4a^2<\left( {2M+N_s}\right)^2$ holds true. When $a=(2M+N_s)/2=a_{\mathrm{ext}}$, an extremal KRBH forms, where $r_+=r_-=M+N_s/2=r_{\mathrm{ext}}$. However, for $4a^2>\left( {2M+N_s}\right) ^2$, a naked singularity emerges. Notably, in this specific case, no cosmological horizons are observed. Furthermore, from Eq. \eqref{eq:zeta}, specific values of $\zeta$ impose constraints on ranges for $w_s$ and $\psi$, determining the maximum number of solutions to the equation $\Delta=0$. The left panel of Fig. \ref{fig:zeta1} illustrates these domains for $w_s$ and $\psi$ within a diagram considering $\zeta=1$.
According to this diagram, the allowed range of the Rastall coupling, within the domain $-1<w_s<-1/3$, is $-\infty<\psi<-0.5$. The plot of the radial profile of the function $\Delta$ in the right panel of Fig.~\ref{fig:zeta1} confirms this anticipated outcome that the number of real solutions remains at either two (black hole), one (extremal black hole), or zero (naked singularity). Thus, despite the presence of a quintessential field around the black hole, a cosmological horizon is not observed for this specific value of $\zeta$.

\begin{figure}[!h]
\centering
 \includegraphics[width=5.3cm]{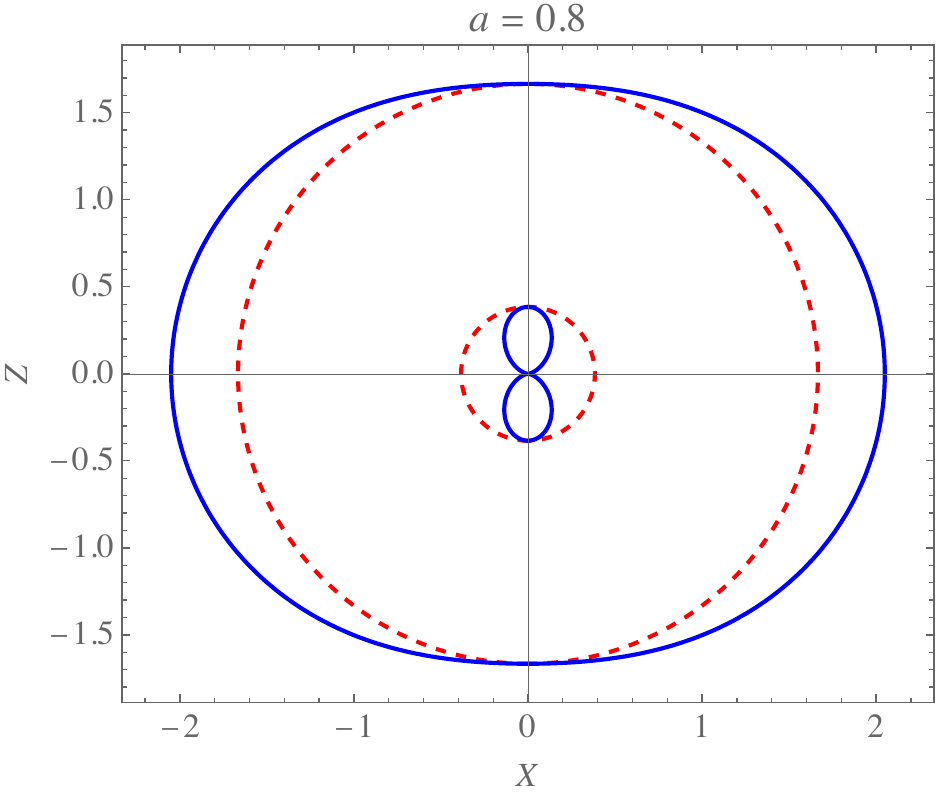}~(a)
 \includegraphics[width=5.3cm]{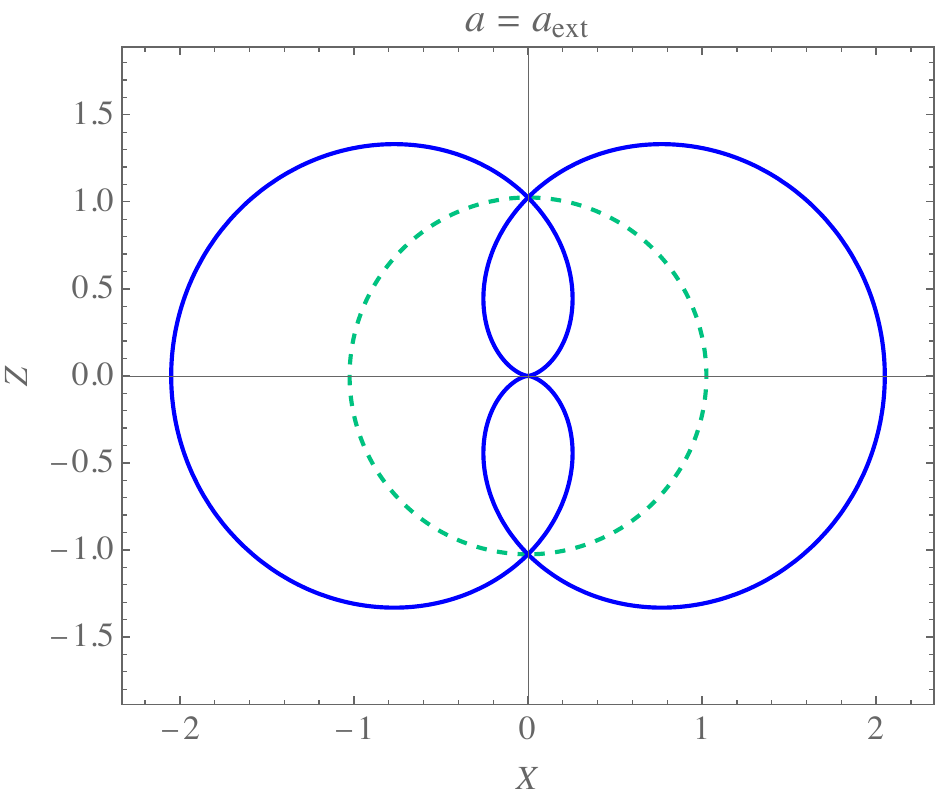}~(b)
 \includegraphics[width=5.3cm]{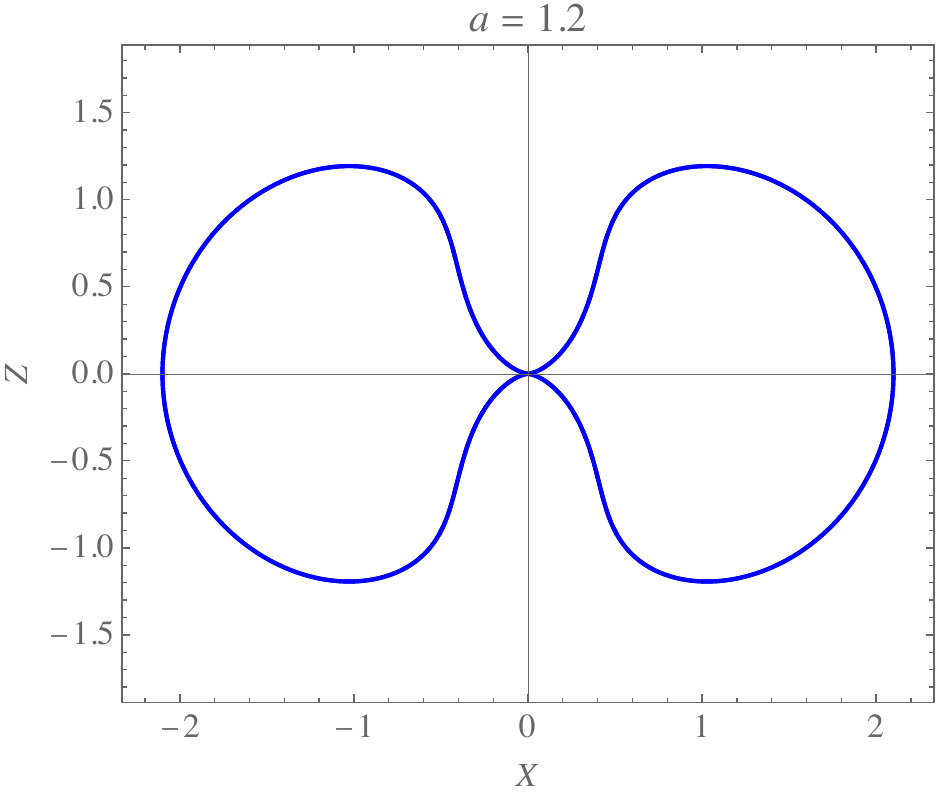}~(c)
	\caption{The cross-section of the ergoregions of the KRBH in the $Z$-$X$ plane, for (a) black hole with $a=0.8$, (b) extremal black hole with $a=a_{\mathrm{ext}}=1.025$, and (c) naked singularity with $a=1.2$, for the case of $\zeta=1$. The red and green dashed circles correspond, respectively, to $r_\mp$ and $r_{\mathrm{ext}}$. }
	\label{fig:rSLzeta1}
\end{figure}

It is important to note that apart from horizons, stationary black holes also feature other hypersurfaces where static observers, defined by the world-lines $\bm{u}=(-g_{tt})^{-1/2}\bm{\xi}_{t}$, cannot exist. These particular hypersurfaces are termed as the \textit{static limits}, where $\bm{u}$ becomes null. The radii of these static limits are determined by the equation $g_{tt}=0$. For the KRBH, utilizing Eq.~\eqref{metr}, the radii of the static limits are derived as 
\begin{equation}\label{eq:rSLzeta1}
r_{\mathrm{SL}_{\mp}}(\theta) = {2M+N_s\over 2}\mp\sqrt{\left( {2M+N_s\over 2}\right) ^2-a^2\cos^2\theta},
\end{equation}
which are located, respectively, inside the Cauchy horizon and outside the event horizon. The spaces between the static limits and the horizons are known as the \textit{ergoregions}, and correspond to the domains, in which, all observers are in the state of \textit{corotation} with the black hole. Considering the common Cartesian coordinates $X=r \sin\theta \cos\phi$, $Y= r \sin\theta\sin\phi$, and $Z=r\cos\theta$, in Fig.~\ref{fig:rSLzeta1}, we have demonstrated the ergoregions for the case of $\zeta=1$.  
Clearly, as the spin parameter escalates, the radial extent of the ergoregion diminishes. In scenarios involving an extremal black hole with a solitary horizon at $r_{\mathrm{ext}}$, it is noteworthy that $r_{\mathrm{SL}_-}^0=r_{\mathrm{SL}_+}^0=r_{\mathrm{SL}_\mp}(0)|_{a=a_{\mathrm{ext}}}$. Despite the horizon vanishing for the naked singularity, the ergoregion persists, albeit bifurcating into two distinct branches.

\subsubsection{The case of $\zeta = 2$}\label{subsub:zeta2}

\begin{figure}[!h]
\centering
 \includegraphics[width=6cm]{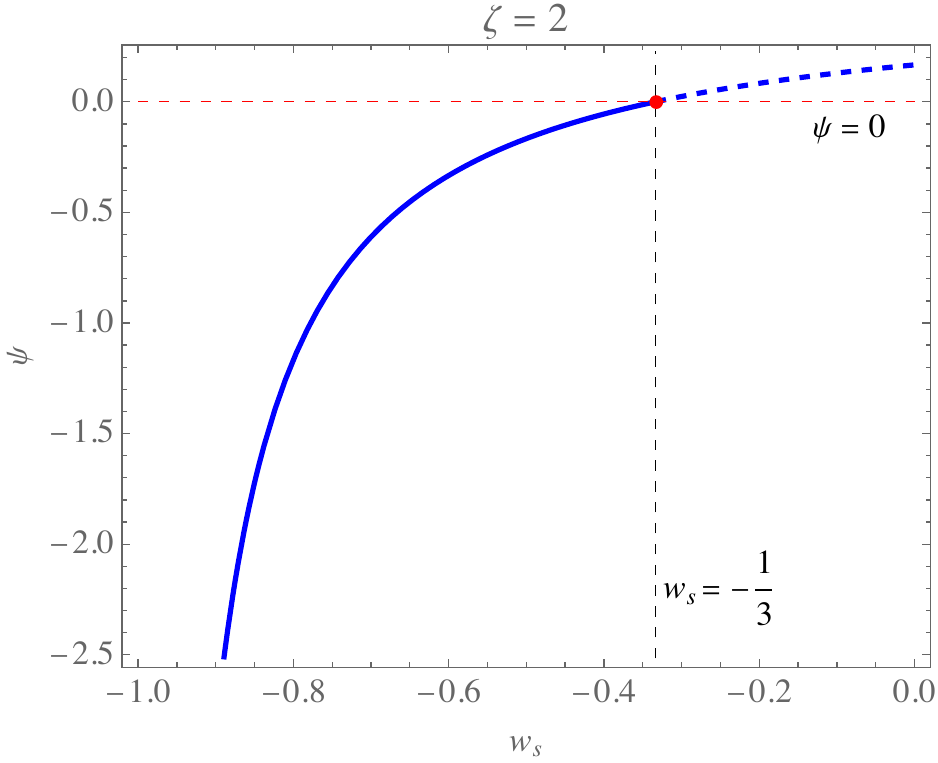}~(a)\qquad\qquad
 \includegraphics[width=6cm]{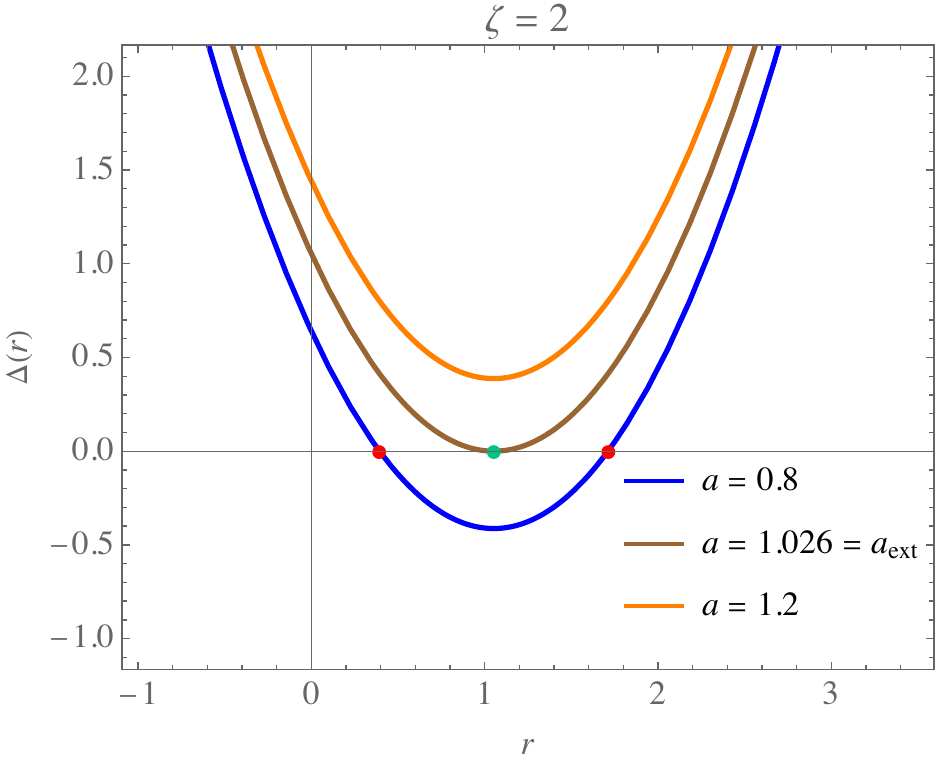}~(b)
	\caption{(a) The relationship between $\psi$ and $w_s$, and (b) the radial structure of $\Delta(r)$ exhibited across three distinct spin parameters, under the setting of $\zeta = 2$ and $N_s = 0.05$.}
	\label{fig:zeta2}
\end{figure}

In this case, the horizons are found at
\begin{equation}
r_\mp={M\over 1-N_s}\mp\sqrt{\left( {M\over 1-N_s}\right) ^2-{a^2\over 1-N_s}},
\label{eq:rHzeta2}
\end{equation}
which implies that the existence of the KRBH corresponds to the condition $a^2(1-N_s)< M^2$, and hence, $a_{\mathrm{ext}}=M/\sqrt{1-N_s}$, implying that $r_{\mathrm{ext}}=M/(1-N_s)$. As demonstrated in Fig.~\ref{fig:zeta2} for a quintessential background fluid, the above choice of $\zeta$ confines the Rastall coupling inside the domain $-\infty<\psi<0$. 
It is evident that there are no cosmological horizons in this particular scenario. In this chosen $\zeta$ configuration, the ergoregions are delineated by the radii of the static limits, expressed as
\begin{equation}\label{eq:rSLzeta2}
r_{\mathrm{SL}_\mp}(\theta) = {M\over 1-N_s}\mp\sqrt{\left( {M\over 1-N_s}\right) ^2-{a^2\cos^2\theta\over 1-N_s}}.
\end{equation}
\begin{figure}[!h]
\centering
 \includegraphics[width=5.3cm]{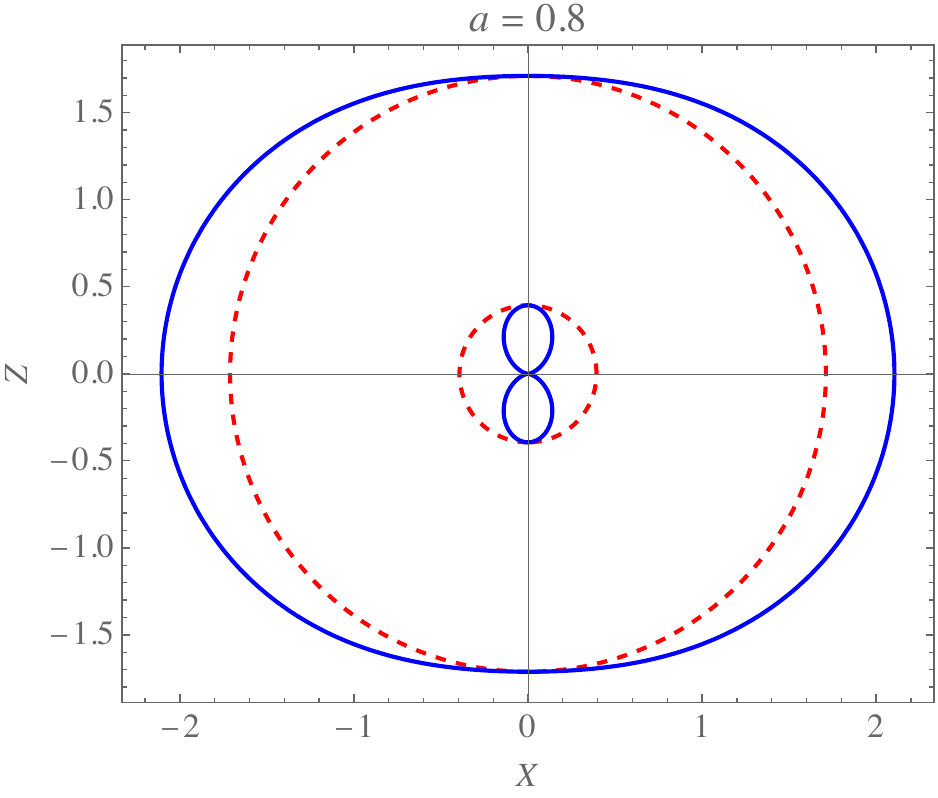}~(a)
 \includegraphics[width=5.3cm]{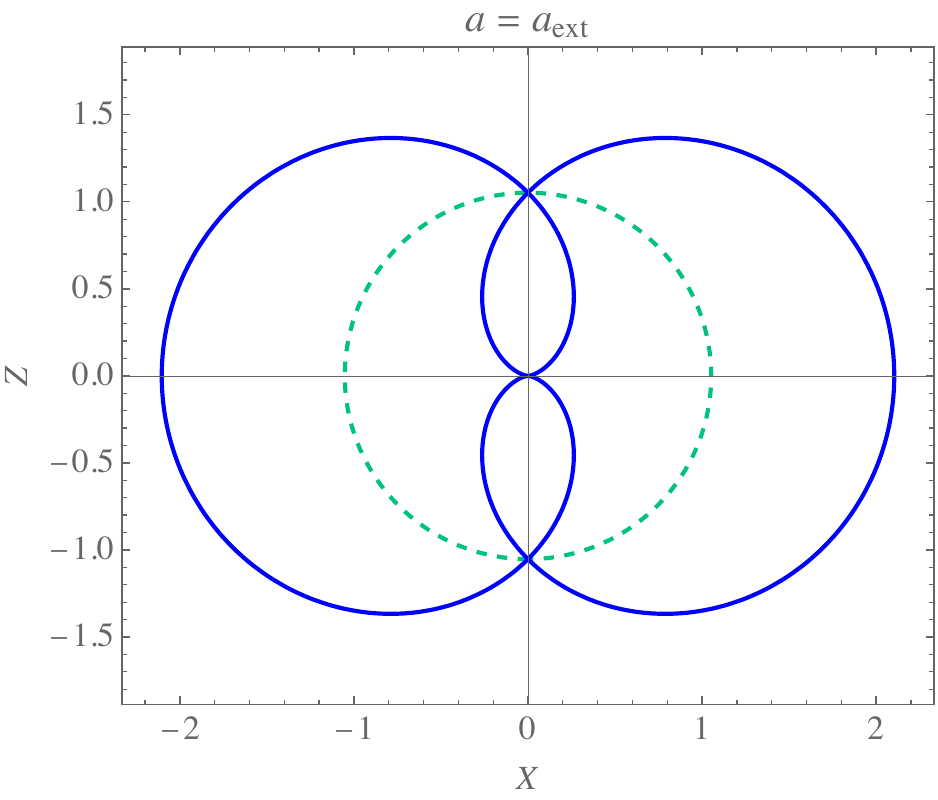}~(b)
 \includegraphics[width=5.3cm]{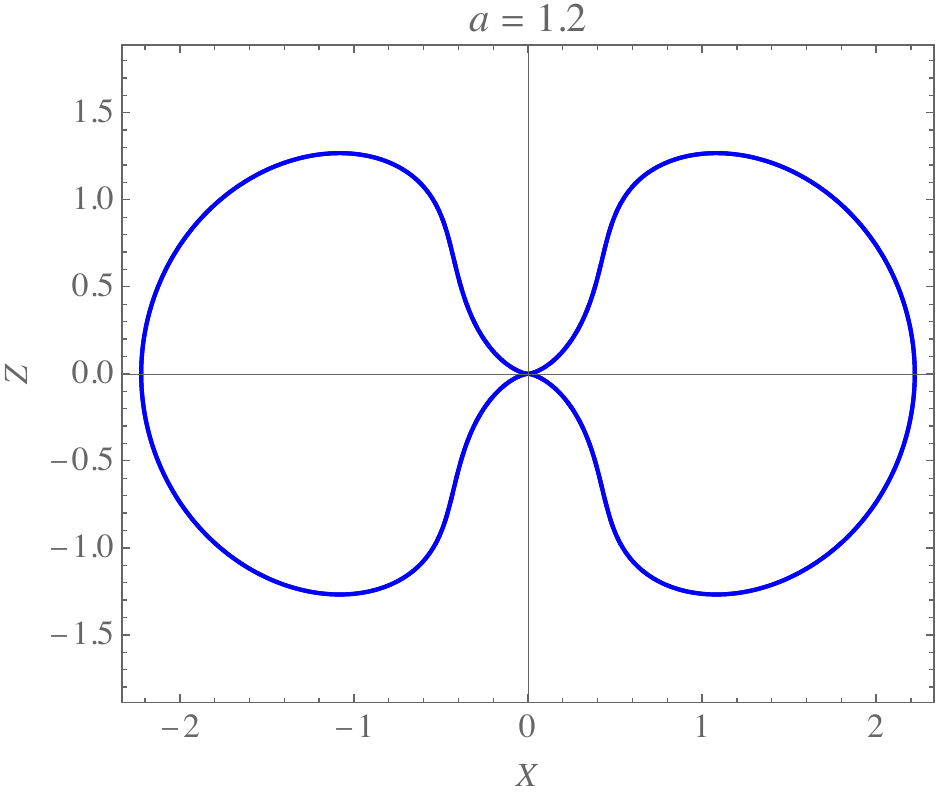}~(c)
	\caption{The cross-section of the ergoregions of the KRBH for three different spin parameters, corresponding to the choice of $\zeta=2$. In this particular case, $a_{\mathrm{ext}}=1.026$.}
	\label{fig:rSLzeta2}
\end{figure} 
In Fig.~\ref{fig:rSLzeta2}, the cross-section of the corresponding ergoregions has been plotted in the $Z$-$X$ plane.

\subsubsection{The case of $\zeta = 3$}\label{subsub:zeta3}

\begin{figure}[!h]
\centering
 \includegraphics[width=6cm]{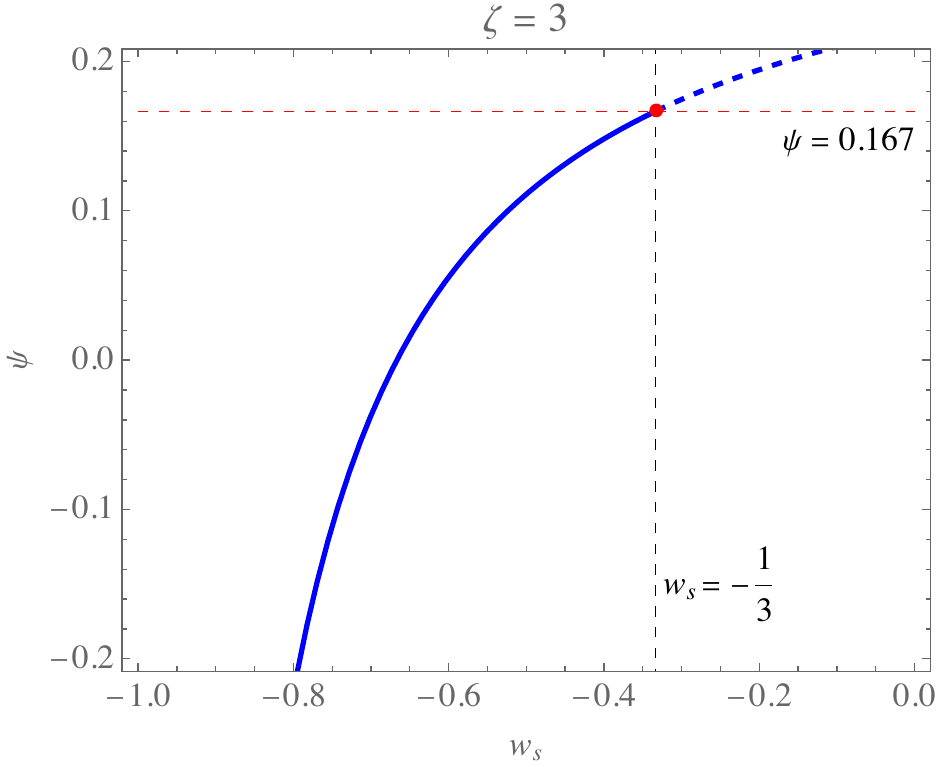}~(a)\qquad\qquad
 \includegraphics[width=6cm]{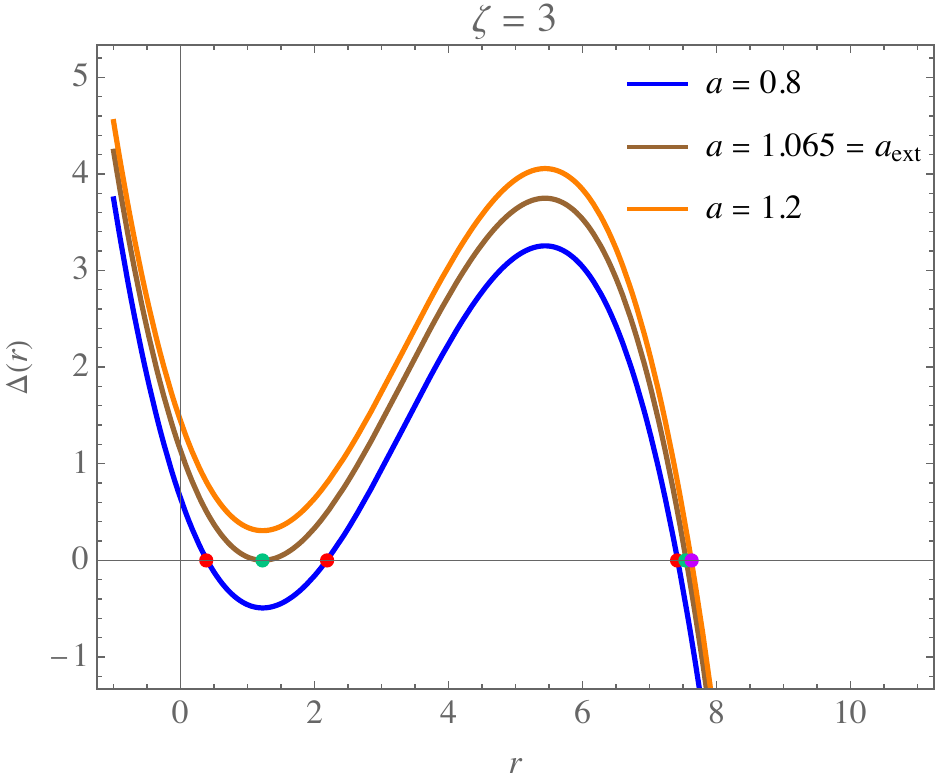}~(b)
	\caption{(a) The correlation between $\psi$ and $w_s$, and (b) the radial profile of $\Delta(r)$ for three distinct spin parameters, for the case of $\zeta = 3$ and $N_s = 0.1$.}
	\label{fig:zeta3}
\end{figure}

\begin{figure}[!h]
\centering
 \includegraphics[width=5.3cm]{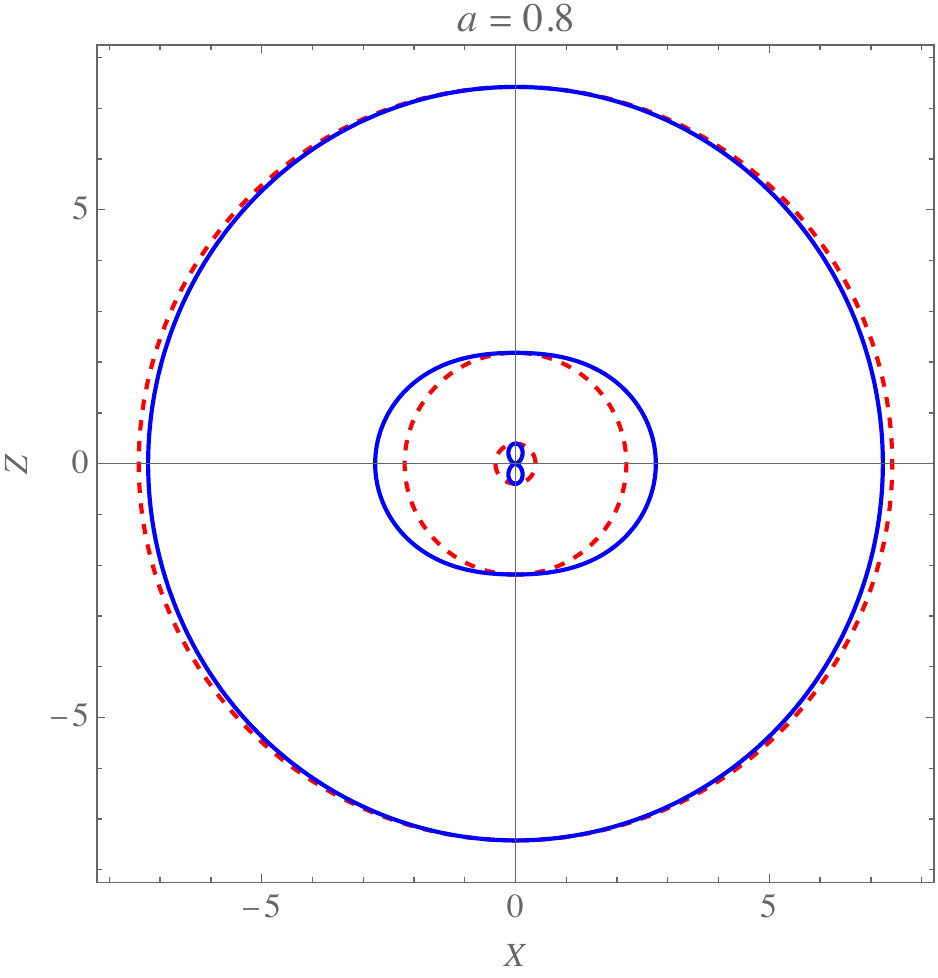}~(a)
 \includegraphics[width=5.3cm]{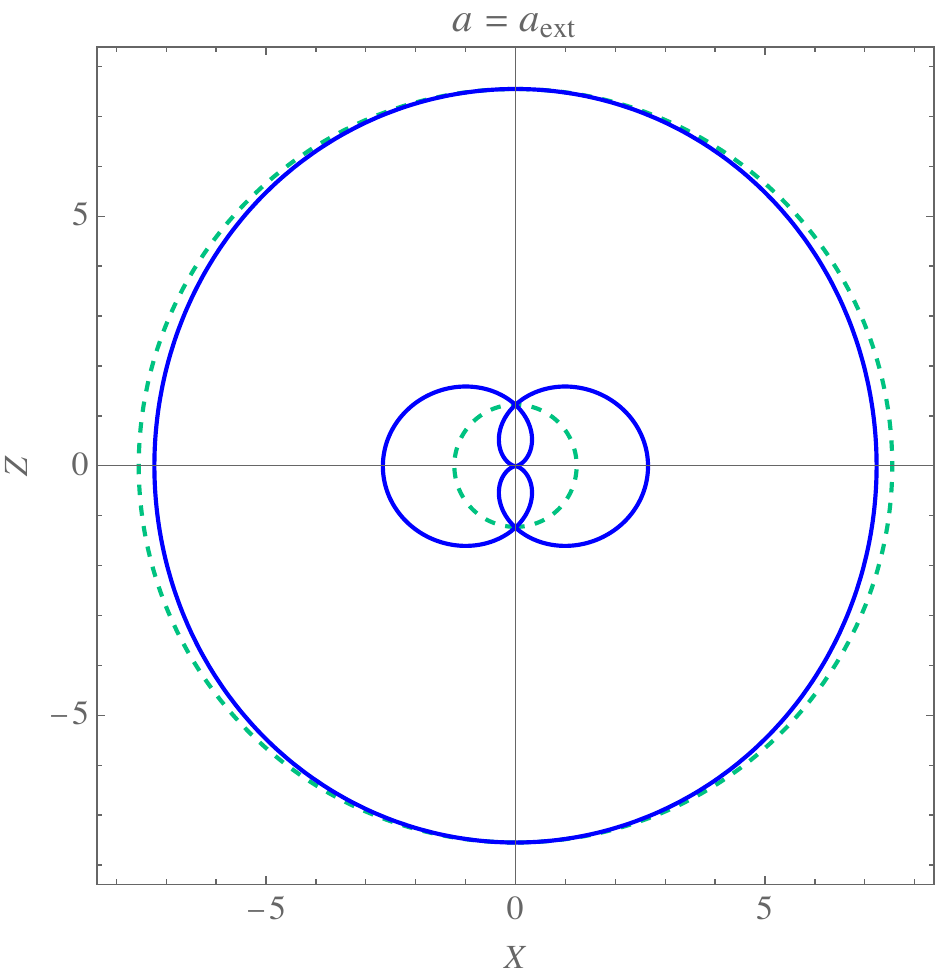}~(b)
 \includegraphics[width=5.3cm]{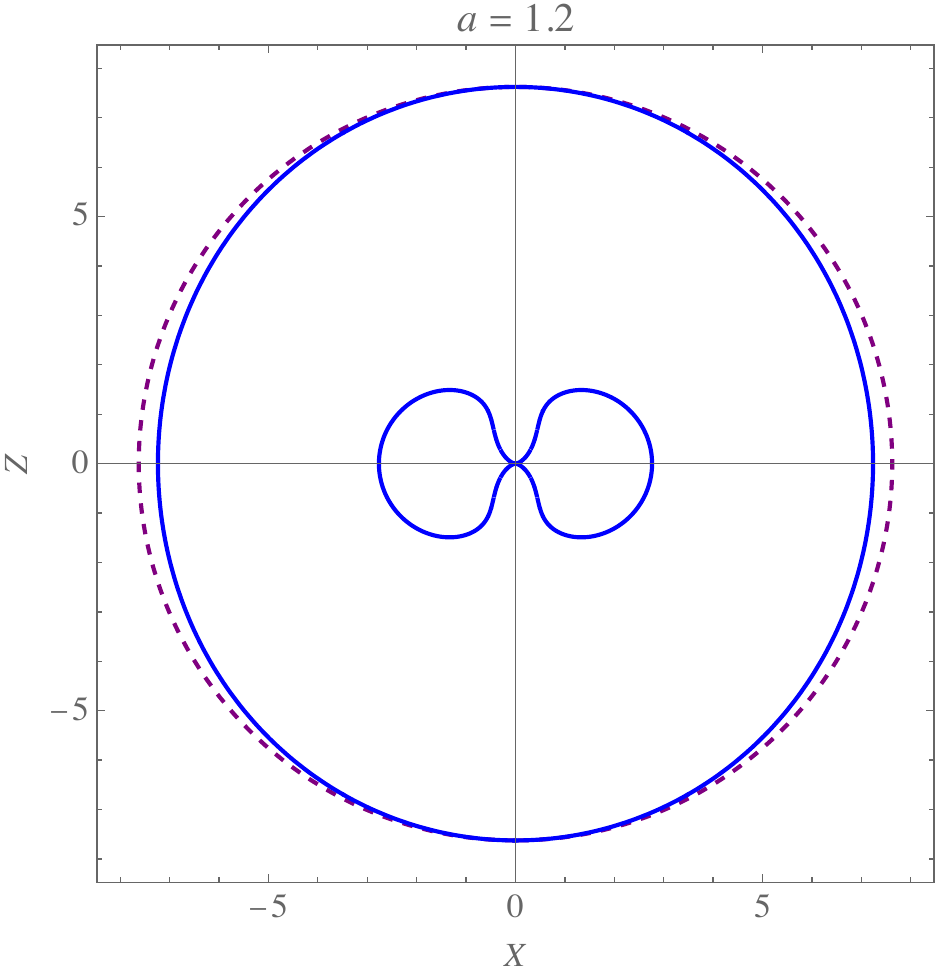}~(c)
	\caption{The cross-section of the ergoregions of the KRBH for three different spin parameters, corresponding to the choice of $\zeta=3$. In panel (a), the three horizons are available separately for the KRBH and have been indicated by red dashed circles. Hence there will be three ergoregions.  The extremal case is shown in panel (b) for $a_{\mathrm{ext}}=1.065$, where $r_-=r_+$ and $r_{++}$ have been indicated by green dashed circles. There will be two ergoregions in this case. Finally, in panel (c) a naked singularity is shown with only $r_{++}$ being available, which is indicated by a purple dashed circle. As in the previous cases, there is an ergoregion connected to the singularity, but in this case, there will be an exterior ergoregion formed together with the cosmological horizon.}
	\label{fig:rSLzeta3}
\end{figure}

{We can expect the emergence of the cosmological (quintessential) horizon when $\zeta=3$, as it allows us to capture all feasible real solutions of the cubic equation $\Delta=0$.} This results in the three solutions
\begin{eqnarray}
&& r_{+}=\sqrt{\frac{g_2}{3}} \cos \left(\frac{1}{3}\arccos \left(3  g_3\sqrt{\frac{3}{g_2^3}}\,\right)+\frac{4 \pi }{3}\right)+\frac{1}{3 N_s},\label{eq:re}\\
&&r_{++}=\sqrt{\frac{g_2}{3} } \cos \left(\frac{1}{3} \arccos\left(3 g_3\sqrt{\frac{3}{g_2^3}} \,\right)\right)+\frac{1}{3 N_s},\label{eq:rc} \\
&&r_{-}=\sqrt{\frac{g_2}{3} } \cos \left(\frac{1}{3} \arccos\left(3 \sqrt{\frac{3}{g_2^3}}\, \right)+\frac{2 \pi  }{3}\right)+\frac{1}{3 N_s},\label{eq:rC}
\end{eqnarray}
where
\begin{subequations}
\begin{align}
& g_2=4 \left(\frac{1}{3 N_s^2}-\frac{2 M}{N_s}\right),\label{eq:g2} \\
& g_3=4 \left(\frac{a^2}{N_s}-\frac{2 M}{3 N_s^2}+\frac{2}{27 N_s^3}\right).\label{eq:g3}
\end{align}
\label{eq:gs}
\end{subequations}
It is straightforward to verify that the discriminant of the cubic is always non-negative, i.e. $\delta_\Delta=16 \left(g_2^3-27 g_3^2\right)\geq0$. Consequently, all the aforementioned roots are real. Moreover, as depicted in Fig.~\ref{fig:zeta3}(a), within a quintessential fluid background, the Rastall coupling is constrained within the domain $-\infty<\psi<1/6=0.167$. Within this range, all three zeros of $\Delta(r)=0$ will be positive (see Fig.~\ref{fig:zeta3}(b)). 
Accordingly, the radii \eqref{eq:re}--\eqref{eq:rc} indicate, respectively, the event, cosmological, and Cauchy horizons of the KRBH for the above choice of $\zeta$. Hence, the extremal KRBH black hole with two horizons (a cosmological horizon and an event horizon) can be identified in terms of the equation $\delta_{\Delta}=0$. This equation provides 
\begin{equation}
a^2_{\mathrm{ext}} = \frac{2}{27} \left(\frac{\sqrt{(1-6 M N_s)^3}}{N_s^2}+\frac{9 M}{N_s}-\frac{1}{N_s^2}\right),
    \label{eq:aext3}
\end{equation}
as the extremal spin parameter. Therefore, the KRBH with three horizons is present when $a\geq a_{\mathrm{ext}}$, while the naked singularity with a solitary cosmological horizon emerges when $a\leq a_{\mathrm{ext}}$ (refer to Fig. \ref{fig:zeta3}(b)). Similar to previous cases, the static limit can also be identified. However, for $\zeta=3$, an additional ergoregion forms within the cosmological horizon. Building upon the findings of the prior two cases, it is evident that the radii of the static limits are $r_{\mathrm{SL}_{++}}$, $r_{\mathrm{SL}_{+}}$, and $r_{\mathrm{SL}_{-}}$. These radii share identical expressions as $r_{++}$, $r_{+}$, and $r_{-}$, respectively, considering the exchange $a^2\rightarrow a^2\cos^2\theta$ within the $g_3$ coefficient in Eq. \eqref{eq:g3}. These expressions are employed in Fig.~\ref{fig:rSLzeta3} to illustrate the ergoregions for the three instances of the KRBH, the extremal KRBH, and the naked singularity.
It is evident that the choice of $\zeta=3$ provides more diversity in the spacetime structure and it is of more completeness. Hence, in the rest of the paper, we consider this choice as our reference and we base our analytical and numerical studies on the metric function 
\begin{equation}
\Delta(r) = r^2+a^2-2Mr-N_s r^3 = N_s \left(r_{++}-r\right)\left(r-r_+\right)\left(r-r_-\right),
    \label{eq:Delta_zeta3}
\end{equation}
in accordance with the characteristic radii given in Eqs.~\eqref{eq:re}--\eqref{eq:rC}.
Moving forward, we turn our attention to the inclusion of a plasmic medium around the KRBH and proceed to present the corresponding equations of motion in the following section.

\section{Light propagation in a non-magnetized plasma}
\label{sec:HJEOM}

In this section, we introduce some salient aspects of plasma effects on the light ray trajectories that pass a black hole spacetime. In fact, we regard the plasma as a medium that is only discernible by its influence on the rays so the wave optics descriptions are neglected here. It is well-known that inside a plasma, light rays do not travel on null geodesics, and in fact, it is the frequency \cite{perlick_calculating_2022}
\begin{equation}\label{eq:omegap}
\omega_p(\bm{x}) = \frac{4 \pi e^2}{m_e} N_p(\bm{x}),
\end{equation}
of an electron-ion plasma, that affects the propagation of light in such a medium. In the above relation, $e$, $m_e$, and $N_p$ are, respectively, the electron charge, electron mass $m_e$, and the electron number density $N_p$, in the coordinate system $\bm{x}\equiv x^{\mu}=(x^0,x^1,x^2,x^3)$. Here, we assume that the plasma is non-magnetized and pressure-less so the only factor that affects the light rays is the electron density. In order to formulate the equations of motion for light rays that propagate inside this medium, we rely on the phase space dynamics and introduce the Hamiltonian
\begin{equation}\label{H1}
\mathcal{H}(\bm{x},\bm{p})=\frac{1}{2}\left[g^{\mu \nu}(\bm{x})p_{\mu}p_{\nu}+
\omega_p^2(\bm{x})\right],
\end{equation}
defined on the cotangent bundle of the Riemannian manifold $(\mathcal{M},\bm{g})$, with $\bm{p}\equiv p_{\mu}=(p_0,p_1,p_2,p_3)$ being the canonical momentum covector. In this regard, the propagation of the light rays is identified as solution to the canonical equations 
\begin{eqnarray}\label{EqsH}
	&& {\ed x^{\mu}\over \ed\lambda}=\frac{\partial \mathcal{H}}{\partial p_{\mu}},\label{EqsH1}\\
	&&{\ed p_{\mu}\over \ed\lambda}=-\frac{\partial \mathcal{H}}{\partial x^{\mu}},\label{EqsH2}\\
	&&\mathcal{H}=0\label{EqsH3},
\end{eqnarray}
in which, $\lambda$ is the curve parameter along the trajectories. In fact, Breuer and Ehler were the first to derive the Hamiltonian \eqref{H1} from Maxwell's equations in a general relativistic spacetime, while using a model of the plasma involving two charged fluids \cite{Breuer:1980}.  

Since the light rays travel on time-like trajectories with respect to spacetime manifold's metric, it is advantageous to consider a comoving observer with the four-velocity $\bm{u}$, who measures the light ray frequency $\omega(\bm{x})$, and characterize the plasma by the refractive index 
\begin{equation}\label{eq:rn2}
n^2(\bm{x}) = 1 + \frac{p_\mu p^\mu}{(p_\nu u^\nu)^2}
= 1-\frac{\omega_p^2(\bm{x})}{\omega^2(\bm{x})},
\end{equation}
in writing which, we have used the identity $p_\nu u^\nu = p_0/\sqrt{g_{00}} = -\omega$, corresponding to an observer located on the $x^0$ curves, having the four-velocity $u^\mu = {\delta_0^\mu}/{\sqrt{-g_{00}}}$. This way, one can recast the Hamiltonian as
\begin{equation}\label{eq:hamilton_1}
\mathcal{H} = \frac{1}{2}\left[
g^{\mu\nu} p_\mu p_\nu + (n^2-1)(p_\nu u^\nu)^2
\right] = \frac{1}{2}\mathfrak{o}^{\mu\nu} p_\mu p_\nu,
\end{equation}
where $\mathfrak{o}^{\mu\nu}$ is the optical metric associated with the the refractive index $n\equiv n(\Vec{x},\omega)$. Contributing the Jacobi action $\mathcal{S}$, with the properties 
\begin{subequations}\label{eq:StoP}
	\begin{align}
	&  p_\mu = \frac{\partial\mathcal{S}}{\partial x^\mu},\label{eq:StoP_1}\\
	&  \mathcal{H} = \frac{\partial\mathcal{S}}{\partial\lambda},\label{eq:StoP_2}
	\end{align}
\end{subequations}
one can rewrite the Hamilton-Jacobi equations as \cite{Atamurotov:2015,Perlick:2017}
\begin{eqnarray}\label{eq:hamilton_2}
\frac{\partial\mathcal{S}}{\partial\lambda} &=& -\frac{1}{2}\left[
g^{\mu\nu} \frac{\partial\mathcal{S}}{\partial x^\mu} \frac{\partial\mathcal{S}}{\partial x^\nu} - \left(n^2-1\right)\omega^2
\right]\nonumber\\
&=& -\frac{1}{2}\left[
g^{\mu\nu} \frac{\partial\mathcal{S}}{\partial x^\mu} \frac{\partial\mathcal{S}}{\partial x^\nu} +\omega_p^2
\right].
\end{eqnarray}
In fact, the study of light propagation in dispersive media with frequency-dependent refractive indices within the framework of general relativity was pioneered by Synge \cite{Synge:1960}, although he did not consider the case of the plasma (see also Refs.~\cite{Bicak:1975,PhysRevD.45.525}). If the spacetime exhibits spherical symmetry, the plasmic frequency will vary with respect to the radial distance $\omega_p\equiv\omega_p(r)$, and hence, the propagation of light can be determined through appropriate integrations \cite{Muhleman:1966,Perlick_ray_2000}. The gravitational deflection of light in homogeneous plasma has been studied in Refs.~\cite{bisnovatyi-kogan_gravitational_2009,bisnovatyi-kogan_gravitational_2010,Tsupko:2013}.  It has been shown in Ref.~\cite{perlick_light_2017} that the separation of the Hamilton-Jacobi equations in the Kerr spacetimes is possible, only if the plasma frequency adopts the form 
\begin{equation}
\omega_p^2\equiv\omega_p^2(r,\theta)= {f_r(r)+f_{\theta}(\theta)\over r^2+a^2\cos^2\theta},
\label{plasma2}
\end{equation}
in which $f_r$ and $f_\theta$ can have arbitrary expressions. As we will see in what follows, this holds, as well, for the case of the KRBH. Such an expression has been considered in Ref.~\cite{fathi_analytical_2021} to analyze the light ray trajectories around a Kerr black hole in the presence of an inhomogeneous plasma. Note that, the form \eqref{plasma2} is also of crucial importance in the analytical determination of the black hole shadow boundary, as shown in Refs.~\cite{perlick_light_2017,Yan:2019}. On the other hand, when the separation of the Hamilton-Jacobi equation is not at hand, one has to make numerical implementation in order to determine the shadow \cite{huang_revisiting_2018}. However, since the current discussion mainly deals with the analytical description of the light ray trajectories rather than the shadow, we do not continue this review and proceed with the determination of the equations of motion.

\subsection{Separation of the Hamilton-Jacobi equation and the equations of motion}\label{subsec:EqMotion}

The method of the separation of the Hamilton-Jacobi equation is based on rewriting the action $\mathcal{S}$ as \cite{Carter:1968,Chandrasekhar:2002}
\begin{equation}\label{eq:S_sep}
\mathcal{S} = -E t + L \phi + \mathcal{S}_r(r) + \mathcal{S}_\theta(\theta) + \frac{1}{2} m^2 \lambda, 
\end{equation}
in which, the energy $E$ and the angular momentum $L$ are the constants of motion specific to the Hamilton equations, and $m$ is the mass of the test particles (here, $m=0$). In fact, applying the line element \eqref{metr} in the Hamiltonian \eqref{H1}, we get 
\begin{equation}\label{H2}
\mathcal{H}(\bm{x},\bm{p}) = {1\over 2\rho^2}\left[  \Delta\, p_r^2+p_{\theta}^2+\left(a p_t\sin \theta+{p_{\phi} \over\sin \theta} \right) ^2 - {1\over\Delta}\Big(p_t\left(r^2+a^2\right)
+ap_{\phi}\Big)^2+\rho^2 \omega_p^2\right],
\end{equation}
in which, the components of the four-momentum are determined in alignment with Eqs.~\eqref{eq:StoP_1} and \eqref{eq:S_sep}, taking into account the expressions 
\begin{subequations}\label{ctesmov}
	\begin{align}
	& p_t = \frac{\partial\mathcal{S}}{\partial t} = - E  =  - \omega_0,\\
	& p_\phi =\frac{\partial\mathcal{S}}{\partial\phi} = L,
	\end{align}
\end{subequations}
for the constants of motion. It is then clear that $p_\phi$ corresponds to the axial symmetry of the spacetime, while $\omega_0$ could be regarded as the light rays' frequency. Hence, the constants of motion $E$ and $L$, together with the eikonal equation $\mathcal{H}=0$, characterize the propagation of light. Now from Eqs.~\eqref{H2} and \eqref{ctesmov}, the eikonal equation takes the form
\begin{equation}\label{H3}
0 =  \Delta \,p_r^2+\,p_{\theta}^2+\left(\omega_0\,a\sin \theta-{L \over\sin \theta} \right) ^2 
- {1\over\Delta}\left[\omega_0\left(r^2+a^2\right)
-aL\right]^2+\rho^2\omega_p^2,
\end{equation}
from which, it is evident that the separability to the $r$-dependent and $\theta$-dependent segments, is only possible when the condition \eqref{plasma2} is respected. By exploiting the identity 
\begin{equation}
\left(\omega_0\,a\sin \theta-{L \over\sin \theta} \right) ^2=
\left(L^2\csc^2\theta-a^2\omega_0^2\right)\cos^2\theta\\
+(L-a\omega_0)^2,
\end{equation}
together with Eqs.~\eqref{H3} and \eqref{plasma2}, we get the separated form
of the Hamilton-Jacobi equation, which reads as
\begin{eqnarray}\label{H4}
\mathscr{Q}&=&  \,p_{\theta}^2+\left(L^2\csc^2\theta-a^2\omega_0^2\right)\cos^2\theta+f_{\theta}(\theta)\nonumber \\
&=&- \Delta \,p_r^2+{1\over\Delta}\left[\omega_0\left(r^2+a^2\right)
-aL\right]^2-(L-a\omega_0)^2-f_{r}(r),
\end{eqnarray}
where $\mathscr{Q}$ is the Carter's constant. Accordingly, one can write
\begin{eqnarray}
&& p_{\theta}^2 = \mathscr{Q}-\left(L^2\csc^2\theta-a^2\omega_0^2\right)\cos^2\theta-f_{\theta}(\theta),\label{H5}\\
&& \Delta\, p_r^2 = {1\over\Delta}\left[\omega_0\left(r^2+a^2\right)
-aL\right]^2-\mathscr{Q}-(L-a\omega_0)^2
-f_{r}(r).\label{H5an}
\end{eqnarray}
Incorporating Eqs. \eqref{H5} and \eqref{H5an} into the Hamiltonian \eqref{H2}, and then applying Eq.~\eqref{EqsH1}, yields the first order differential equations of motion as follows:
\begin{eqnarray}\label{basiceqs}
&& \rho^2\frac{\ed r}{\ed\lambda}=\sqrt{\mathcal{R}(r)},\label{basiceqsr}\\ &&\rho^2\frac{\ed\theta}{\ed\lambda}=\sqrt{\Theta (\theta)},\label{basiceqtheta}\\
&&\rho^2\frac{\ed\phi}{\ed\lambda}={ L\left(\Delta-a^2\sin^2\theta\right)\csc^2\theta+a\omega_0\left(r^2+a^2-\Delta\right)\over \Delta}, \label{basiceqphi} \\
&&\rho^2\frac{\ed t}{\ed\lambda}={ \omega_0\Sigma^2-aL\left(r^2+a^2-\Delta\right)\over \Delta}, \label{basiceqst}
\end{eqnarray}
where
\begin{subequations}\label{basiceqs2}
	\begin{align}
	& \mathcal{R}(r) = \left[\omega_0\left( r^2+a^2\right) -aL\right] ^2-\Delta\left[ \mathscr{Q}+f_r(r)+(L-a\omega_0)^2\right],\label{basiceqs2b}\\
	& \Theta(\theta) = \mathscr{Q}-f_{\theta}(\theta)-\cos^2\theta\left(L^2\csc^2\theta-a^2\omega_0^2\right),\label{basiceqs2c} \\
	& \Sigma^2=\left( r^2+a^2\right)^2-\Delta \,a^2\sin^2\theta.\label{basiceqs2d}
	\end{align}
\end{subequations}
However, in order to facilitate the mathematical manipulations in what follows, we work with the dimension-less \textit{Mino time}, $\gamma$, which is defined in terms of the relation $\rho^2\ed\gamma=M\ed\lambda$ \cite{Mino:2003}. In this context, the equations of motion adopt the forms
\begin{eqnarray}
\label{basiceqsR}
M{\ed r\over \ed \gamma}&=&\sqrt{\mathcal{R}(r)},\\ 
\label{basiceqstheta}
M{\ed\theta\over \ed \gamma}&=&\sqrt{\Theta (\theta)},\\
\label{basiceqsphi}
M{\ed\phi\over \ed\gamma}&=&{ L\left(\Delta-a^2\sin^2\theta\right)\csc^2\theta+a\omega_0\left(r^2+a^2-\Delta\right)\over \Delta}, \\
\label{basiceqste}
M{\ed t\over \ed \gamma}&=&{ \omega_0\Sigma^2-aL\left(r^2+a^2-\Delta\right)\over \Delta}. 
\end{eqnarray}
We extend our discussion in the next section by implementing a specific criterion for the frequency $\omega_p$. In line with the form \eqref{plasma2}, we construct an inhomogeneous anisotropic plasma. Subsequently, we analyze the aforementioned equations of motion to obtain precise analytical solutions for the evolution of the coordinates. These solutions determine the trajectories of light rays around the KRBH, considering the case where $\zeta=3$ as previously mentioned in Sect. \ref{sec:BHsol}.

\section{General analysis of the equations of motion}
\label{sec:eqofmot}

The separation condition stated in Eq.~\eqref{plasma2} defines the plasma and its geometric distribution in the spacetime surrounding the black hole. Consequently, the characteristic functions $f_r(r)$ and $f_\theta(\theta)$ assume significant roles in delineating the configuration of the plasma. In this section, consistent with the approach outlined in Ref. \cite{fathi_analytical_2021}, we constrain the light rays to traverse through an inhomogeneous anisotropic plasma by making the specific choices
\begin{eqnarray}
&& f_r(r)\equiv f_r= \mathrm{constant},\label{H61}\\
&& f_{\theta}(\theta)\equiv f_\theta=\mathrm{constant.}\label{H62}
\end{eqnarray} 
Furthermore, within the text, we use, frequently, the conventions \cite{Chandrasekhar:2002}
\begin{eqnarray}
&& \xi={L\over \omega _0},\label{eq:xi}\\
&& \eta= {\mathscr{Q}\over\omega _0^2},\label{eq:eta}
\end{eqnarray}
in order to simplify the analysis. In what follows, the temporal evolution of the spacetime coordinates are analyzed separately for this plasmic medium, and the relevant exact solutions are presented.

\subsection{Latitudinal motion}\label{subsec:thetamotion}

\begin{figure}
    \centering
    \includegraphics[width=8cm]{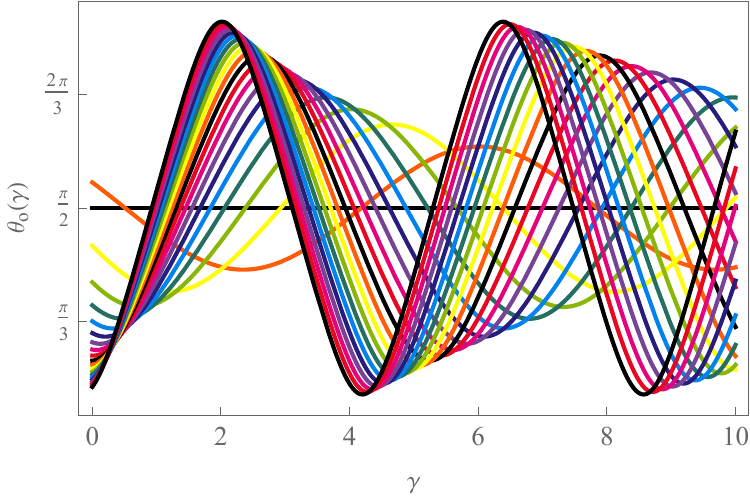} (a)\quad
    \includegraphics[width=8cm]{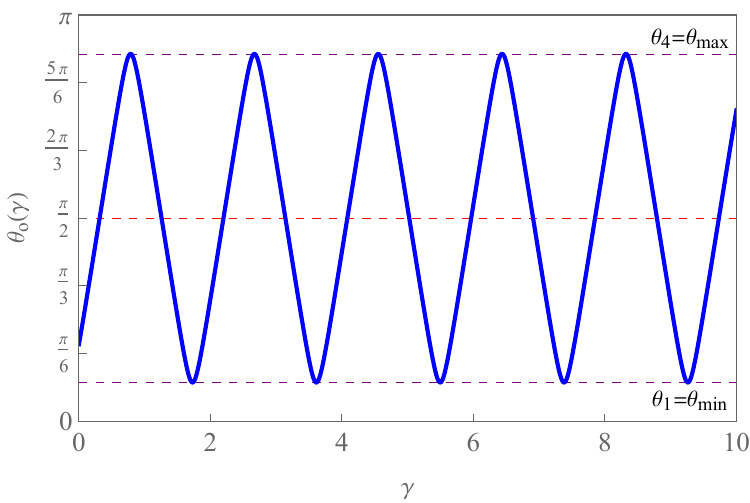} (b)
    \caption{The oscillating evolution of the $\theta$-coordinate in the Mino time, by considering $a=0.8$, $L=1$ and $\omega_0=0.75$ (corresponding to $\xi=1.33$), and $\theta_\s=\pi/2$. Panel (a) shows a table of graphs for $\theta_o(\gamma)$ for the impact parameter range $\tet\in[0,2]$ by $0.1$ steps. As it is evident, by decreasing $\tet$, the oscillations approach the $\theta=\pi/2$ line until they finally stop oscillating (planar orbits). In panel (b) a typical graph has been plotted by taking into account $\Q=9$ and $f_\theta=1$ (corresponding to $\tet=17.77$). In this diagram, the roots of the angular potential have been shown as the maximum and minimum values of the $\theta$-coordinate oscillations.}
    \label{fig:theta_o}
\end{figure}

To analyze the evolution of the polar coordinate $\theta$, we need to examine the angular potential $\Theta(\theta)$. To proceed, we apply the change of variable $z=\cos^2\theta$, so that the potential adopts the form
\begin{equation}
\Theta_z(z)=\frac{\omega_0^2}{1-z}\left[-a^2 z^2+\tilde{\chi} z +\tilde\eta\right],
    \label{eq:Thetaz}
\end{equation}
in which we have notated $\tilde\eta = \eta-f_\theta/\omega_0^2$ and $\tilde\chi=a^2-\tilde{\eta}-\xi^2<0$. This potential is consequently second order in $z$, and its roots are obtained as
\begin{equation}
z_\pm=\frac{\tilde{\chi}\pm\sqrt{4a^2\tilde\eta+\tilde{\chi}^2}}{2 a^2},
    \label{eq:zpm}
\end{equation}
providing $z_+z_-=-\tilde\eta/a^2$. This way, one can recast the angular potential as
\begin{equation}
    \Theta_z(z) = \frac{\omega_0^2}{1-z}\left[a^2(z_+-z)(z-z_-)\right].
    \label{eq:Thetaz_1}
\end{equation}
Accordingly, the evolution of the polar angle may encounter four turning points, which are given by
\begin{eqnarray}
&&\theta_1=\arccos(\sqrt{z_+}),\label{eq:theta1}\\
&&\theta_2=\arccos(\sqrt{z_-}),\label{eq:theta2}\\
&&\theta_3=\arccos(-\sqrt{z_-}),\label{eq:theta3}\\
&&\theta_4=\arccos(-\sqrt{z_+}).\label{eq:theta4}
\end{eqnarray}
The nature of the roots becomes evident through the positivity or negativity of $z_\pm$ or the sign of $\tilde\eta$. Therefore, in the subsequent discussion, we categorize the latitudinal motion into three cases.
\begin{itemize}
    \item[(i)] \underline{$\tet>0$ (ordinary trajectories)}: In this case $z_+ z_-<0$. On the other hand, it is immediately inferred from Eq. \eqref{eq:zpm} that $z_+>0$ and $z_-<0$. Accordingly, $\theta_{2,3}\in\Bbb{C}$ and $\theta_{1,4}\in\Bbb{R}$. By analyzing the latter case, we find that $\theta_1<\pi/2<\theta_4$, and hence, the polar coordinate oscillates between these two limits in the course of its evolution, and passes the equatorial plane in each oscillation.

   \item [(ii)] \underline{$\tet=0$ (planar trajectories)}: It is straightforward to see that, in this case, $z_+ = 0$ and $z_-=\tch/a^2<0$. Hence $\theta_1=\theta_4=\pi/2$ and $\theta_{2,3}\in\Bbb{C}$. In this regard, the orbits are confined in the equatorial plane.

   \item [(iii)] \underline{$\tet<0$ (vortical trajectories)}: Under this condition, $z_+ z_->0$, which based on the fact that $\tch<0$, then it holds $z_\pm<0$, and all the roots $\theta_i$ are imaginary. Hence, this case is not physically valid and is ruled out from consideration.

\end{itemize}
Accordingly, to obtain the analytical solution to the evolution of the $\theta$-coordinate, we consider the case of $\tet\geq0$. Assuming a source located at the angular position $\theta_\mathrm{s}$ and an observer at $\theta_\mathrm{o}$, the Mino time can be obtained by means of the equation of motion \eqref{basiceqstheta} and the angular potential \eqref{eq:Thetaz_1}, which after proper inversion, results in the analytical evolution relation
\begin{equation}
\theta_\ob(\gamma)=\arccos\bigg(-\sqrt{z_+}\sn\Big(
\frac{z_0}{M}\gamma-\bF(\varphi_\s|\m)\Big|\m
\Big)
\bigg),
    \label{eq:thetao_tau}
\end{equation}
in which $z_0=a\omega_0\sqrt{-z_-}$, $\m=z_+/z_-$ is the elliptic modulus, $\sn(x|\m)$ is the Jacobian elliptic sine function with argument $x$, $\bF(\varphi_s|\m)$ is the incomplete elliptic integral of the first kind with argument $\varphi_s$ \cite{byrd_handbook_1971}, and
\begin{equation}
        \varphi_\s = \arcsin\left(\frac{\cos{\theta_\s}}{\sqrt{z_+}}\right).
        \label{eq:varphi_s}
\end{equation}
Based on the periodicity of the elliptic sine function, as expected, the $\theta$-coordinate will oscillate between its maximum and minimum values in the course of full trajectories. This has been shown in Fig. \ref{fig:theta_o} for some specific examples.

\subsection{Radial motion}\label{subsec:r}

In the analysis of particle trajectories in curved spacetimes, a fundamental aspect involves comprehending the approach and recession of particles influenced by gravity. Traditionally, this exploration revolves around computing the radial effective gravitational potential governing these particles. Consequently, when scrutinizing the motion of light rays in our specified context, our focus shifts to the radial equation of motion in Eq. \eqref{basiceqsR}. To advance our study, this subsection concentrates on solving the aforementioned evolution equation under different conditions, leading to various types of motion. We will discuss, separately, the deflecting trajectories (both bound and unbound), captured trajectories, and the critical orbits (both bound and unbound). However before delving into the analytical solution of the radial equation of motion, let us consider some aspects of \textit{spherical} photon orbits around the black hole.

\subsubsection{Orbits of constant radius around the KRBH}

Around black holes with a non-zero spin parameter, a multitude of light rays follow trajectories where the radial distance $r$, remains constant. These rays essentially skim the surface of a spherical structure, and carry a critical nature, given that they either escape from  the black hole or plunge into it (as further explained below). Consequently, they constitute a major portion of a more general form of orbits, termed as the \textit{critical orbits} (see below). Within these orbits, the $\theta$-coordinate oscillates between the turning points of the angular potential (as elaborated in the last subsection). For Kerr black holes, these particular orbits have been extensively examined in standard texts such as Refs.~\cite{Chandrasekhar:2002,Bardeen:1972a,Bardeen:1973b}. These photon orbits are inherently unstable, and aligned with the number of half orbits around the black hole, they collectively constitute an infinite number of photon rings in the observer's sky that converge to a critical curve confining the black hole shadow, as indicated in \cite{Gralla:2019}. From a theoretical standpoint, determining spherical orbits for static spacetimes (like Schwarzschild) is straightforward, as they are planar and essentially circles on the equatorial plane. However, this simplicity changes for stationary (rotating) black holes due to the frame-dragging effect. This effect creates a region within a specific radial range around the black hole, bound by the innermost and outermost planar orbits that follow circular paths. Essentially, the remaining  orbits are non-planar. The region populated by spherical photon orbits is appropriately termed the \textit{photon region}, representing the set of points that these orbits traverse, forming a closed boundary. For Kerr black holes and by applying the geodesic equations for light rays, the radii of planar and polar orbits were initially provided in Refs.~\cite{Bardeen:1972a,Bardeen:1973b}, and discussed further in the context of Kerr geodesics in Ref.~\cite{Chandrasekhar:2002}. Ever since, numerous publications have been devoted to the study of spherical photon orbits, photon regions and photon rings in Kerr and Kerr-like black hole spacetimes (see for example Refs.~\cite{stoghianidis_polar_1987,cramer_using_1997,Teo:2003,Johannsen:2013,Grenzebach:2014,Perlick:2017,charbulak_spherical_2018,Johnson_universal_2020,Himwich:2020,Gelles:2021,Ayzenberg:2022,Das:2022,fathi_spherical_2023,ANJUM2023101195,Chen:2023,andaru_spherical_2023}). 

In fact, the spherical photon orbits form under specific conditions which are established by the equations $\mathcal{R}(r)=0=\mathcal{R}'(r)$ \cite{Teo:2003}. Given that from Eq. \eqref{basiceqs2b} we can rewrite 
\begin{equation}
\mathcal{R}(r)=\omega_0^2 \Big(a^2+r^2-a \xi \Big)^2-\omega_0^2 \Delta (r) \Big[(\xi -a)^2+\eta_r\Big],
    \label{eq:R}
\end{equation}
with $\eta_r=\eta+f_r/\omega_0^2$, these mutual conditions provide the criteria
\begin{eqnarray}
    && \xi^{c} = \frac{a^2+r^2}{a},\label{eq:xic_1}\\
    && \eta_r^{c} = -\frac{r^4}{a^2},\label{eq:etarc_1}  
\end{eqnarray}
or
\begin{eqnarray}
    && \xi^{c} =\frac{a^2 [2 M+r (3 N_s r+2)]-r^2 [6 M+r (N_s r-2)]}{a [2 M+r (3 N_s r-2)]},\label{eq:xic_2}\\
    && \eta_r^{c} =-\frac{r^3 \left[8 a^2 \left(N_s r^2-2 M\right)+r \left[6 M+r (N_s r-2)\right]^2\right]}{a^2 [2 M+r (3 N_s r-2)]^2},\label{eq:etarc_2}
\end{eqnarray}
for the case of $\zeta=3$ in Eq. \eqref{eq:Delta}. These expressions indicate the critical impact parameters for light rays on $r$-constant orbits. For the case of Eq. \eqref{eq:etarc_1}, it is clear that $\eta^c_r<0$, and in accordance with the statements in the classification (iii) of the $\theta$-motion in the previous subsection, confronts us with a non-physical problem. Hence, we disregard this case and consider the second set of the critical impact parameters in Eqs. \eqref{eq:xic_2} and \eqref{eq:etarc_2}.

\begin{figure}[t]
    \centering
    \includegraphics[width=5.3cm]{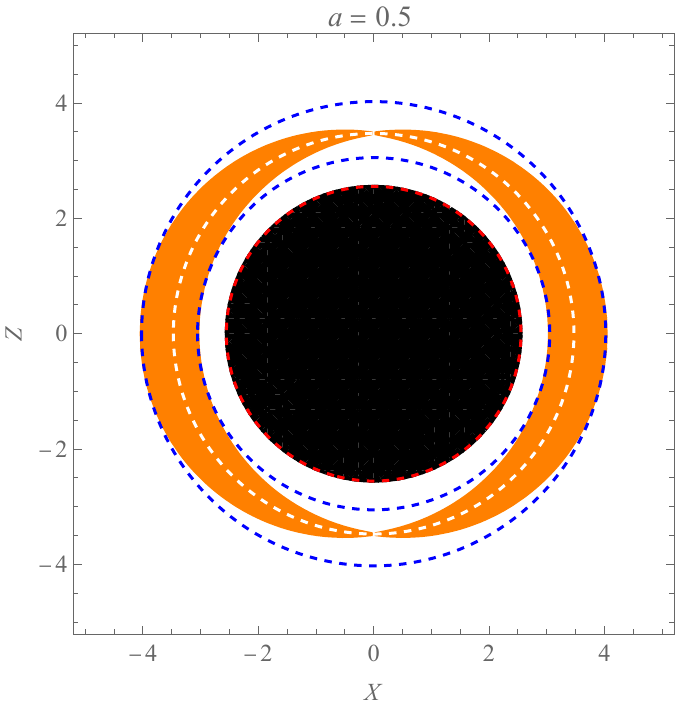} (a)
    \includegraphics[width=5.3cm]{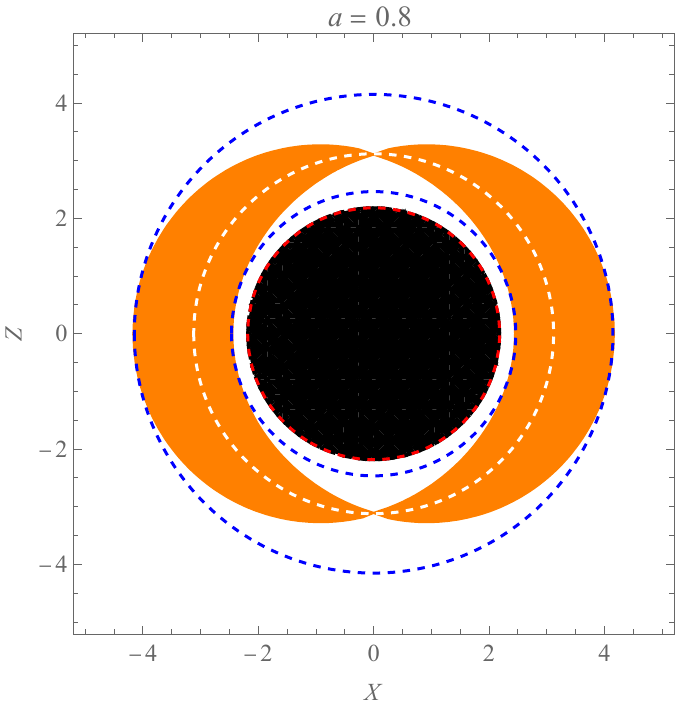} (b)
    \includegraphics[width=5.3cm]{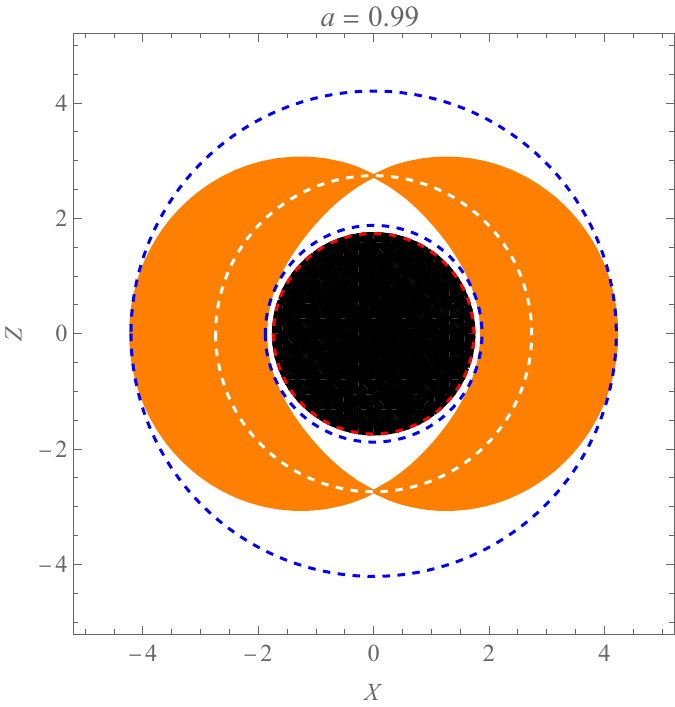} (c)
    \caption{{The orange color shows the photon region around the KRBH plotted for three different spin parameters}, considering $N_s=0.1$ and $f_r=f_\theta=1$. In these diagrams, the blue dashed curves indicate $r_{\ph-}$ and $r_{\ph+}$, the white dashed curve is the radius of polar orbits $r_\po$, and the black disk shows the region $r\leq r_+$.}
    \label{fig:photonregion}
\end{figure}

Recalling that for planar orbits, the condition $\theta=0$ holds, which together with the condition of spherical orbits, leads us to the condition $\eta_r^{c}=(f_r+f_\theta)/{\omega_0^2}$. This condition corresponds to spherical planar orbits (i.e., circular orbits) for photons. Utilizing Eq. \eqref{eq:etarc_2}, this leads to the octic equation.
\begin{equation}
P_8(r)=\sum_{j=0}^8 b_j r^j=0,
    \label{eq:octic}
\end{equation}
where
\begin{subequations}
    \begin{align}
       & b_0 = 4 a^2 M^2\left(f_r + f_\theta\right),\\
       & b_1 = 8 a^2 M \left(f_\theta-f_r\right),\\
       & b_2 = 4a^2\left(f_r + f_\theta\right)\left(1  + 3 M N_s \right),\\
       & b_3 = 4a^2\left[3 N_s \left(f_\theta-f_r\right) - 4 M \omega_0^2\right],\\
       & b_4 = 9 a^2 N_s^2\left(f_r+f_\theta\right) + 36 M^2 \omega_0^2,\\
       & b_5 = 8\omega_0^2\left(a^2 N_s-3 M\right)
       ,\\
       & b_6 = 4 \omega_0^2\left(1 + 3 M N_s \right),\\
       & b_7 =- 4 N_s \omega_0^2,\\
       & b_8 = N_s^2 \omega_0^2.
    \end{align}
    \label{eq:bj-octic}
\end{subequations}
Although such polynomial equations cannot be solved in terms of simple radicals, they however have been treated by means of approximate methods (see for example Ref. \cite{Tavlayan:2020} for the case of spherical photon orbits in Kerr spacetime). On the other hand, since this study is aimed at general analytical solutions to the equations of motion, we do not deal with solutions to the above octic, since its numerical solutions suffice for our purpose. This way, one can verify that the octic \eqref{eq:octic} has only two positive solutions $r_{\ph\mp}$ which present, respectively, the inner prograde, and the outer retrograde radii of circular photon orbits. These two radii constitute the boundary of the radii of spherical orbits, in the way that the photon region is confined within the range $r_{\ph-}\leq r \leq r_{\ph+}$. The photon region is indeed determined by the condition $\Theta(\theta)\geq 0$ for spherical photon orbits, which is specified by applying the critical impact parameters \eqref{eq:xic_2} and \eqref{eq:etarc_2} in the angular potential \eqref{basiceqs2c}. In Fig. \ref{fig:photonregion}, the photon region around the KRBH has been demonstrated for three different spin parameters. 
Note that, photons with no initial angular momentum are able to move through the whole polar angle range, and hence, travel on polar orbits. The radius of such orbits can be determined by the equation $\xi^c=0$, which by means of Eq. \eqref{eq:xic_2} results in the quartic equation $N_s r^4-2 r^3+3r^2 \left(2M- a^2 N_s\right)-2 a^2 r-2 a^2 M=0$, whose the only positive solution in the domain $r\in[r_+,r_{++}]$  can be presented regularly as $r_{\po}$ (see appendix \ref{app:A}). One should, however, note that the presence of plasma has no effect on the radial position of polar orbits. 
%

\subsubsection{Deflecting trajectories}

\begin{figure}[t]
	\begin{center}
		\includegraphics[width=7 cm]{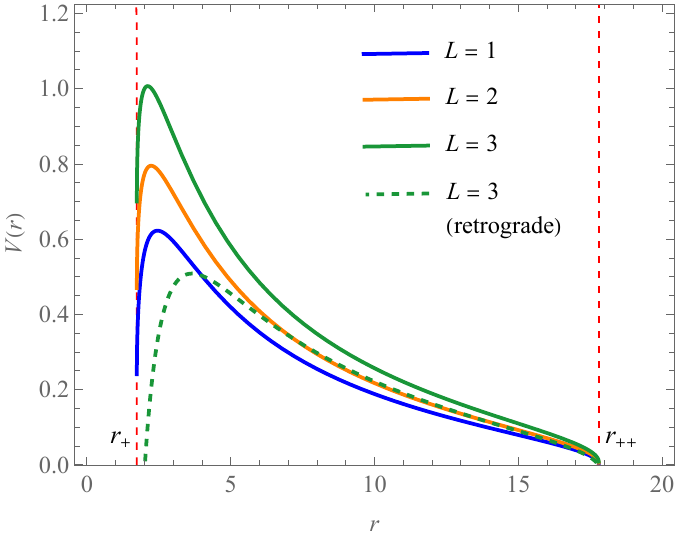}~(a)\quad
		\includegraphics[width=7 cm]{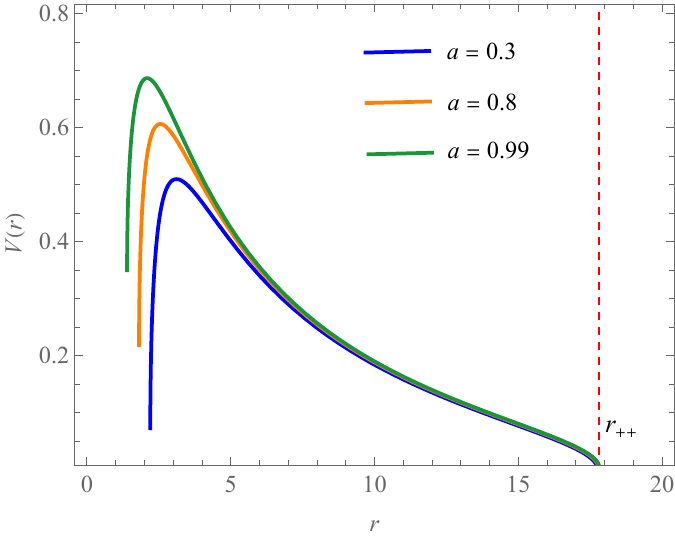}~(b)
	\end{center}
	\caption{The radial profiles of the effective potential plotted for $\Q=9$, $f_r=1$, and $N_s=0.05$. The diagrams show (a) the profiles for three different angular momenta and $a=0.85$, where the retrograde profile is obtained by letting $a=-0.85$, and (b) three different profiles for $L=1$ and different spin parameters. }
	\label{fig:EffectivePotential_general}
\end{figure}

The radial potential \eqref{eq:R} can be recast as
\begin{equation}\label{i.12}
\mathcal{R}(r) = \mathcal{p}(r)\big[\omega_0-V_-(r)\big]\big[\omega_0-V_+(r)\big],
\end{equation}
in which $\mathcal{p}(r)=(r^2+a^2)^2-a^2\,\Delta$, and 
\begin{equation}\label{i.13}
V_{\mp}(r) = \frac{1}{\mathcal{p}(r)}\left[
a^3L+aL\left(r^2-\Delta\right)
\mp\sqrt{({\Q+f_r}){\Delta}
\left[
r^4\left(1+\frac{L^2}{\Q+f_r}\right)
+a^2\left(2r^2-\Delta\right)+a^4
\right]
}\,
\right],
\end{equation}
indicate the radial gravitational potentials. Since the negative branch has no classical interpretations, we rule it out from consideration and assign $V_{\mathrm{eff}}=V_+(r)\equiv V(r)$, as the effective potential for photons affected by the gravitational field of the KRBH. In Fig. \ref{fig:EffectivePotential_general}, the above relation has been used to demonstrate this effective potential for certain values for the constants of motion and the plasma. 
As it can be observed from the diagrams, the effective potential possesses no minimums, and hence, no stable photon orbits are expected. Furthermore, effective potentials for retrograde orbits (i.e. for $a<0$) have lower maximums in their energy levels, compared to those for prograde orbits. Hence, photons traveling in the opposite direction of the black hole's spin, experience a remarkably smoother fall in their energies. 

\begin{figure}[t]
	\begin{center}
		\includegraphics[width=8cm]{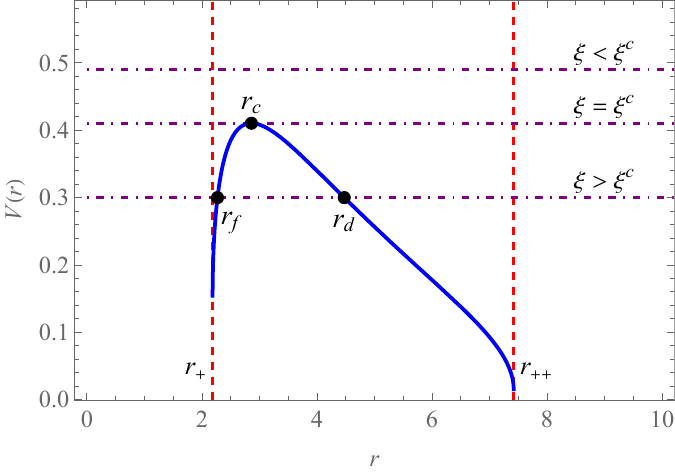}
	\end{center}
	\caption{The radial profile of a typical effective potential, plotted for $a=0.8$, $N_s=0.1$, $L=1$, $\Q=9$, and $f_r=1$. This way, the radius of critical orbits is $r_c = 2.86$, which corresponds to the critical frequency $\omega_c=0.41$, or the critical impact parameter $\xi^c=2.437$, in accordance with Eq. \eqref{eq:xic_2}. The deflecting trajectories occur when $\omega_0<\omega_c$ (or $\xi>\xi^c$), at $r_d$ or $r_f$. The case of $\omega_0>\omega_c$ (or $\xi<\xi^c$) corresponds to the capture zone.
 }
	\label{fig:EffectivePotential_typical}
\end{figure}

\begin{figure}[t]
    \centering
    \includegraphics[width=5cm]{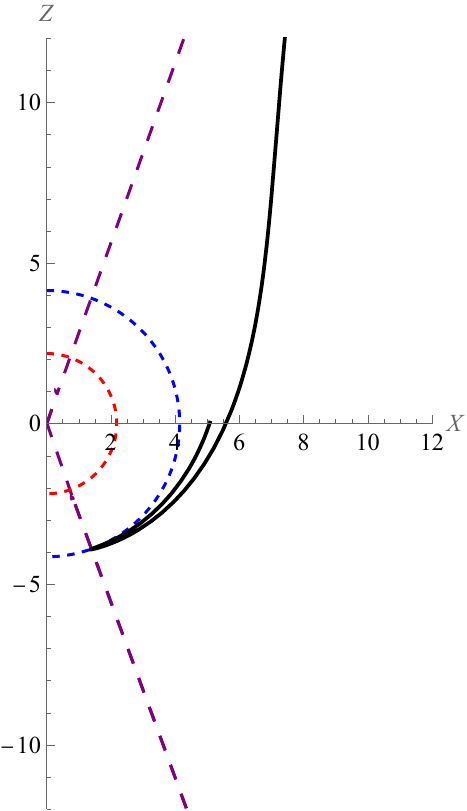} (a)\qquad\quad
    \includegraphics[width=5cm]{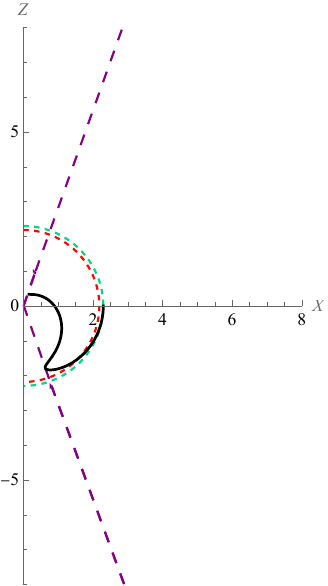} (b)
    \caption{Examples of (a) DFK, and (b) DSK, inside the cone of $\theta$-confinement (confined between $\theta_1$ and $\theta_4$ in accordance with Eqs. \eqref{eq:theta1} and \eqref{eq:theta4}), plotted for $a=0.8$, $N_s=0.1$, $\Q=9$, $f_r=f_\theta=1$, $L=1$, $\omega_0=0.3$ (corresponding to $\tilde{\eta}=88.89$, $\eta_r=111.11$ and $\xi=3.33$), $\theta_s=\pi/2$ and $r_s=5$. The minimum and maximum polar angles are $\theta_1=19.90^{\circ}$ and $\theta_4=160.10^{\circ}$ from the $Z$-axis. The red, blue, and green dashed circles correspond, respectively, to $r_+$, $r_d$ and $r_f$. }
    \label{fig:DFK2D}
\end{figure}

Now, in order to study the possible photon orbits at the vicinity of the KRBH, let us consider the typical effective potential as demonstrated in Fig.~\ref{fig:EffectivePotential_typical}. The orbits are then categorized according to the position of the initial frequency $\omega_0$ with respect to $\omega_c$, which is the frequency of photons on critical orbits (CO). 
The analytical study of the trajectories, however, requires direct analysis of the equation of motion \eqref{basiceqsR}, and hence, the radial potential $\mathcal{R}(r)$. In fact, using the form given in Eq. \eqref{eq:R}, one can recast the radial equation of motion as
\begin{equation}\label{tl11a}
M{\ed r\over \ed \gamma}=\omega_0\sqrt{\mathcal{P}(r)},
\end{equation}
given in terms of the characteristic polynomial
\begin{equation}\label{eq:P(r)_char}
\mathcal{P}(r) = r^4+ \sum_{j=0}^{3}\mathrm{a}_j r^j,
\end{equation}
where
\begin{subequations}
\label{eq:C12}
\begin{align}
 &  \mathrm{a}_0 =a^4 -2 a^3 \xi-a^2 \left[\left(\xi-a \right)^2+\eta_r
 \right]+a^2 \xi^2,\\
 &  \mathrm{a}_1 = 2 M \left[\left(\xi-a\right)^2+\eta_r
 \right],\\
 &  \mathrm{a}_2 =2 a^2 -2 a \xi-(\xi-a )^2-\eta_r
 ,\\ 
 & \mathrm{a}_3= N_s \left[\left(\xi-a \right)^2+\eta_r
 \right].
 \end{align}
\end{subequations}
As is clear from Fig. \ref{fig:EffectivePotential_typical}, deflecting trajectories correspond to $\omega_0<\omega_c$ (or $\xi>\xi^c$), for which the characteristic polynomial possesses four real roots $r_d>r_+$, $r_+<r_f<r_d$, $0<r_3<r_-$ and $r_4<0$ (see appendix \ref{app:A}). In fact, for the aforementioned initial frequency (or impact parameter), Photons that approach from $r\geq r_d$ will deflect from the black hole, escape towards a distant observer, and are detected. This kind of deflecting trajectory remains dominant up to the energy level $\omega_0\lesssim\omega_c$ and the initial radius $r_d\gtrsim r_c $, and is known as the deflection of the first kind (DFK). Based on the fact that the solutions to the characteristic polynomial respect the condition $r_4<r_3<r_f<r_d<r_\s<r_\ob$, the solution to the radial equation of motion \eqref{tl11a} for the DFK, after the integration and performing the proper inversion in accordance with the observer at $r_\ob$, is obtained as
\begin{equation}\label{eq:r(gamma)_dfk}
r_\ob(\gamma) = {r_d\left(r_f-r_4\right)-r_f\left(r_d-r_4\right)\sn^2\Big(\tilde{\kappa}\gamma+\bF(\tilde{\varphi}_\s|\tilde{\m})\Big|\tilde{\m}\Big)\over \left(r_f-r_4\right)-\left(r_d-r_4\right)\sn^2\Big(\tilde{\kappa}\gamma+\bF(\tilde{\varphi}_\s|\tilde{\m})\Big|\tilde{\m}\Big)},
\end{equation}
in which 
\begin{subequations}\label{eq:tg2tg3}
	\begin{align}
 & \tilde{\kappa} = \frac{\omega_0\sqrt{(r_d-r_3)(r_f-r_4)}}{2M },\\
 & \tilde{\varphi}_\s=\arcsin\left(\sqrt{\frac{(r_f-r_4)(r_\s-r_d)}{(r_d-r_4)(r_\s-r_f)}}\right),\label{eq:tvarphi_s}\\
 & \tilde{\m} = \frac{(r_f-r_3)(r_d-r_4)}{(r_d-r_3)(r_f-r_4)}.\label{eq:tildek}
	\end{align}
\end{subequations}
On the other hand, photons with the same initial frequencies experience a plunge into the black hole, if they approach the black hole from the radial distances $r\leq r_f$. This way, they deflect into the event horizon and are never detected by a distant observer. This kind of deflection is dominant in the same energy levels as in the DFK and is radially valid for the initial radius $r_f\lesssim r_c $, and is known as the deflection of the second kind (DSK). In this case, the inequality $r_4<r_3<r_\ob<r_f<r_d<r_\s$ holds, and hence, the evolution of the $r$-coordinate for the DSK is obtained as
\begin{equation}\label{eq:r(gamma)_dsk}
r_\ob(\gamma) = {r_f\left(r_d-r_3\right)-r_d\left(r_f-r_3\right)\sn^2\Big(\tilde{\kappa}\gamma-\bF(\tsup{\varphi}_\s|\tilde\m)\Big|\tilde\m\Big)\over \left(r_d-r_3\right)-\left(r_f-r_3\right)\sn^2\Big(\tilde{\kappa}\gamma-\bF(\tsup{\varphi}_\s|\tilde\m)\Big|\tilde\m\Big)},
\end{equation}
in which 
\begin{equation}\label{eq:vaphis_dsk}
  \tsup{\varphi}_\s=\arcsin\left(\sqrt{\frac{(r_d-r_3)(r_f-r_\s)}{(r_f-r_3)(r_d-r_s)}}\right).
\end{equation}
In Fig. \ref{fig:DFK2D}, an example of the DFK and DSK is presented in the polar plane, aligning with the data provided in Fig. \ref{fig:EffectivePotential_typical}.
%

\subsubsection{Critical trajectories}\label{subsubsec:critical}

\begin{figure}[t]
 	\begin{center}
 		\includegraphics[width=5cm]{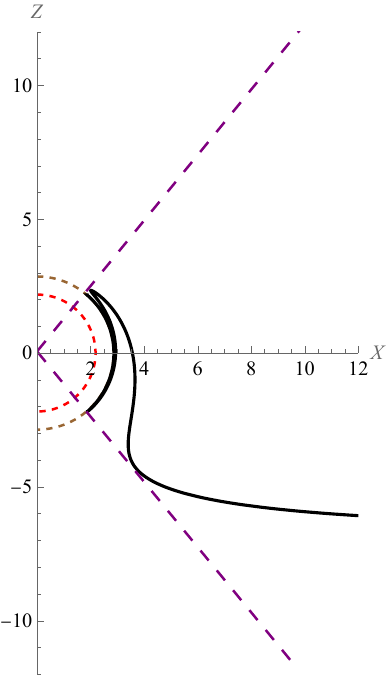}~(a)\qquad\quad
 		\includegraphics[width=5cm]{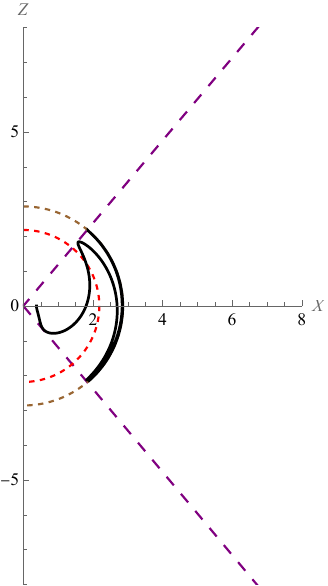}~(b)
 	\end{center}
 	\caption{Examples of (a) COFK, and (b) COSK, inside the cone of confinement for the $\theta$-coordinate ($\theta_1=49.55^{\circ}$ and $\theta_4=130.44^{\circ}$), plotted for $a=0.8$, $N_s=0.1$, $\Q=9$, $f_r=f_\theta=1$, $L=1$ and $\omega_0=\omega_c=0.41$ (corresponding to $\tilde{\eta}=47.52$, $\eta_r=59.40$ and $\xi=2.44=\xi^c$).
  The red and brown dashed circles correspond, respectively, to $r_+$ and $r_c = 2.86$.}
  \label{fig:COFSK_2D}
 \end{figure}

As stated above, photons on the orbits of constant radius, are indeed traveling on CO with the initial frequency $\omega_0=\omega_c$. According to the effective potential in Fig. \ref{fig:EffectivePotential_typical}, such orbits are indeed unstable. However, based on the point from which the photons approach the black hole, this instability may result in different fates; the photons on CO may finally plunge into the black hole or escape from it. Hence, the critical orbit of the first kind (COFK) occurs when the photons approach the black hole from $r\geq r_c$, and after performing spherical orbits for a certain duration, they finally deflect and escape from the black hole. On the other hand, photons perform a critical orbit of the second kind (COSK), when they approach from $r\leq r_c$, which finally leads to a plunge into the black hole after performing spherical orbits. To obtain the analytical solutions for the evolution of the $r$-coordinate for such orbits, let us highlight the fact that the characteristic polynomial \eqref{eq:P(r)_char} in this case can be recast as $\mathcal{P}(r)=(r-r_c)^2(r-r_3)(r-r_4)$. This way, direct integration from the radial equation of motion yields
\begin{equation}\label{eq:r(gamma)_cofk}
r_{\ob}(\gamma) = r_c+{2\left(r_c-r_3\right)\left(r_c-r_4\right)\over \left(r_3-r_4\right)\cosh{\left(\tilde{\kappa}_c \gamma +\tilde{\varphi}_{c}\right)}-\left(2r_c-r_3-r_4\right)},
\end{equation}
for the COFk, and 
\begin{equation}\label{eq:r(gamma)_cosk}
r_{\ob}(\gamma) = r_c-{2\left(r_c-r_3\right)\left(r_c-r_4\right)\over \left(r_3-r_4\right)\cosh{\left(\tilde{\kappa}_c \gamma \right)}+\left(2r_c-r_3-r_4\right)},
\end{equation}
for the COSK, where
\begin{subequations}\label{eq:kappac}
	\begin{align}
 & \tilde{\kappa}_c = \frac{\omega_c\sqrt{\left(r_c-r_3\right)\left(r_c-r_4\right)}}{M },\\
 & \cosh{(\tilde{\varphi}_{c})} = \frac{2r_c-r_3-r_4}{r_3-r_4}.
	\end{align}
\end{subequations}
In Fig. \ref{fig:COFSK_2D}, an example for each of the above kinds of orbits has been demonstrated in the polar plane, in accordance with the data given in Fig. \ref{fig:EffectivePotential_typical}
%

\subsubsection{The capture zone}\label{subsubsec:capture1}

\begin{figure}[t]
	\begin{center}
		\includegraphics[width=8cm]{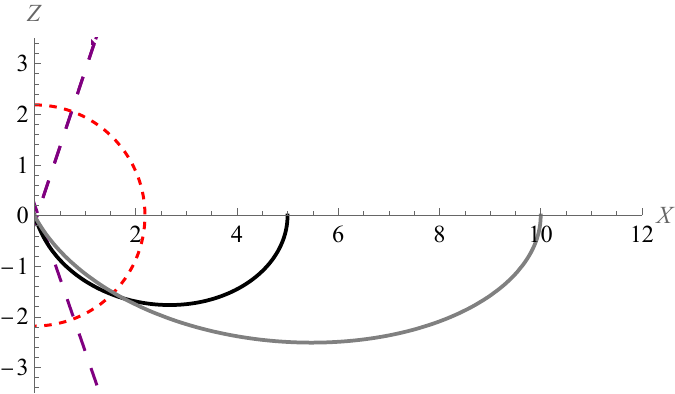}
	\end{center}
	\caption{two examples of captured orbits for $a=0.8$, $N_s=0.1$, $\Q=9$, $f_r=f_\theta=1$, $L=1$, $\omega_0=\omega_c+0.3=0.71$, $\theta_s=\pi/2$, and the initial radii $r_s=5, 10$. These orbits are confined inside the angular domain $\theta_1=19.15$ and $\theta_4=160.85$.  The red dashed circle corresponds to $r_+$.
 }
	\label{fig:rc}
\end{figure}

In the case that $\omega_0>\omega_c$ (or $\xi<\xi^c$), the photons do not encounter any turning points, and hence, become completely unstable in the exterior region of the black hole. Such photons do travel on particular bound orbits which finally plunge into the black hole, and form the capture zone. To obtain the analytical solution for the evolution of the radial coordinate for this type of motion, we should bear in mind that in this case, the characteristic polynomial takes the form $\mathcal{P}(r)=(r-r_1)(r-r_1^*)(r-r_3)(r-r_4)$, in which $r_1\in\Bbb{C}$ and is obtained by means of the relation $r_1=r_d$, together with the extra condition $r_1^*=r_f$, in accordance with the values given in Eqs. \eqref{eq:A12} and \eqref{eq:A13}. According to the fact that in this case the inequalities $r_4<r_3<r_\ob<r_\s$ hold, the direct integration of Eq. \eqref{tl11a} results in 
\begin{equation}\label{eq:r(gamma)_po}
r_\ob(\gamma) = \frac{\tilde{B}r_3-\tilde{A}r_4+(
\tilde{B}r_3+\tilde{A}r_4
)\cn\Big(
\tilde{\kappa}_p\gamma+\bF(\tilde{\varphi}_{p,\s}|\tilde{\m}_p)\Big|\tilde{\m}_p
\Big)}{(\tilde{B}-\tilde{A})+(\tilde{A}+\tilde{B})\cn\Big(
\tilde{\kappa}_p\gamma+\bF(\tilde{\varphi}_{p,\s}|\tilde{\m}_p)\Big|\tilde{\m}_p
\Big)},
\end{equation}
in which $\cn(x|\tilde{\m}_p)$ is the Jacobian elliptic cosine function with argument $x$ \cite{byrd_handbook_1971}, and
\begin{subequations}\label{eq:kappa_p,etc.}
	\begin{align}
 & \tilde{A}=\sqrt{\left(r_d-r_3\right)\left(r_f-r_3\right)},\\
 & \tilde{B}=\sqrt{\left(r_d-r_4\right)\left(r_f-r_4\right)},\\
 & \tilde{\kappa}_p = \frac{\omega_0\sqrt{\tilde{A} \tilde{B}}}{M},\\
 & \tilde{\m}_p=\frac{(\tilde{A}+\tilde{B})^2-\left(r_3-r_4\right)^2}{4\tilde{A} \tilde{B}},\\
 & \tilde{\varphi}_{p,\s}=\arccos\left(\frac{r_3\tilde{B}-r_4\tilde{A}+(\tilde{A}-\tilde{B})r_s}{-r_3 \tilde{B}-r_4 \tilde{A}+(\tilde{A}-\tilde{B})r_s}\right).
	\end{align}
\end{subequations}
In Fig. \ref{fig:rc}, two examples of this kind of orbit have been plotted, which correspond to the capture zone of the KRBH.

\subsection{Azimuth motion}\label{subsec:azimuth}

In order to obtain the evolution of the $\phi$-coordinate, let us divide the integral of motion \eqref{basiceqsphi} into the two parts
\begin{equation}\label{eq:h.4}
    \phi(\gamma) =  \Phi_\theta(\gamma)+ \Phi_r(\gamma),
\end{equation}
in which
\begin{subequations}\label{h.4}
\begin{align}
 &  \Phi_\theta(\gamma) = \frac{L}{M}\int^{\pi/2}_{\theta_{i}(\gamma)}
\frac{\ed\theta}{\sin^{2}\theta\sqrt{\Theta(\theta)}}, \label{h.4a}\\
 &   \Phi_r(\gamma) = {\frac{1}{M}}\int^{r(\gamma)}_{r_{i}}
\frac{\left(2Ma\omega_0 r+N_s a\omega_0r^3-a^2L\right)\ed r}{\Delta\sqrt{\mathcal{R}(r)}}. \label{h.4b}
\end{align}
\end{subequations}
Without loss of generality, we have taken the minimum value of the $\theta$-coordinate to coincide with the initial point $r_i$, which can be set as either of the turning points, in accordance with the considered type of orbit. For $\tilde{\eta}>0$, direct integration of the integral \eqref{h.4a} results in 
\begin{equation}\label{eq:h.42}
\Phi_{\theta}(\gamma) = \frac{L}{M a\sqrt{-z_-}}\Big[\bPi(z_+;\varphi_\s|\m)- \bPi(z_+;\varphi_\ob(\gamma)|\m)\Big],
\end{equation}
with 
\begin{equation}
\varphi_\ob(\gamma) = \arcsin\left(\frac{\cos\left(\theta_\ob(\gamma)\right)}{\sqrt{z_+}}\right),
    \label{eq:varphi_ogamma}
\end{equation}
in which $\theta_\ob(\gamma)$ has been derived in Eq. \eqref{eq:thetao_tau}, and all the other included parameters are the same as those in that equation.

Now, the $r$-dependent integral yields the following categories, corresponding to the different possible types of orbits:
\begin{itemize}
    \item [(i)] \underline{DFK}: In this case, after proper manipulation of the integral, we get to the solution
\begin{equation}
\Phi_r(\gamma) = -\frac{a\omega_0}{M}g_d\left[
\bF(\tilde{\varphi}_\ob(\gamma)|\tilde{\m})-
\bF(\tilde{\varphi}_\s|\tilde{\m})
\right]
+\frac{g_d}{M N_s}
\sum_{j=1}^{3}A_j^{\mathrm{DFK}}I_j^\DFK(\gamma),
    \label{eq:phi_rDFK}
\end{equation}
where
\begin{subequations}
    \begin{align}
        & g_d = \frac{2}{\sqrt{(r_d-r_3)(r_f-r_4)}},\label{eq:gd}\\
        & \tilde{\varphi}_\ob(\gamma) = \arcsin\left(\sqrt{\frac{(r_f-r_4)(r_\ob(\gamma)-r_d)}{(r_d-r_4)(r_\ob(\gamma)-r_f)}}\right),\label{eq:tvarphi_ogamma}
    \end{align}
\end{subequations}
in which $r_\ob(\gamma)$ is identified in Eq. \eqref{eq:r(gamma)_dfk} for the DFK, $\tilde{\varphi}_\s$ and $\tilde{\m}$ are given in Eqs. \eqref{eq:tvarphi_s} and \eqref{eq:tildek}, and the coefficients $A_j^\DFK$ and the elliptic integrals $I_j^\DFK$ are presented in appendix \ref{app:B}. With these values considered, Fig. \ref{fig:DFK3D} illustrates two examples of the DFK in the context of the effective potential shown in Fig. \ref{fig:EffectivePotential_typical}, assuming a thin disk in the black hole's equatorial plane. These diagrams highlight two significant scenarios. Firstly, direct emission occurs when the initial frequency  differs significantly from $\omega_c$ (panel (a)). In such cases, photons emitted from the disk directly deflect at $r_d$ and escape the black hole without intersecting the disk. Conversely, as $\omega_0$ approaches $\omega_c$ (panel (b)), the emitted photons approach the turning point $r_d$ and then curve away from the black hole after crossing the disk. As a result, observers perceive a demagnified lensed image of the disk. This phenomenon generates an image of higher order compared to that from direct emission. The diagrams also display both initial frequency scenarios, but for the case where the KRBH exists in a vacuum. In this instance, direct emission exhibits more variation in the $\theta$-coordinate, whereas the additional lensed image yields a demagnified representation of the disk's backside.
\begin{figure}[t]
	\begin{center}
		\includegraphics[width=8cm]{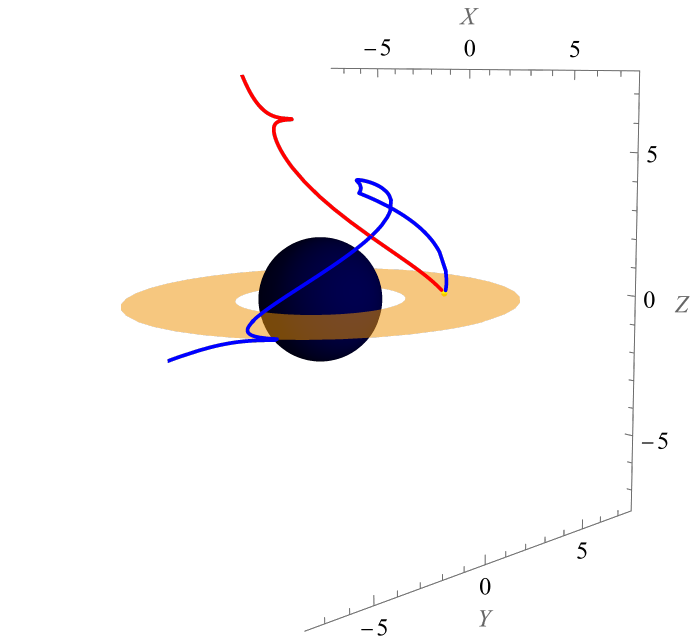} (a)\qquad\quad
 \includegraphics[width=8cm]{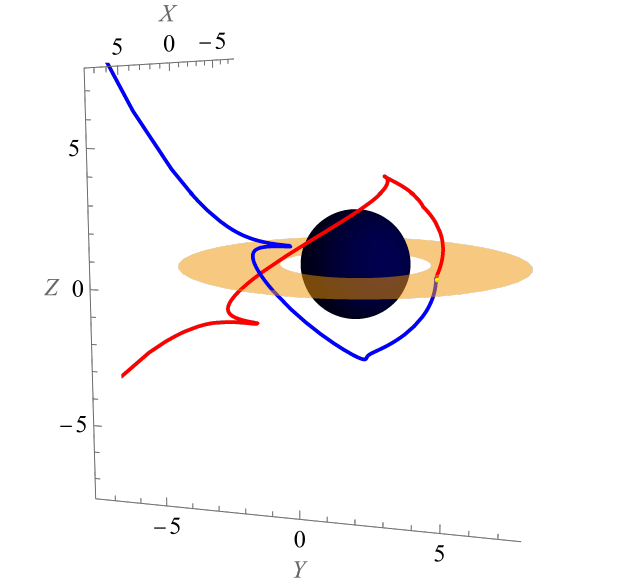} (b)
	\end{center}
	\caption{Examples of the DFK plotted for the specific parameter values $a=0.8$, $N_s=0.1$, $\Q=9$, $L=1$, initial radii $r_\s=4$ (indicated as the yellow point), $\theta_s=\pi/2$, and two distinct initial frequencies, namely, (a) $\omega_0=0.3$ and (b) $\omega_0=0.375$. The black sphere in the diagrams represents the black hole with a radius of $r_+$, and the orange region indicates the thin emission disk situated in the equatorial plane, illuminating the black hole. The red and blue curves in the diagrams correspond, respectively, to the paths of light rays propagating around the KRBH in a plasmic medium with $f_r=f_\theta=1$, and in a vacuum with $f_r=f_\theta=0$. In diagram (a), the DFK for both cases results in simple lensed images produced by the direct emission from the disk. In diagram (b), both scenarios involve light rays crossing the disk before escaping the black hole. This leads to the observer detecting demagnified lensed images of points on the disk, stemming from the rays completing half orbits around the black hole before their ultimate deflection. In the scenario involving the KRBH surrounded by a plasmic medium, this mechanism yields higher-order images originating from the front side of the disk. Conversely, for the KRBH in a vacuum, these images correspond to the backside of the disk.
 }
	\label{fig:DFK3D}
\end{figure}
In fact, by having at hand the analytical solutions for the azimuth motion in the course of the DFK, one can also calculate the deflection angle, as detected by an observer at a finite radial distance $r_\ob$. Confining ourselves to the equatorial plane and exploiting Eqs. \eqref{basiceqphi} and \eqref{basiceqsR}, one yields the equation
\begin{equation}
\phi_i=\int_{r_d}^{r_{i}}
\frac{\left(2Ma\omega_0 r+N_s a\omega_0r^3-a^2L\right)\ed r}{\Delta\sqrt{\mathcal{R}(r)}}  
-L\int_{r_d}^{r_i}\frac{\ed r}{\sqrt{\mathcal{R}(r)}},
\label{eq:phi_i}
\end{equation}
which provides the extent of azimuth angle covered as the light traverses the distance between the turning point $r_d$ and the fixed radial position $r_i$. Accordingly, the lens equation can be written as \cite{fathi_analytical_2021,fathi_study_2022}
\begin{equation}
\hat{\vartheta} = (\phi_\ob+\phi_\s)-\pi.
    \label{eq:lens_0}
\end{equation}
This relation calculates the deflection angle $\hat{\vartheta}$ of the light rays around the KRBH, where
\begin{subequations}
    \begin{align}
    & \phi_\ob = \left.-(a\omega_0+L)g_d\bF(\tilde{\varphi}_i|\tilde{\m})\right|_{r_d}^{r_\ob}+\frac{g_d}{N_s}\sum_{j=1}^{3} A_j^\DFK I_j^{\mathrm{L}_\ob},\\
    & \phi_\s = \left.-(a\omega_0+L)g_d\bF(\tilde{\varphi}_i|\tilde{\m})\right|_{r_d}^{r_\s}+\frac{g_d}{N_s}\sum_{j=1}^{3} A_j^\DFK I_j^{\mathrm{L}_\s},
    \end{align}
    \label{eq:phio-phis}
\end{subequations}
in which $I_j^{\mathrm{L}_\ob}$ and $I_j^{\mathrm{L}_\s}$ are obtained, respectively, by doing the pair of exchanges $\left\{r_\s\rightarrow r_d,r_\ob(\gamma)\rightarrow r_\ob=\mathrm{const.}\right\}$, and $\left\{r_\s\rightarrow r_d,r_\ob(\gamma)\rightarrow r_\s=\mathrm{const.}\right\}$ in $I^{\DFK}_j$. In Fig. \ref{fig:vartheta}, the deflection angle profile $\hat{\vartheta}$ has been plotted against the impact parameter $\xi$ for an illustrative case. As observed in the figure, the deflection angle diverges at the critical impact parameter $\xi_c$, and subsequently decreases rapidly until it reaches a minimum value. It then gradually increases until peaking at a specific impact parameter, followed by a rapid decline once more. From these diagrams, it's evident that when light propagates around the KRBH in a vacuum, the deflection angle exhibits a lower minimum but a higher maximum compared to its behavior within a plasmic medium. Generally, the deflection angle values in a plasmic medium tend to be larger than those in a vacuum, an anticipated outcome due to the nonunity refractive index of the plasma. Furthermore, gravitational lensing by the KRBH spans a broader range of impact parameters when the black hole exists in a vacuum as opposed to being immersed in plasma. This concept will be explored further in an upcoming study, delving into the extensive gravitational lensing by the KRBH within a plasmic medium.
\begin{figure}[t]
	\begin{center}
		\includegraphics[width=8cm]{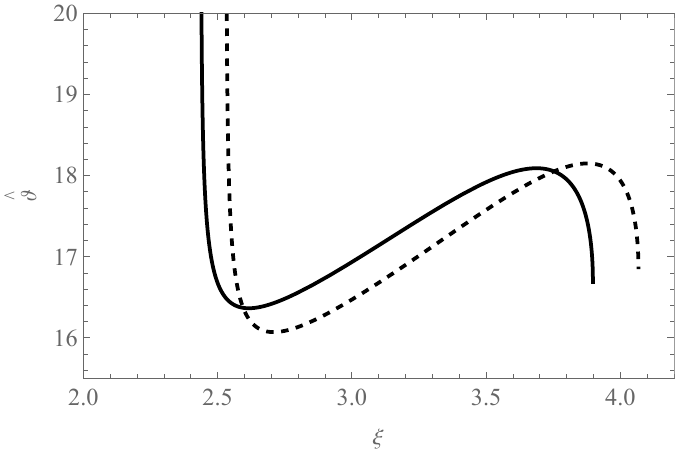}
	\end{center}
	\caption{The $\xi$-profile of the deflection angle $\hat{\vartheta}$ plotted for $a=0.8$, $N_s=0.1$, $\Q=9$, $r_\s=5$ and $r_\ob=100$. The solid curve corresponds to the plasmic medium with $f_r=1$, whereas the dashed curve corresponds to the vacuum case. }
	\label{fig:vartheta}
\end{figure}
%
%
%
%
%
 \item [(ii)] \underline{DSK}: Based on the same methods, the direct integration of Eq. \eqref{h.4b} for the case of the DSK, results in 
 \begin{equation}
 \Phi_r(\gamma) = -\frac{a\omega_0}{M}g_d\Big[
 \bF(\tsup{\varphi}_\ob(\gamma)|\tilde{\m})-\bF(\tsup{\varphi}_\s|\tilde{\m})
 \Big]
 +
 \frac{g_d}{M N_s}\sum_{j=0}^{3} A^\DSK_j I_j^\DSK,
       \label{eq:phi_rDSK}
 \end{equation}
for which $A_j^\DSK=A_j^\DFK$ as given in appendix \ref{app:B}, $\tsup{\varphi}_\s$ is defined in Eq. \eqref{eq:vaphis_dsk}, and
\begin{equation}
\tsup{\varphi}_\ob(\gamma) = 
\arcsin\left(\sqrt{\frac{(r_d-r_3)(r_f-r_\ob(\gamma))}{(r_f-r_3)(r_d-r_\ob(\gamma))}}\right),
    \label{eq:tsupvarphi_o_dsk}
\end{equation}
where $r_\ob(\gamma)$ has been expressed in Eq. \eqref{eq:r(gamma)_dsk}. The relevant expressions for the integrals $I_j^{\DSK}$ have been given in Eqs. \eqref{eq:B5} and \eqref{eq:B6} in appendix \ref{app:B}. Based on the fact that the DSK occurs in the vicinity of the event horizon, it does not have any particular latitudinal behavior, as illustrated in Fig. \ref{fig:DFK2D}(b). We have therefore omitted any diagrams representing this type of orbit.
%
%
 \item [(iii)] \underline{COFK}: Based on the discussion provided in Subsect. \ref{subsubsec:critical}, the integral \eqref{h.4b} for this type of orbit results in 
 \begin{equation}
 \Phi_r(\gamma)=\frac{2}{M N_s}\sum_{j=1}^{4}A_j^{\COFK} I_j^{\COFK}(\gamma),
     \label{eq:Phir_COFK}
 \end{equation}
for which, the coefficients $A_j^{\COFK}$ and the hyperbolic expressions $I_j^{\COFK}(\gamma)$ have been given in Eqs. \eqref{eq:B7} and \eqref{eq:B8}. Some examples of such orbits are depicted in Fig. \ref{fig:COFK}. In each scenario, the photon emission initiates at the radial distance $r_c$, where the photons execute unstable spherical orbits before escaping the black hole. As observed from the diagrams, a higher initial angular momentum results in a narrower range of oscillations of the $\theta$-coordinate. Additionally, as a general trend, the number of times the orbits traverse the equatorial plane (i.e., the number of nodes) is greater in a vacuum environment compared to that within a plasmic medium.
\begin{figure}[t]
	\begin{center}
 \includegraphics[width=8cm]{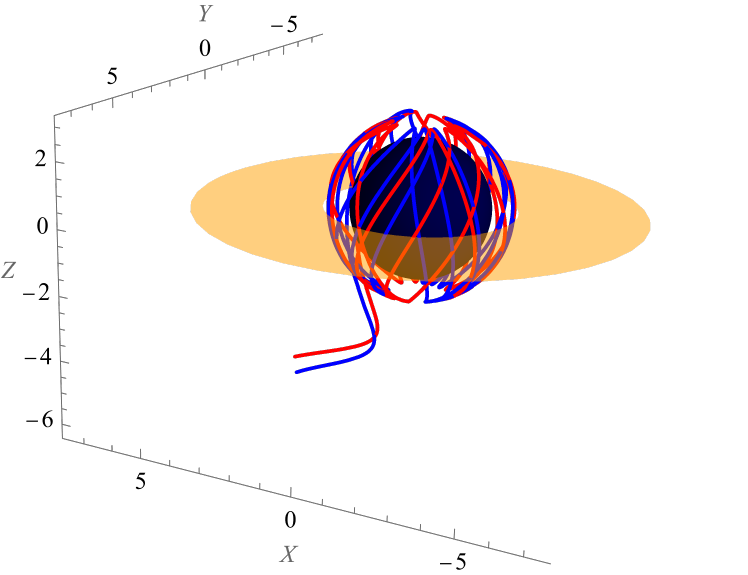} (a)\qquad\quad
 \includegraphics[width=8cm]{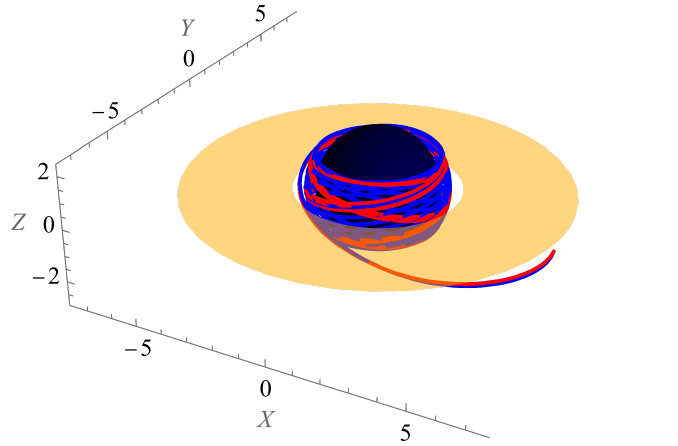} (b)\qquad\quad
 \includegraphics[width=8cm]{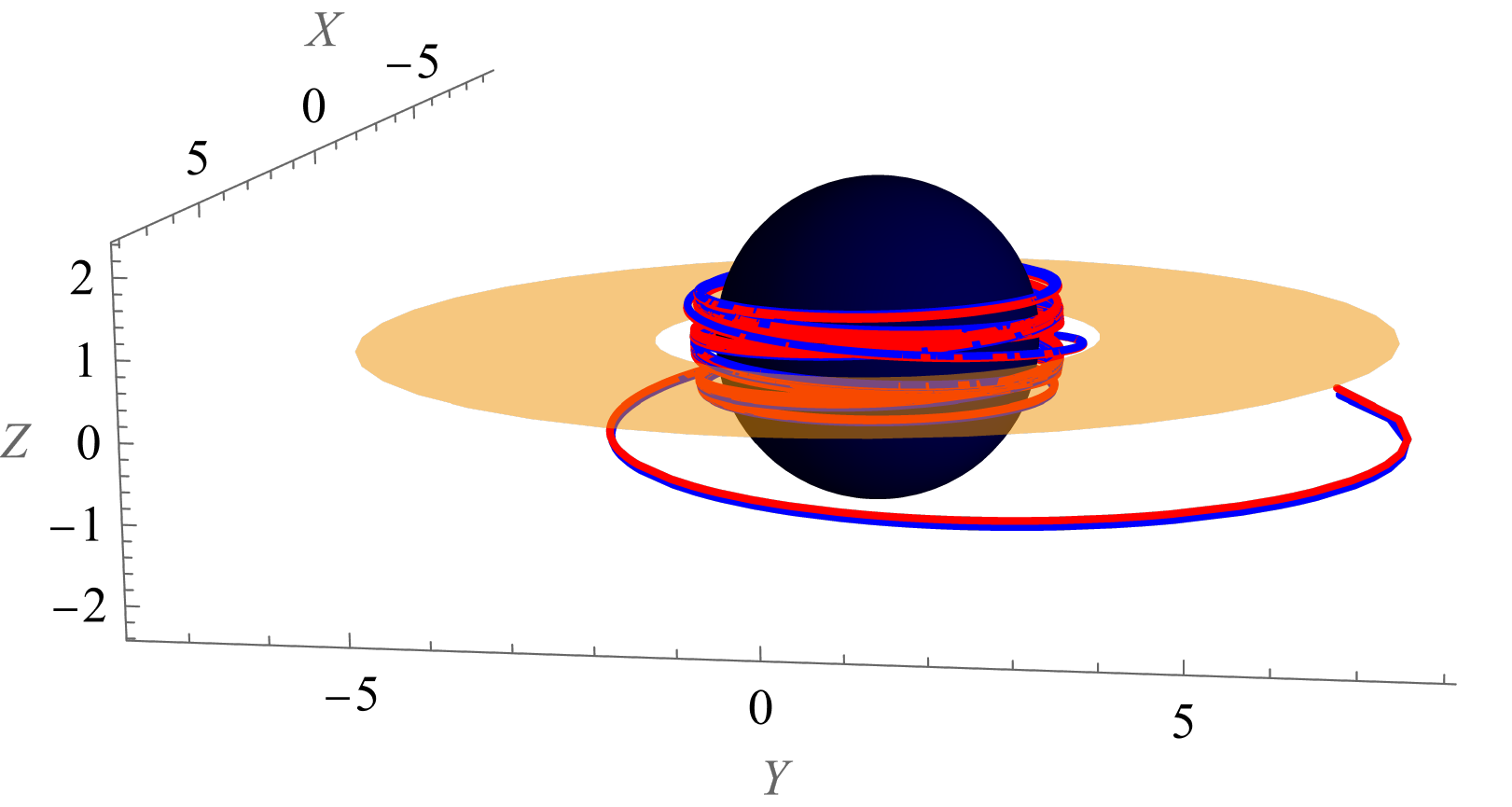} (c)\qquad\quad
 \includegraphics[width=8cm]{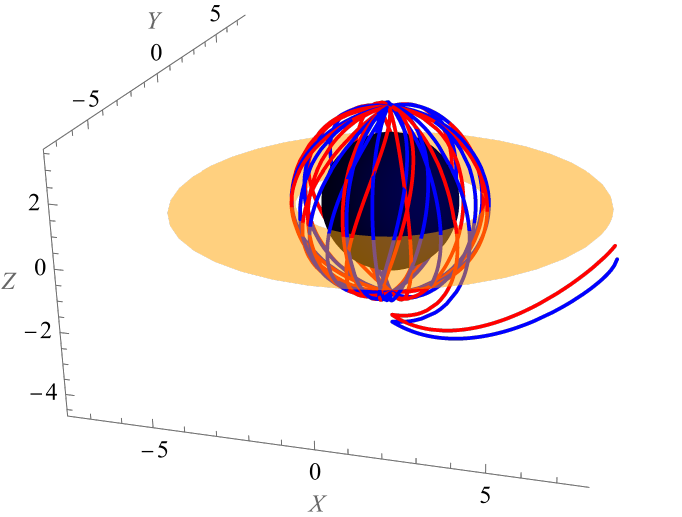} (d)
	\end{center}
	\caption{Examples of the COFK for specific parameter values: $a=0.8$, $N_s=0.1$, $\Q=9$, and $L=1$. The red and blue curves represent light propagation in plasma with $f_r=f_\theta=1$ and in a vacuum, respectively. The diagrams showcase the following cases: (a) $L=1$: $r_c=2.86$ and $\omega_c=0.41$ in plasma, and $r_c=2.85$ and $\omega_c=0.39$ in vacuum, (b) $L=5$: $r_c=2.51$, $\omega_c=1.02$ in plasma, and $r_c=2.50$, $\omega_c=1.01$ in vacuum, (c) $L=10$: $r_c=2.458$, $\omega_c=1.914$ in plasma, and $r_c=2.456$, $\omega_c=1.909$ in vacuum, (d) $L=0$: $r_c=3.12$, $\omega_c=0.32$ in plasma, and $r_c=3.12$, $\omega_c=0.30$ in vacuum. In each case, photons start from the peak of the effective potential, perform spherical orbits at the radial distance $r_c$, and eventually escape the black hole.}
	\label{fig:COFK}
\end{figure}
Due to the similarity in the evolution of the radial coordinates between COFK and COSK, as seen in Eqs. \eqref{eq:r(gamma)_cofk} and \eqref{eq:r(gamma)_cosk}, their analytical azimuth motion solutions show minimal distinctions. Consequently, the shapes of the orbits remain fundamentally unchanged. As a result, we will omit this scenario and move forward with the plunge orbit analysis
%
%
 \item [(iv)] \underline{The capture zone}: For plunge orbits discussed in Subsect. \ref{subsubsec:capture1}, the integral equation \eqref{h.4b} provides the analytical solution
\begin{equation}
\Phi_r(\gamma) = -\frac{a\omega_0}{M}\Big[\bF(\tilde{\varphi}_{p,\ob}|\tilde{\m}_p)-\bF(\tilde{\varphi}_{p,\s}|\tilde{\m}_p)\Big] + \frac{1}{M N_s}\sum_{j=1}^3 A_j^\PO I_j^\PO(\gamma),
    \label{eq:Phir_PO}
\end{equation}
for which the coefficients $A_j^{\PO}$ and the elliptic integrals $I_j^{\PO}$ have been given in Eqs. \eqref{eq:B9}--\eqref{eq:B15}. Some examples of this kind of orbit have been demonstrated in Fig. \ref{fig:plunge3D}. 
\begin{figure}
    \centering
    \includegraphics[width=8cm]{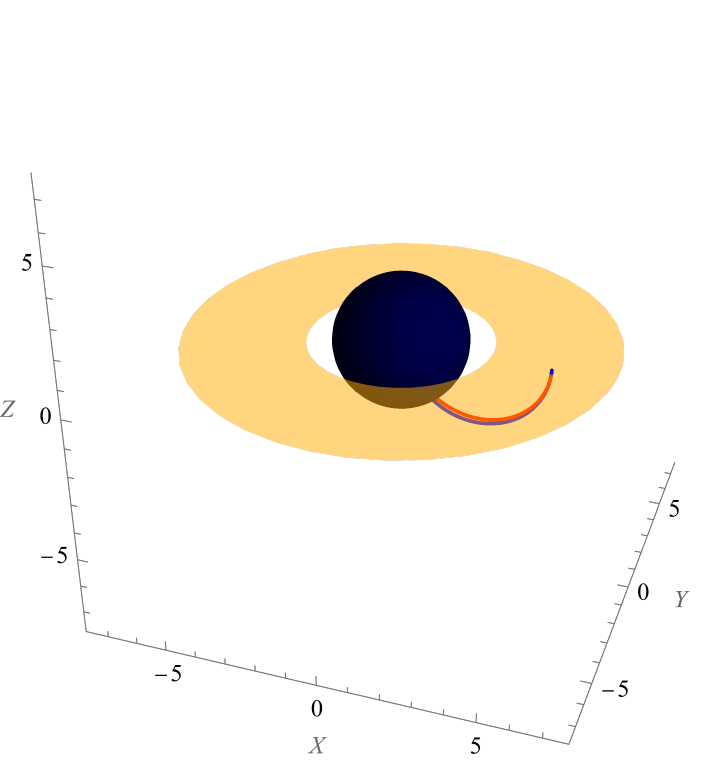} (a)\qquad\quad
    \includegraphics[width=8cm]{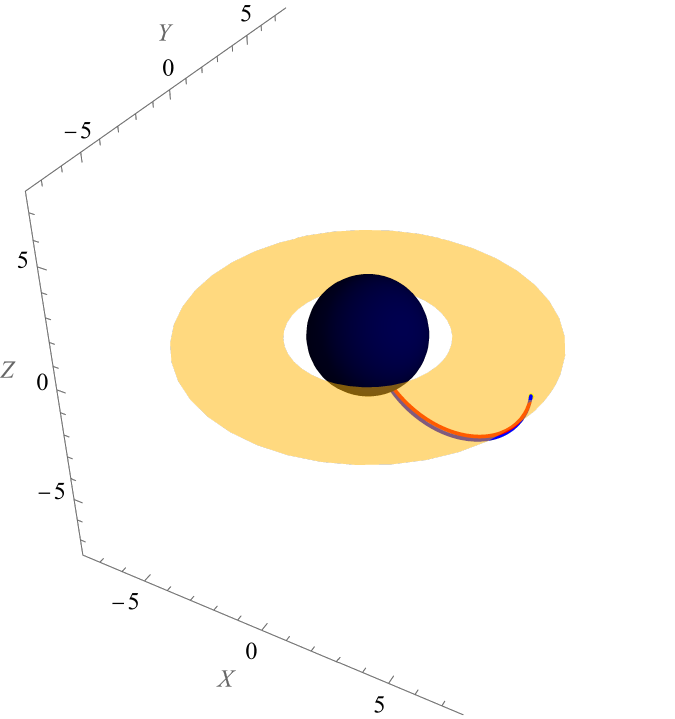} (b)
    \caption{Two examples of plunge orbits, plotted by considering $a=0.8$, $N_s=0.1$, $\Q=9$, $L=1$ and $\theta_s=\pi/2$. The red and blue curves correspond respectively to the propagation of light in plasma with $f_r=f_\theta=1$ and $\omega_0=\omega_c+0.02=0.43$, and in vacuum with $\omega_0=0.41$. The diagrams correspond to (a) $r_s=5$, and (b) $r_s=6.5$.}
    \label{fig:plunge3D}
\end{figure}
As evident from the diagrams, in both scenarios of light propagation through plasma and in a vacuum, the light rays initiate their trajectory towards the black hole from the rear side of the emission disk. This occurrence is solely a result of the spacetime geometry, which dictates this behavior in the rays for plunge orbits (see also Fig. \ref{fig:rc}).
    
\end{itemize}
%

\subsection{The $t$-coordinate evolution}\label{subsec:t}

By employing the same methodology utilized in computing integrals for the $\phi$-motion, and considering Eq. \eqref{basiceqste} alongside Eqs.~\eqref{basiceqsR} and \eqref{basiceqstheta}, the integral governing the evolution of the $t$-coordinate can be isolated as
\begin{equation}\label{eq:h.5}
    t(\gamma) =  t_\theta(\gamma)+ t_r(\gamma),
\end{equation}
with 
\begin{subequations}
\begin{align}
 &  t_\theta(\gamma) =-\omega_0 a^2\int^{\pi/2}_{\theta_{i}(\gamma)}
\frac{\sin^2\theta\,\ed\theta}{\sqrt{\Theta(\theta)}}, \label{h.5a}\\
 &   t_r(\gamma) = \frac{1}{M}\int^{r(\gamma)}_{r_{i}}
\frac{\left[\omega_0(r^2+a^2)^2-2MaL r-N_saLr^3\right]\ed r}{\Delta\sqrt{\mathcal{R}(r)}}. \label{h.5b}
\end{align}
\end{subequations}
Direct integration of the angular part in Eq. \eqref{h.5a}, then results in the analytical solution 
\begin{equation}\label{eq:h.42}
t_{\theta}(\gamma) = -\frac{a}{\sqrt{-z_-}}
\left[\left(1-{z_-\over \m}\right)\Big(\bF(\varphi_\s|\m)-\bF(\varphi_\ob(\gamma)|\m)\Big)+{z_-\over \m}\Big(\bE(\varphi_\s|\m)-\bE(\varphi_\ob(\gamma)|\m)\Big)\right],
\end{equation}
in which $\varphi_\ob(\gamma)$ has been given in Eq. \eqref{eq:varphi_ogamma}, and $\bE(x|\m)$ is the elliptic integral of the second kind with argument $x$ and modulus $\m$ \cite{byrd_handbook_1971}. Moreover, when considering the most general case for the radial integral \eqref{h.5b}, which corresponds to the DFK, the solution is given by
\begin{equation}
t_r(\gamma) = -\frac{a L N_s +\omega_0(r_{++}+r_++r_-)}{M N_s}g_d\left[\bF(\tilde{\varphi}_\ob(\gamma)|\tilde{\m})
-\bF(\tilde{\varphi}_\s|\tilde{\m})\right] + \sum_{j=1}^{3} B_j^t I^\DFK_j(\gamma)
+\frac{\omega_0u_f\sqrt{-u_d u_f u_3 u_4}}{u_d^2}g_t I^t(\gamma),
    \label{eq:t_rgamma}
\end{equation}
where $\tilde{\varphi}_\ob(\gamma)$ and $\tilde{\varphi}_\s$ are provided in Eqs. \eqref{eq:tvarphi_ogamma} and \eqref{eq:tvarphi_s}, respectively, while $I_j^\DFK(\gamma)$ is defined in Eqs. \eqref{eq:B3}, and the coefficients $B_j^t$ are outlined in Eqs. \eqref{eq:B16}. Additionally, we've introduced $u_i=1/r_i$, along with
\begin{equation}
g_t=\frac{2}{\sqrt{(u_f-u_d)(u_f-u_4)}},
    \label{eq:gt}
\end{equation}
and 
\begin{equation}
I^t(\gamma) = \left.- \frac{(u_d-u_4)(u_f-u_d)}{u_d(u_f-u_4)}\bF(\tilde{\varphi}^t_i|\tilde{\m}_t)\right|_{\tilde{\varphi}^t_\ob(\gamma)}^{\tilde{\varphi}^t_\s}
\left.-\frac{u_f-u_d}{u_f}\bPi\left(\frac{u_d-u_4}{u_f-u_4},\tilde{\varphi}_i^t|\tilde{\m}_t\right)\right|_{\tilde{\varphi}^t_\ob(\gamma)}^{\tilde{\varphi}^t_\s},
    \label{eq:It(gamma)}
\end{equation}
in which 
\begin{equation}
\tilde{\m}_t=\sqrt{\frac{(u_3-u_f)(u_d-u_4)}{(u_3-u_d)(u_f-u_4)}},
    \label{eq:tildekt}
\end{equation}
and 
\begin{subequations}
\begin{align}
    & \tilde{\varphi}^t_\s = \arcsin\left(
    \sqrt{\frac{(u_f-u_4)(u_d-u_\s)}{(u_d-u_4)(u_f-u_\s)}}\,
    \right),\label{eq:varphit_s}\\
    & \tilde{\varphi}^t_\ob(\gamma) = \arcsin\left(
    \sqrt{\frac{(u_f-u_4)\left(u_d-u_\ob(\gamma)\right)}{(u_d-u_4)\left(u_f-u_\ob(\gamma)\right)}}\,
    \right),\label{eq:varphit_o}
\end{align}
\end{subequations}
where $u_\ob(\gamma)=1/r_\ob(\gamma)$, and $r_\ob(\gamma)$ is outlined in Eq. \eqref{eq:r(gamma)_dfk}. It is worth noting that our orbit simulations exclusively utilized the evolution of the three spatial coordinates. Hence, the inclusion of the $t$-coordinate motion analysis is only for the sake of completeness in our study.

\section{More on the spherical orbits}\label{sec:spherical}

As highlighted in point (iii) of Subsect. \ref{subsec:azimuth}, spherical orbits are pivotal within photon trajectories navigating critical paths. These orbits notably contribute to forming the black hole shadow, representing the closest orbits that escape the black hole and become observable to the observer. However, photons traversing these orbits possess two crucial characteristics requiring careful consideration. First, photons take a finite time to complete each half orbit around the black hole, thereby inducing a time delay for these photons contributing to the total time for subsequent images reaching the observer. Second, the azimuth angle experiences distinct changes during each half orbit, influencing the rotation of the photon ring and ensuing subrings. These parameters significantly impact the determination of higher-order images generated by the black hole. Hence, this section focuses on calculating the azimuth angle changes and time delays occurring during critical orbits at $r_c$. Given these orbits remain at a constant radius from the black hole, the radial part of the integrals of motion remains constant. The determination of each parameter involves integration over half orbits, between the angular turning points of the angular potential, $\theta_1$ and $\theta_4$, as discussed in Subsect. \ref{subsec:thetamotion}. 

We begin by computing the changes in the azimuth angle, $\phi_\ob-\phi_\s$, throughout the orbits. By incorporating this constant radius into Eq. \eqref{basiceqsphi}, we arrive at the integral
\begin{equation}
\delta\phi_c\equiv\phi_\ob-\phi_\s=\frac{1}{M}\left[
\frac{a\omega_0\left(r_c^2+a^2-\Delta_c\right)-L a^2}{\Delta_c}\right]\int_{\theta_1}^{\theta_4}\frac{\ed\theta}{\sqrt{\Theta(\theta)}}+\frac{L}{M}\int_{\theta_1}^{\theta_4}\frac{\ed\theta}{\sin^2\theta\sqrt{\Theta(\theta)}},
\label{eq:deltaphic_int}
\end{equation}
in which, we have considered the identity $\gamma=M\int\frac{\ed r}{\sqrt{\mathcal{R}(r)}}=M\int\frac{\ed\theta}{\sqrt{\Theta(\theta)}}$, and $\Delta_c\equiv\Delta(r_c)$. Direct integration of Eq. \eqref{eq:deltaphic_int} results in 
\begin{equation}
\delta\phi_c=\frac{2}{z_0}\left[
\frac{a\omega_0\left(r_c^2+a^2-\Delta_c\right)-L a^2}{\Delta_c}\right]\bK(\m)+\frac{2L\omega_0}{Mz_0}\bPi(z_+|\m),
    \label{eq:deltaphic_1}
\end{equation}
where $\bK(\m)=\bF(\pi/2|\m)$ is the complete elliptic integral of the first kind, and
$\bPi(z_+|\m) = \bPi(z_+;\pi/2|\m)$ is the complete elliptic integral of the third kind \cite{byrd_handbook_1971}. Similarly, to determine the time delay during these orbits, we analyze Eq. \eqref{basiceqste} for $r=r_c$, which yields the relationship
\begin{equation}
\delta t_c\equiv t_\ob-t_\s = \frac{1}{M}\left[
\frac{\omega_0\left(r_c^2+a^2\right)^2-a\left(r_c^2+a^2-\Delta_c\right)}{\Delta_c}
\right]\int_{\theta_1}^{\theta_4}\frac{\ed\theta}{\sqrt{\Theta(\theta)}}-a^2\omega_0\int_{\theta_1}^{\theta_4}\frac{\sin^2\theta\,\ed\theta}{\sqrt{\Theta(\theta)}},
    \label{eq:deltatc_int}
\end{equation}
and has the solution
\begin{equation}
\delta t_c=\frac{2}{z_0}\left[
\frac{\omega_0\left(r_c^2+a^2\right)^2-a\left(r_c^2+a^2-\Delta_c\right)}{\Delta_c}
\right] \bK(\m)-\frac{2a^2\omega_0^2}{z_0}\left[
\bK(\m)-\frac{4z_+}{\m}\bJ(\m)
\right],
    \label{eq:deltatc_1}
\end{equation}
where we have defined
\begin{equation}
\bJ(z|\m)=\bF(z|\m)-\bE(z|\m).
    \label{eq:J}
\end{equation}
Hence, $\bJ(\m)=\bJ(\pi/2|\m)$ represents the complete form of this function. These findings hold promise for computing alterations in crucial parameters during light propagation around black holes. However, for the KRBH in a plasmic medium, the application of these outcomes to examine photon rings and subrings as images of the black hole's accretion disk remains a task for future research. Consequently, we conclude our current discussion here and provide a summary of our results in the next section.

\section{Summary and conclusions}\label{sec:conclusions}

This study delved into the behavior of light near a rotating black hole within Rastall gravity, considering a structure filled with cold, non-magnetized electronic plasma. Our investigation commenced with an overview of Rastall gravity, introducing its static spherically symmetric black hole solution involving an anisotropic cosmic fluid. We extended this solution to the rotating scenario (KRBH) using a modified Newman-Janis algorithm and investigated diverse values of the parameter $\zeta$ in the cosmological term. Throughout, we emphasized the ergoregion and identified $\zeta=3$ as a favorable choice due to the absence of imaginary horizons. Shifting focus to Hamiltonian dynamics, we constructed a framework for light propagation within a non-magnetized electronic plasma. Assuming constant plasma parameters, we derived analytical solutions for the temporal evolution of spacetime coordinates in terms of the Mino time. Our attention turned to motion along the latitude, revealing two turning points for the polar angle and expressing the solution using elliptic integrals and Jacobi elliptic functions. We highlighted the solution's oscillatory nature within allowable polar angle ranges. Analyzing the radial motion, we concentrated on spherical orbits, pivotal for black hole imaging. Calculations involved critical impact parameters and photon regions, and we discerned between deflecting trajectories, distinguishing orbits easily escaping the black hole (DFK) from those drawn into it (DSK). We derived analytical solutions for both cases and discussed critical trajectories and plunge orbits. Simulating two-dimensional light ray trajectories around the KRBH in plasma, our study encompassed all orbit types. We calculated azimuth angle evolution, expressing analytical solutions using elliptic integrals and highlighting plasma versus vacuum differences in light propagation. Illustrative examples with an equatorial disk as the illumination source showcased the impact of plasma on light. Exploring the role of DFK in gravitational lensing, we presented the lens equation and calculated deflection angles for light passing the black hole, observing larger deflection angles in plasma compared to vacuum. Additionally, we explored the COFK, focusing on spherical orbits and escape, demonstrating how higher initial impact parameters led to smaller polar ranges. Analytical solutions for plunge orbits were also provided. In conclusion, we revisited spherical orbits, offering expressions for azimuth angle change and time delay during each half orbit. These findings and preceding solutions lay the foundation for future studies. This paper marks the initial step in forthcoming research exploring strong lensing and accretion disk imaging for KRBH in plasmic media, to be detailed in future papers.

\section*{Acknowledgements}
M.F. acknowledges financial support from Vicerrector\'{i}a de Investigaci\'{o}n, Desarrollo e Innovaci\'{o}n - Universidad de Santiago de Chile (USACH), Proyecto DICYT, C\'{o}digo 042331CM$\_$Postdoc. J.R.V. is partially supported by the Centro de Astrof\'isica de Valpara\'iso (CAV). M.F. would like to thank the hospitality of the faculty of Instituto de F\'{i}sica y Astronom\'{i}a (IFA) of Universidad de Valpara\'{i}so during his visit to this institute, where this work was accomplished. \\\\

{\textbf{Data Availability Statement:} No Data associated in the manuscript.}

\appendix

\section{Derivation of the roots of a quartic equation}\label{app:A}

Here, we provide the method of solving the quartic equation $\xi^c=0$ that generates the radii of polar orbits $r_\po$. This method is also used to find the turning points in the radial motion. In fact, the equation $\xi^c=0$ can be rewritten as 
\begin{equation}
P_4(r)=r^4-\frac{2}{N_s}r^3+\frac{3}{N_s}\left(M-a^2N_s\right)r^2-\frac{2a^2}{N_s}r-\frac{2a^2}{M}=0.
    \label{eq:A1}
\end{equation}
Applying the change of variable $r=u-\frac{1}{2N_s}$, the above polynomial equation reduces to 
\begin{equation}
P_4(u)=u^4+\mathrm{p}_2 u^2+\mathrm{p}_1 u+\mathrm{p}_0,
    \label{eq:A2}
\end{equation}
where
\begin{subequations}
    \begin{align}
        & \mathrm{p}_2=\frac{3(4MN_s-1)}{2N_s^2}-3a^2,\\
        & \mathrm{p}_1=\frac{6MN_s-5a^2N_s^2-1}{N_s^3},\\
        & \mathrm{p}_0=-\frac{3+4N_s\left[7a^2N_s+2M\left(4a^2N_s^2-3\right)\right]}{16N_s^4}.
    \end{align}
    \label{eq:A3}
\end{subequations}
The solutions to this quartic are then obtained as (see the appendices of Refs. \cite{fathi_gravitational_2021,fathi_spherical_2023})
\begin{eqnarray}
    && u_1 = \bar{\mathrm{a}}+\sqrt{\bar{\mathrm{a}}^2-\bar{\mathrm{b}}},\label{eq:A4}\\
    && u_2 = \bar{\mathrm{a}}-\sqrt{\bar{\mathrm{a}}^2-\bar{\mathrm{b}}},\label{eq:A5}\\
    && u_3 = -\bar{\mathrm{a}}+\sqrt{\bar{\mathrm{a}}^2-\bar{\mathrm{c}}},\label{eq:A6}\\
    && u_4 = -\bar{\mathrm{a}}-\sqrt{\bar{\mathrm{a}}^2-\bar{\mathrm{c}}},\label{eq:A7}
\end{eqnarray}
where 
\begin{subequations}
    \begin{align}
        & \bar{\mathrm{a}} = \sqrt{\bar{\mathrm{u}}-\frac{\mathrm{p}_2}{6}},\\
        & \bar{\mathrm{b}}=2\bar{\mathrm{a}}^2+\frac{\mathrm{p}_2}{2}+\frac{\mathrm{p}_1}{4\bar{\mathrm{a}}},\\
         & \bar{\mathrm{c}}=2\bar{\mathrm{a}}^2+\frac{\mathrm{p}_2}{2}-\frac{\mathrm{p}_1}{4\bar{\mathrm{a}}},
    \end{align}
    \label{eq:A8}
\end{subequations}
and
\begin{equation}
\bar{\mathrm{u}}=\sqrt{\frac{\bar{\epsilon}_2}{3}}\cosh\left(\frac{1}{3}\arccosh\left(3\bar{\epsilon}_3\sqrt{\frac{3}{\bar{\epsilon}_2^3}}\right)\right),
    \label{eq:A9}
\end{equation}
with
\begin{subequations}
    \begin{align}
& \bar{\epsilon}_2 = \frac{\mathrm{p}_2^2}{12}+\mathrm{p}_0,\\
& \bar{\mathrm{\epsilon}}_3 = \frac{\mathrm{p}_2^3}{216}-\frac{\mathrm{p}_2\mathrm{p}_0}{6}+\frac{\mathrm{p}_1^2}{16}.
    \end{align}
    \label{eq:A10}
\end{subequations}
The positive radii of polar photon orbits outside the event horizon are then given in terms of the values in Eqs. \eqref{eq:A4} and \eqref{eq:A5}. However, the only reliable solution inside the domain $r\in[r_+,r_{++}]$ is based on Eq. \eqref{eq:A5}, and is obtained as
\begin{equation}
       r_{\po} = u_2 - \frac{1}{2N_s}.
       \label{eq:A11}
\end{equation}
Employing the same methods, the roots of the characteristic polynomial \eqref{eq:P(r)_char} are obtained as
\begin{eqnarray}
r_d &=& \tilde{\alpha}+\sqrt{\tilde{\alpha}^2-\tilde{\beta}}-{\bar{A}\over 4},\label{eq:A12}\\
r_f &=& \tilde{\alpha}-\sqrt{\tilde{\alpha}^2-\tilde{\beta}}-{\bar{A}\over 4},\label{eq:A13}\\
r_3 &=& -\tilde{\alpha}+\sqrt{\tilde{\alpha}^2-\tilde{\gamma}}-{\bar{A}\over 4},\label{eq:A14}\\
r_4 &=& -\tilde{\alpha}-\sqrt{\tilde{\alpha}^2-\tilde{\gamma}}-{\bar{A}\over 4},\label{eq:A15}
\end{eqnarray}
where $\bar{A}=N_s \left[\left(\xi-a \right)^2+\eta_r
 \right]$, and 
\begin{subequations}\label{eq:A16}
	\begin{align}
	&  \tilde{\alpha} = \sqrt{\Xi-{\tilde{p}\over 3}},\\
	&  \tilde{\beta} = 2\tilde{\alpha}^2 +{\mathcal{A}\over 2}+{\mathcal{B}\over 4\tilde{\alpha}},\\
 &  \tilde{\gamma} = 2\tilde{\alpha}^2 +{\mathcal{A}\over 2}-{\mathcal{B}\over 4\tilde{\alpha}},
	\end{align}
\end{subequations}
with
\begin{equation}\label{eq:A17}
\Xi = \sqrt{{\chi_2\over 3}} \cosh\left(\frac{1}{3}\arccosh\left(3\chi_3\sqrt{{3\over \chi_2^3}}  \right) \right),
\end{equation}
in which
\begin{subequations}\label{eq:A18}
	\begin{align}
	& \chi_2 = 4 \left( {\tilde{p}^2\over 3}-\tilde{q}\right), \\
	& \chi_3 =4 \left( {2\tilde{p}^3\over 27}+{\tilde{p}\,\tilde{q}\over 3}- \tilde{r}\right),
	\end{align}
\end{subequations}
where 
\begin{subequations}\label{eq:A19}
	\begin{align}
	&  \tilde{r} = -{\mathcal{B}^2\over 64},\\
	&  \tilde{q} = {1\over 4} \left( {\mathcal{A}^2\over 4}-\mathcal{C}\right),\\
	&  \tilde{p} ={\mathcal{A}\over 2},
	\end{align}
\end{subequations}
and
\begin{subequations}\label{eq:A20}
	\begin{align}
	&  \mathcal{A} =\mathrm{a}_2-{3\mathrm{a}_3^2\over 8},\\
	&  \mathcal{B} = \mathrm{a}_1+{\mathrm{a}_3^3\over 8}-{\mathrm{a}_3\,\mathrm{a}_2\over 2},\\
	&  \mathcal{C} =\mathrm{a}_0-{3\mathrm{a}_3^4\over 256}+{\mathrm{a}_3^2 \,\mathrm{a}_2\over 16}-{\mathrm{a}_3\,\bar{C}\over 4},
	\end{align}
\end{subequations}
with $\bar{C}=2 M \left[\left(\xi-a\right)^2+\eta_r
 \right].$
%

\section{Coefficients and the analytical relations for the integrals of $\Phi_r(\gamma)$  and $t_r(\gamma)$}\label{app:B}

For the DFK, we have defined in Eq. \eqref{eq:phi_rDFK} that
\begin{subequations}
    \begin{align}
  &  A_1^\DFK = - \frac{
  L a^2-2Ma\omega_0r_--N_sa\omega_0r_-^3
  }{(r_+-r_-)(r_{++}-r_-)(r_d-r_-)(r_f-r_-)},\\
  &  A_2^\DFK =  \frac{
  L a^2-2Ma\omega_0r_+ -N_sa\omega_0r_+^3
  }{(r_+-r_-)(r_{++}-r_+)(r_d-r_+)(r_f-r_+)},\\
  &  A_3^\DFK =  \frac{
  -L a^2+2Ma\omega_0r_{++}+N_sa\omega_0r_{++}^3
  }{(r_{++}-r_+)(r_{++}-r_-)(r_{++}-r_d)(r_{++}-r_f)}.
    \end{align}
    \label{eq:B1}
\end{subequations}
Furthermore, the elliptic integrals in the summation in Eq. \eqref{eq:phi_rDFK} are expressed as
\begin{subequations}
    \begin{align}
& I_1^\DFK(\gamma) = \Big[
(r_d-r_f)\bPi_i^- + (r_d-r_-)\bF_i
\Big] \Big{|}_{r_\s}^{r_\ob(\gamma)},\\
& I_2^\DFK(\gamma) = \Big[
(r_d-r_f)\bPi_i^+ + (r_d-r_+)\bF_i
\Big] \Big{|}_{r_\s}^{r_\ob(\gamma)},\\
& I_3^\DFK(\gamma) = \Big[
(r_d-r_f)\bPi_i^{++} - (r_{++}-r_d)\bF_i
\Big] \Big{|}_{r_\s}^{r_\ob(\gamma)},
    \end{align}
    \label{eq:B3}
\end{subequations}
with
\begin{subequations}
    \begin{align}
        & \bPi_i^-=\bPi\left(
        \frac{(r_d-r_4)(r_f-r_-)}{(r_d-r_-)(r_f-r_4)};\tilde{\varphi}_i\Big|\tilde{\m}
        \right),\\
        & \bPi_i^+=\bPi\left(
        \frac{(r_d-r_4)(r_f-r_+)}{(r_d-r_+)(r_f-r_4)};\tilde{\varphi}_i\Big|\tilde{\m}
        \right),\\
         & \bPi_i^{++}=\bPi\left(
        \frac{(r_d-r_4)(r_{++}-r_f)}{(r_{++}-r_d)(r_f-r_4)};\tilde{\varphi}_i\Big|\tilde{\m}
        \right),\\
        & \bF_i = \bF(\tilde{\varphi}_i|\tilde{\m}),
    \end{align}
    \label{eq:B4}
\end{subequations}
in which $\tilde{\varphi}_i$ changes from $\tilde{\varphi}_\s$ to $\tilde{\varphi}_\ob(\gamma)$. 
For the DSK, the integrals $I_j^{\DSK}$ are defined as
\begin{subequations}
    \begin{align}
        & I_1^\DSK(\gamma) = \Big[
(r_d-r_f)\tilde{\bPi}_i^- + (r_f-r_-)\tilde{\bF}_i
\Big] \Big{|}_{r_\s}^{r_\ob(\gamma)},\\
& I_2^\DSK(\gamma) = \Big[
(r_d-r_f)\tilde{\bPi}_i^+ + (r_f-r_+)\tilde{\bF}_i
\Big] \Big{|}_{r_\s}^{r_\ob(\gamma)},\\
& I_3^\DSK(\gamma) = \Big[
(r_d-r_f)\tilde{\bPi}_i^{++} - (r_{++}-r_f)\tilde{\bF}_i
\Big] \Big{|}_{r_\s}^{r_\ob(\gamma)},
    \end{align}
    \label{eq:B5}
\end{subequations}
in which
\begin{subequations}
    \begin{align}
        & \tilde{\bPi}_i^-=\bPi\left(
        \frac{(r_f-r_3)(r_d-r_-)}{(r_d-r_3)(r_f-r_-)};\tsup{\varphi}_i\Big|\tilde{\m}
        \right),\\
        & \tilde{\bPi}_i^+=\bPi\left(
        \frac{(r_f-r_3)(r_d-r_+)}{(r_d-r_3)(r_f-r_+)};\tsup{\varphi}_i\Big|\tilde{\m}
        \right),\\
         & \tilde{\bPi}_i^{++}=\bPi\left(
        \frac{(r_f-r_3)(r_{++}-r_d)}{(r_d-r_3)(r_{++}-r_f)};\tsup{\varphi}_i\Big|\tilde{\m}
        \right),\\
        & \tilde{\bF}_i = \bF(\tsup{\varphi}_i|\tilde{\m}),
    \end{align}
    \label{eq:B6}
\end{subequations}
where $\tsup{\varphi}_i$ changes from $\tsup{\varphi}_\s$ to $\tsup{\varphi}_\ob(\gamma)$. 

The parameters involved in the $r$-dependent expression for the COFK, as provided in Eq. \eqref{eq:Phir_COFK}, are
\begin{subequations}
    \begin{align}
        & A_1^\COFK = \frac{La^2-2Ma\omega_0r_c-N_sa\omega_0 r_c^3}{(r_c-r_-)(r_c-r_+)(r_{++}-r_c)\sqrt{(r_c-r_3)(r_c-r_4)}},\\
        & A_2^\COFK = \frac{La^2-2Ma\omega_0r_--N_sa\omega_0 r_-^3}{(r_c-r_-)(r_+-r_-)(r_{++}-r_-)\sqrt{(r_--r_3)(r_--r_4)}},\\
        & A_3^\COFK = \frac{-La^2+2Ma\omega_0r_++N_sa\omega_0 r_+^3}{(r_c-r_+)(r_+-r_-)(r_{++}-r_+)\sqrt{(r_+-r_3)(r_+-r_4)}},\\
        & A_4^\COFK = \frac{-La^2+2Ma\omega_0r_{++}+N_sa\omega_0 r_{++}^3}{(r_{++}-r_c)(r_{++}-r_-)(r_{++}-r_+)\sqrt{(r_{++}-r_3)(r_{++}-r_4)}},
    \end{align}
    \label{eq:B7}
\end{subequations}
together with the inverse hyperbolic functions
\begin{subequations}
    \begin{align}
& I_1^\COFK(\gamma)=\left.\arccosh\left(\sqrt{\frac{(r_c-r_4)(r-r_3)}{(r_3-r_4)(r-r_c)}}\,\right)\right|_{r_\s}^{r_\ob(\gamma)},\\
& I_2^\COFK(\gamma)=\left.\arccosh\left(\sqrt{\frac{(r_- -r_4)(r-r_3)}{(r_3-r_4)(r-r_-)}}\,\right)\right|_{r_\s}^{r_\ob(\gamma)},\\
& I_3^\COFK(\gamma)=\left.\arccosh\left(\sqrt{\frac{(r_+-r_4)(r-r_3)}{(r_3-r_4)(r-r_+)}}\,\right)\right|_{r_\s}^{r_\ob(\gamma)},\\
& I_4^\COFK(\gamma)=\left.\arccosh\left(\sqrt{\frac{(r_{++}-r_4)(r-r_3)}{(r_3-r_4)(r_{++}-r)}}\,\right)\right|_{r_\s}^{r_\ob(\gamma)},
    \end{align}
    \label{eq:B8}
\end{subequations}
in which, $r_\ob(\gamma)$ corresponds to the analytical solution \eqref{eq:r(gamma)_cofk} for the COFK. 

In the solution \eqref{eq:Phir_PO} we have defined 
\begin{subequations}
    \begin{align}
& g_p=\frac{1}{\sqrt{\tilde{A} \tilde{B}}},\\
& \tilde{\varphi}_{p,\ob} = \arccos\left(\frac{r_3\tilde{B}-r_4\tilde{A}+(\tilde{A}-\tilde{B})r_\ob(\gamma)}{-r_3 \tilde{B}-r_4 \tilde{A}+(\tilde{A}-\tilde{B})r_\ob(\gamma)}\right),
    \end{align}
    \label{eq:B9}
\end{subequations}
and the expressions for $\tilde{A}$, $\tilde{B}$, $\tilde{\m}_p$ and $\tilde{\varphi}_{p,\s}$ have been given in Eqs. \eqref{eq:kappa_p,etc.}, and $r_\ob(\gamma)$ corresponds to the analytical solution \eqref{eq:r(gamma)_po} for plunge orbits. Furthermore, the coefficients $A_j^{\PO}$ in Eq. \eqref{eq:Phir_PO} are defined as 
\begin{subequations}
    \begin{align}
&  A_1^\PO = \frac{L a^2-2Ma\omega_0 r_- - N_s a \omega_0 r_-^3}{(r_+-r_-)(r_{++}-r_-)},\\
&  A_2^\PO = \frac{-L a^2+2Ma\omega_0 r_+ +N_s a \omega_0 r_+^3}{(r_+-r_-)(r_{++}-r_+)},\\
&  A_3^\PO = \frac{-L a^2+2Ma\omega_0 r_{++} +N_s a \omega_0 r_{++}^3}{(r_{++}-r_-)(r_{++}-r_+)}.
    \end{align}
    \label{eq:B10}
\end{subequations}
The included elliptic integrals in this solution are given by 
\begin{subequations}
    \begin{align}
& I_1^\PO(\gamma)=\left.\frac{g_p(\tilde{B}-\tilde{A})\left[
(\oalpha{}_- - \alpha_p)\boR{}_i^{-}+\alpha_p\obF{}_i
\right]}{r_3\tilde{B}-r_4\tilde{A}-(\tilde{A}+\tilde{B})r_-}\right|_{r_\s}^{r_\ob(\gamma)},\\
& I_2^\PO(\gamma)=\left.\frac{g_p(\tilde{B}-\tilde{A})\left[
(\oalpha{}_+ - \alpha_p)\boR{}_i^{+}+\alpha_p\obF{}_i
\right]}{r_3\tilde{B}-r_4\tilde{A}-(\tilde{A}+\tilde{B})r_+}\right|_{r_\s}^{r_\ob(\gamma)},\\
& I_3^\PO(\gamma)=\left.\frac{g_p(\tilde{B}-\tilde{A})\left[
(\oalpha{}_{++} - \alpha_p)\boR{}_i^{++}+\alpha_p\obF{}_i
\right]}{r_3\tilde{B}-r_4\tilde{A}-(\tilde{A}+\tilde{B})r_{++}}\right|_{r_\s}^{r_\ob(\gamma)},
    \end{align}
    \label{eq:B11}
\end{subequations}
in which $\obF_i=\bF(\tilde{\varphi}_{p,i}|\tilde{\m})$, $\alpha_p=(\tilde{A}+\tilde{B})/(\tilde{B}-\tilde{A})$, and 
\begin{subequations}
    \begin{align}
        & \oalpha{}_- = \frac{\tilde{B}r_3+\tilde{A}r_4-(\tilde{A}+\tilde{B})r_-}{\tilde{B}r_3-\tilde{A}r_4+(\tilde{A}+\tilde{B})r_-},\\
        & \oalpha{}_+ = \frac{\tilde{B}r_3+\tilde{A}r_4-(\tilde{A}+\tilde{B})r_+}{\tilde{B}r_3-\tilde{A}r_4+(\tilde{A}+\tilde{B})r_+},\\
        & \oalpha{}_{++} = \frac{\tilde{B}r_3+\tilde{A}r_4-(\tilde{A}+\tilde{B})r_{++}}{\tilde{B}r_3-\tilde{A}r_4+(\tilde{A}+\tilde{B})r_{++}},
    \end{align}
    \label{eq:B12}
\end{subequations}
so that
\begin{subequations}
    \begin{align}
 & \boR{}_i^{-}=\frac{1}{1-\oalpha{}_-^2}\left[
  \obPi{}_i^--\oalpha{}_- \obf{}_i^-
  \right],\\
 & \boR{}_i^{+}=\frac{1}{1-\oalpha{}_+^2}\left[
  \obPi{}_i^+-\oalpha{}_+ \obf{}_i^+
  \right],\\
 & \boR{}_i^{++}=\frac{1}{1-\oalpha{}_{++}^2}\left[
  \obPi{}_i^{++}-\oalpha{}_{++} \obf{}_i^{++}
  \right],
    \end{align}
    \label{eq:B13}
\end{subequations}
where
\begin{subequations}
    \begin{align}
& \obPi{}_i^{-} = \bPi\left(
\frac{\oalpha{}_-^2}{\oalpha{}_-^2-1};\tilde{\varphi}_{p,i}\Big|\tilde{\m}_p
\right),\\
& \obPi{}_i^{+} = \bPi\left(
\frac{\oalpha{}_+^2}{\oalpha{}_+^2-1};\tilde{\varphi}_{p,i}\Big|\tilde{\m}_p
\right),\\
& \obPi{}_i^{++} = \bPi\left(
\frac{\oalpha{}_{++}^2}{\oalpha{}_{++}^2-1};\tilde{\varphi}_{p,i}\Big|\tilde{\m}_p
\right),
    \end{align}
    \label{eq:B14}
\end{subequations}
with
\begin{subequations}
    \begin{align}
& \obf{}_i^{-} = \sqrt{\frac{\oalpha{}_{-}^2-1}{\tilde{\m}_p+(1-\tilde{\m}_p)\oalpha{}_{-}^2}}\,\ln\left|
\frac{\sqrt{\tilde{\m}_p+(1-\tilde{\m}_p)\oalpha{}_{-}^2}\,\dn(\tilde{\mathbf{u}}_{p,i}|\tilde{\m}_p)+\sqrt{\oalpha{}_{-}^2-1}\,\sn(\tilde{\mathbf{u}}_{p,i}|\tilde{\m}_p)}{\sqrt{\tilde{\m}_p+(1-\tilde{\m}_p)\oalpha{}_{-}^2}\,\dn(\tilde{\mathbf{u}}_{p,i}|\tilde{\m}_p)-\sqrt{\oalpha{}_{-}^2-1}\,\sn(\tilde{\mathbf{u}}_{p,i}|\tilde{\m}_p)}
\right|,\\
& \obf{}_i^{+} = \sqrt{\frac{\oalpha{}_{+}^2-1}{\tilde{\m}_p+(1-\tilde{\m}_p)\oalpha{}_{+}^2}}\,\ln\left|
\frac{\sqrt{\tilde{\m}_p+(1-\tilde{\m}_p)\oalpha{}_{+}^2}\,\dn(\tilde{\mathbf{u}}_{p,i}|\tilde{\m}_p)+\sqrt{\oalpha{}_{+}^2-1}\,\sn(\tilde{\mathbf{u}}_{p,i}|\tilde{\m}_p)}{\sqrt{\tilde{\m}_p+(1-\tilde{\m}_p)\oalpha{}_{+}^2}\,\dn(\tilde{\mathbf{u}}_{p,i}|\tilde{\m}_p)-\sqrt{\oalpha{}_{+}^2-1}\,\sn(\tilde{\mathbf{u}}_{p,i}|\tilde{\m}_p)}
\right|,\\
& \obf{}_i^{++} = \sqrt{\frac{\oalpha{}_{++}^2-1}{\tilde{\m}_p+(1-\tilde{\m}_p)\oalpha{}_{++}^2}}\,\ln\left|
\frac{\sqrt{\tilde{\m}_p+(1-\tilde{\m}_p)\oalpha{}_{++}^2}\,\dn(\tilde{\mathbf{u}}_{p,i}|\tilde{\m}_p)+\sqrt{\oalpha{}_{++}^2-1}\,\sn(\tilde{\mathbf{u}}_{p,i}|\tilde{\m}_p)}{\sqrt{\tilde{\m}_p+(1-\tilde{\m}_p)\oalpha{}_{++}^2}\,\dn(\tilde{\mathbf{u}}_{p,i}|\tilde{\m}_p)-\sqrt{\oalpha{}_{++}^2-1}\,\sn(\tilde{\mathbf{u}}_{p,i}|\tilde{\m}_p)}
\right|,
    \end{align}
    \label{eq:B15}
\end{subequations}
where $\dn(x|\tilde{\m}_p)$ is the Jacobian elliptic delta amplitude of the argument $x$, and $\tilde{\mathbf{u}}_{p,i}=\cn^{-1}(\cos(\tilde{\varphi}_{p,i})|\tilde{\m}_p)$, in which $\cn^{-1}(x|\tilde{\m}_p)$ is the inverse Jacobian elliptic cosine function of the argument $x$ \cite{byrd_handbook_1971}.

The coefficients $B_j^t$ in the solution \eqref{eq:t_rgamma} are given as
\begin{subequations}
    \begin{align}
 &B_1^t=\frac{-a^4\omega_0+r_-\left[2aLM-r_-\left(2a^2\omega_0-aLN_sr_-+\omega_0r_-^2\right)\right]}{M N_s(r_{++}-r_-)(r_+-r_-)(r_d-r_-)(r_f-r_-)\sqrt{(r_d-r_3)(r_f-r_4)}},\\
&B_2^t=\frac{a^4\omega_0+r_+\left[-2aLM+r_+\left(2a^2\omega_0-aLN_sr_+ +\omega_0r_+^2\right)\right]}{M N_s(r_{++}-r_+)(r_+-r_-)(r_d-r_+)(r_f-r_+)\sqrt{(r_d-r_3)(r_f-r_4)}},\\
&B_3^t=-\frac{a^4\omega_0+r_{++}\left[-2aLM+r_{++}\left(2a^2\omega_0-aLN_sr_{++} +\omega_0r_{++}^2\right)\right]}{M N_s(r_{++}-r_+)(r_{++}-r_-)(r_{++}-r_d)(r_{++}-r_f)\sqrt{(r_d-r_3)(r_f-r_4)}}.
    \end{align}
    \label{eq:B16}
\end{subequations}

\bibliographystyle{ieeetr}
\bibliography{biblio_v1.bib}

\begin{thebibliography}{100}

\bibitem{riess_type_2004}
A.~G. Riess, L.~Strolger, J.~Tonry, S.~Casertano, H.~C. Ferguson, B.~Mobasher,
  P.~Challis, A.~V. Filippenko, S.~Jha, W.~Li, R.~Chornock, R.~P. Kirshner,
  B.~Leibundgut, M.~Dickinson, M.~Livio, M.~Giavalisco, C.~C. Steidel,
  T.~Benitez, and Z.~Tsvetanov, ``Type {Ia} {Supernova} {Discoveries} at
  \textit{z} {\textgreater} 1 from the \textit{{Hubble} {Space} {Telescope}} :
  {Evidence} for {Past} {Deceleration} and {Constraints} on {Dark} {Energy}
  {Evolution},'' {\em The Astrophysical Journal}, vol.~607, pp.~665--687, June
  2004.

\bibitem{komatsu_seven-year_2011}
E.~Komatsu, K.~M. Smith, J.~Dunkley, C.~L. Bennett, B.~Gold, G.~Hinshaw,
  N.~Jarosik, D.~Larson, M.~R. Nolta, L.~Page, D.~N. Spergel, M.~Halpern, R.~S.
  Hill, A.~Kogut, M.~Limon, S.~S. Meyer, N.~Odegard, G.~S. Tucker, J.~L.
  Weiland, E.~Wollack, and E.~L. Wright, ``{SEVEN}-{YEAR} \textit{{WILKINSON}
  {MICROWAVE} {ANISOTROPY} {PROBE}} ( \textit{{WMAP}} ) {OBSERVATIONS}:
  {COSMOLOGICAL} {INTERPRETATION},'' {\em The Astrophysical Journal Supplement
  Series}, vol.~192, p.~18, Feb. 2011.

\bibitem{turner_coherent_1983}
M.~S. Turner, ``Coherent scalar-field oscillations in an expanding universe,''
  {\em Physical Review D}, vol.~28, pp.~1243--1247, Sept. 1983.

\bibitem{press_single_1990}
W.~H. Press, B.~S. Ryden, and D.~N. Spergel, ``Single mechanism for generating
  large-scale structure and providing dark missing matter,'' {\em Physical
  Review Letters}, vol.~64, pp.~1084--1087, Mar. 1990.

\bibitem{dubinski_structure_1991}
J.~Dubinski and R.~G. Carlberg, ``The structure of cold dark matter halos,''
  {\em The Astrophysical Journal}, vol.~378, p.~496, Sept. 1991.

\bibitem{sin_late-time_1994}
S.-J. Sin, ``Late-time phase transition and the galactic halo as a {Bose}
  liquid,'' {\em Physical Review D}, vol.~50, pp.~3650--3654, Sept. 1994.

\bibitem{navarro_structure_1996}
J.~F. Navarro, C.~S. Frenk, and S.~D.~M. White, ``The {Structure} of {Cold}
  {Dark} {Matter} {Halos},'' {\em The Astrophysical Journal}, vol.~462, p.~563,
  May 1996.

\bibitem{navarro_universal_1997}
J.~F. Navarro, C.~S. Frenk, and S.~D.~M. White, ``A {Universal} {Density}
  {Profile} from {Hierarchical} {Clustering},'' {\em The Astrophysical
  Journal}, vol.~490, pp.~493--508, Dec. 1997.

\bibitem{hu_fuzzy_2000}
W.~Hu, R.~Barkana, and A.~Gruzinov, ``Fuzzy {Cold} {Dark} {Matter}: {The}
  {Wave} {Properties} of {Ultralight} {Particles},'' {\em Physical Review
  Letters}, vol.~85, pp.~1158--1161, Aug. 2000.

\bibitem{peebles_fluid_2000}
P.~J.~E. Peebles, ``Fluid {Dark} {Matter},'' {\em The Astrophysical Journal},
  vol.~534, pp.~L127--L129, May 2000.

\bibitem{peebles_cosmological_2003}
P.~J.~E. Peebles and B.~Ratra, ``The cosmological constant and dark energy,''
  {\em Reviews of Modern Physics}, vol.~75, pp.~559--606, Apr. 2003.

\bibitem{kiselev2003quintessential}
V.~V. {Kiselev}, ``{Quintessential solution of dark matter rotation curves and
  its simulation by extra dimensions},'' {\em arXiv e-prints},
  pp.~gr--qc/0303031, Mar. 2003.

\bibitem{viel_constraining_2005}
M.~Viel, J.~Lesgourgues, M.~G. Haehnelt, S.~Matarrese, and A.~Riotto,
  ``Constraining warm dark matter candidates including sterile neutrinos and
  light gravitinos with {WMAP} and the {Lyman}-$\alpha$ forest,'' {\em Physical
  Review D}, vol.~71, p.~063534, Mar. 2005.

\bibitem{amendola_dark_2006}
L.~Amendola and R.~Barbieri, ``Dark matter from an ultra-light
  pseudo-{Goldsone}-boson,'' {\em Physics Letters B}, vol.~642, pp.~192--196,
  Nov. 2006.

\bibitem{copeland_dynamics_2006}
E.~J. Copeland, M.~Sami, and S.~Tsujikawa, ``{DYNAMICS} {OF} {DARK} {ENERGY},''
  {\em International Journal of Modern Physics D}, vol.~15, pp.~1753--1935,
  Nov. 2006.

\bibitem{de_la_macorra_dark_2006}
A.~De~La~Macorra and T.~Matos, ``Dark {Energy} and {Dark} {Matter},'' in {\em
  {AIP} {Conference} {Proceedings}}, vol.~857, (Morelia, Michoacan (Mexico)),
  pp.~191--204, AIP, 2006.

\bibitem{rahaman_perfect_2010}
F.~Rahaman, K.~Nandi, A.~Bhadra, M.~Kalam, and K.~Chakraborty, ``Perfect fluid
  dark matter,'' {\em Physics Letters B}, vol.~694, pp.~10--15, Oct. 2010.

\bibitem{rahaman_modeling_2011}
F.~Rahaman, P.~K.~F. Kuhfittig, K.~Chakraborty, M.~Kalam, and D.~Hossain,
  ``Modeling {Galactic} {Halos} with {Predominantly} {Quintessential}
  {Matter},'' {\em International Journal of Theoretical Physics}, vol.~50,
  pp.~2655--2665, Sept. 2011.

\bibitem{li_dark_2011}
M.~Li, X.-D. Li, S.~Wang, and Y.~Wang, ``Dark {Energy},'' {\em Communications
  in Theoretical Physics}, vol.~56, pp.~525--604, Sept. 2011.

\bibitem{schive_cosmic_2014}
H.-Y. Schive, T.~Chiueh, and T.~Broadhurst, ``Cosmic structure as the quantum
  interference of a coherent dark wave,'' {\em Nature Physics}, vol.~10,
  pp.~496--499, July 2014.

\bibitem{marsh_axion_2016}
D.~J. Marsh, ``Axion cosmology,'' {\em Physics Reports}, vol.~643, pp.~1--79,
  July 2016.

\bibitem{hui_ultralight_2017}
L.~Hui, J.~P. Ostriker, S.~Tremaine, and E.~Witten, ``Ultralight scalars as
  cosmological dark matter,'' {\em Physical Review D}, vol.~95, p.~043541, Feb.
  2017.

\bibitem{wang_holographic_2017}
S.~Wang, Y.~Wang, and M.~Li, ``Holographic dark energy,'' {\em Physics
  Reports}, vol.~696, pp.~1--57, June 2017.

\bibitem{Rastall19723357}
P.~Rastall, ``Generalization of the einstein theory,'' {\em Physical Review D},
  vol.~6, no.~12, p.~3357 – 3359, 1972.
\newblock Cited by: 291.

\bibitem{rastall_theory_1976}
P.~Rastall, ``A theory of gravity,'' {\em Canadian Journal of Physics},
  vol.~54, pp.~66--75, Jan. 1976.

\bibitem{batista_rastall_2012}
C.~E.~M. Batista, M.~H. Daouda, J.~C. Fabris, O.~F. Piattella, and D.~C.
  Rodrigues, ``Rastall cosmology and the $\lambda${CDM} model,'' {\em Physical
  Review D}, vol.~85, p.~084008, Apr. 2012.

\bibitem{fabris_rastall_2012}
J.~C. Fabris, O.~F. Piattella, D.~C. Rodrigues, C.~E.~M. Batista, and M.~H.
  Daouda, ``{RASTALL} {COSMOLOGY},'' {\em International Journal of Modern
  Physics: Conference Series}, vol.~18, pp.~67--76, Jan. 2012.

\bibitem{fabris_rastalls_2015}
J.~C. Fabris, O.~F. Piattella, D.~C. Rodrigues, and M.~H. Daouda, ``Rastall’s
  cosmology and its observational constraints,'' (Valparaiso, Chile),
  pp.~50--53, 2015.

\bibitem{heydarzade_black_2017-1}
Y.~Heydarzade, H.~Moradpour, and F.~Darabi, ``Black hole solutions in {Rastall}
  theory,'' {\em Canadian Journal of Physics}, vol.~95, pp.~1253--1256, Dec.
  2017.

\bibitem{heydarzade_black_2017}
Y.~Heydarzade and F.~Darabi, ``Black hole solutions surrounded by perfect fluid
  in {Rastall} theory,'' {\em Physics Letters B}, vol.~771, pp.~365--373, Aug.
  2017.

\bibitem{darabi_einstein_2018}
F.~Darabi, K.~Atazadeh, and Y.~Heydarzade, ``Einstein static universe in the
  {Rastall} theory of gravity,'' {\em The European Physical Journal Plus},
  vol.~133, p.~249, July 2018.

\bibitem{lobo_thermodynamics_2018}
I.~P. Lobo, H.~Moradpour, J.~P. Morais~Graça, and I.~G. Salako,
  ``Thermodynamics of black holes in {Rastall} gravity,'' {\em International
  Journal of Modern Physics D}, vol.~27, p.~1850069, May 2018.

\bibitem{guo_shadow_2021}
S.~Guo, K.-J. He, G.-R. Li, and G.-P. Li, ``The shadow and photon sphere of the
  charged black hole in {Rastall} gravity,'' {\em Classical and Quantum
  Gravity}, vol.~38, p.~165013, Aug. 2021.

\bibitem{guo_observable_2022}
S.~Guo, G.-R. Li, and E.-W. Liang, ``Observable characteristics of the charged
  black hole surrounded by thin disk accretion in {Rastall} gravity,'' {\em
  Classical and Quantum Gravity}, vol.~39, p.~135004, July 2022.

\bibitem{kottler_uber_1918}
F.~Kottler, ``Über die physikalischen {Grundlagen} der {Einsteinschen}
  {Gravitationstheorie},'' {\em Annalen der Physik}, vol.~361, pp.~401--462,
  Jan. 1918.

\bibitem{Carter:1973rla}
B.~Carter, ``{Black holes equilibrium states},'' in {\em {Les Houches Summer
  School of Theoretical Physics}: {Black Holes}}, pp.~57--214, 1973.

\bibitem{kiselev_quintessence_2003}
V.~V. Kiselev, ``Quintessence and black holes,'' {\em Classical and Quantum
  Gravity}, vol.~20, pp.~1187--1197, Mar. 2003.

\bibitem{toshmatov_rotating_2017}
B.~Toshmatov, Z.~Stuchlík, and B.~Ahmedov, ``Rotating black hole solutions
  with quintessential energy,'' {\em The European Physical Journal Plus},
  vol.~132, p.~98, Feb. 2017.

\bibitem{kumar_rotating_2018}
R.~Kumar and S.~G. Ghosh, ``Rotating black hole in {Rastall} theory,'' {\em The
  European Physical Journal C}, vol.~78, p.~750, Sept. 2018.

\bibitem{xu_kerrnewman-ads_2018}
Z.~Xu, X.~Hou, X.~Gong, and J.~Wang, ``Kerr–{Newman}-{AdS} black hole
  surrounded by perfect fluid matter in {Rastall} gravity,'' {\em The European
  Physical Journal C}, vol.~78, p.~513, June 2018.

\bibitem{Eddington:1920}
F.~W. Dyson, A.~S. Eddington, and C.~Davidson, ``Ix. a determination of the
  deflection of light by the sun's gravitational field, from observations made
  at the total eclipse of may 29, 1919,'' {\em Philosophical Transactions of
  the Royal Society of London. Series A, Containing Papers of a Mathematical or
  Physical Character}, vol.~220, no.~571-581, pp.~291--333, 1920.

\bibitem{Einstein:1911}
A.~Einstein, ``Über den einfluß der schwerkraft auf die ausbreitung des
  lichtes,'' {\em Annalen der Physik}, vol.~340, no.~10, pp.~898--908, 1911.

\bibitem{Akiyama:2019}
K.~Akiyama {\em et~al.}, ``{First M87 Event Horizon Telescope Results. IV.
  Imaging the Central Supermassive Black Hole},'' {\em Astrophys. J. Lett.},
  vol.~875, no.~1, p.~L4, 2019.

\bibitem{Akiyama:2022}
K.~Akiyama {\em et~al.}, ``{First Sagittarius A{$^{*}$} Event Horizon Telescope
  results. I. The shadow of the supermassive black hole in the center of the
  Milky Way.},'' {\em Astrophys. J. Lett.}, vol.~930, no.~2, p.~L12, 2022.

\bibitem{Bardeen:1972a}
J.~M. {Bardeen}, W.~H. {Press}, and S.~A. {Teukolsky}, ``{Rotating Black Holes:
  Locally Nonrotating Frames, Energy Extraction, and Scalar Synchrotron
  Radiation},'' {\em Astrophysical Journal}, vol.~178, pp.~347--370, Dec. 1972.

\bibitem{Bardeen:1973a}
J.~Bardeen, ``{Timelike and null geodesics in the Kerr metric},'' in {\em {Les
  Houches Summer School of Theoretical Physics}: {Black Holes}}, pp.~215--240,
  1973.

\bibitem{Bardeen:1973b}
J.~Bardeen, ``{Timelike and null geodesics in the Kerr metric},'' in {\em {Les
  Houches Summer School of Theoretical Physics}: {Black Holes}}, pp.~215--240,
  1973.

\bibitem{Chandrasekhar:2002}
S.~Chandrasekhar, {\em The mathematical theory of black holes}.
\newblock Oxford classic texts in the physical sciences, Oxford: Oxford Univ.
  Press, 2002.

\bibitem{Virbhadra:2000}
K.~S. Virbhadra and G.~F.~R. Ellis, ``Schwarzschild black hole lensing,'' {\em
  Physical Review D}, vol.~62, p.~084003, Sep 2000.

\bibitem{kraniotis_precise_2004}
G.~V. Kraniotis, ``Precise relativistic orbits in {Kerr} and {Kerr}–(anti) de
  {Sitter} spacetimes,'' {\em Classical and Quantum Gravity}, vol.~21,
  pp.~4743--4769, Oct. 2004.

\bibitem{beckwith_extreme_2005}
K.~Beckwith and C.~Done, ``Extreme gravitational lensing near rotating black
  holes,'' {\em Monthly Notices of the Royal Astronomical Society}, vol.~359,
  pp.~1217--1228, June 2005.

\bibitem{kraniotis_frame_2005}
G.~V. Kraniotis, ``Frame dragging and bending of light in {Kerr} and
  {Kerr}–(anti) de {Sitter} spacetimes,'' {\em Classical and Quantum
  Gravity}, vol.~22, pp.~4391--4424, Nov. 2005.

\bibitem{hackmann_complete_2008}
E.~Hackmann and C.~Lämmerzahl, ``Complete {Analytic} {Solution} of the
  {Geodesic} {Equation} in {Schwarzschild}–({Anti}-)de {Sitter}
  {Spacetimes},'' {\em Physical Review Letters}, vol.~100, p.~171101, May 2008.

\bibitem{hackmann_geodesic_2008}
E.~Hackmann and C.~Lämmerzahl, ``Geodesic equation in
  {Schwarzschild}-(anti-)de {Sitter} space-times: {Analytical} solutions and
  applications,'' {\em Physical Review D}, vol.~78, p.~024035, July 2008.

\bibitem{bisnovatyi-kogan_strong_2008}
G.~S. Bisnovatyi-Kogan and O.~Y. Tsupko, ``Strong gravitational lensing by
  {Schwarzschild} black holes,'' {\em Astrophysics}, vol.~51, pp.~99--111, Jan.
  2008.

\bibitem{kagramanova_analytic_2010}
V.~Kagramanova, J.~Kunz, E.~Hackmann, and C.~Lämmerzahl, ``Analytic treatment
  of complete and incomplete geodesics in {Taub}-{NUT} space-times,'' {\em
  Physical Review D}, vol.~81, p.~124044, June 2010.

\bibitem{hackmann_analytical_2010}
E.~Hackmann, C.~Lämmerzahl, V.~Kagramanova, and J.~Kunz, ``Analytical solution
  of the geodesic equation in {Kerr}-(anti-) de {Sitter} space-times,'' {\em
  Physical Review D}, vol.~81, p.~044020, Feb. 2010.

\bibitem{hackmann_complete_2010}
E.~Hackmann, B.~Hartmann, C.~Lämmerzahl, and P.~Sirimachan, ``Complete set of
  solutions of the geodesic equation in the space-time of a {Schwarzschild}
  black hole pierced by a cosmic string,'' {\em Physical Review D}, vol.~81,
  p.~064016, Mar. 2010.

\bibitem{enolski_inversion_2011}
V.~Enolski, E.~Hackmann, V.~Kagramanova, J.~Kunz, and C.~Lämmerzahl,
  ``Inversion of hyperelliptic integrals of arbitrary genus with application to
  particle motion in general relativity,'' {\em Journal of Geometry and
  Physics}, vol.~61, pp.~899--921, May 2011.

\bibitem{kraniotis_precise_2011}
G.~V. Kraniotis, ``Precise analytic treatment of {Kerr} and {Kerr}-(anti) de
  {Sitter} black holes as gravitational lenses,'' {\em Classical and Quantum
  Gravity}, vol.~28, p.~085021, Apr. 2011.

\bibitem{enolski_inversion_2012}
V.~Enolski, B.~Hartmann, V.~Kagramanova, J.~Kunz, C.~Lämmerzahl, and
  P.~Sirimachan, ``Inversion of a general hyperelliptic integral and particle
  motion in {Hořava}–{Lifshitz} black hole space-times,'' {\em Journal of
  Mathematical Physics}, vol.~53, p.~012504, Jan. 2012.

\bibitem{gibbons_application_2012}
G.~W. Gibbons and M.~Vyska, ``The application of {Weierstrass} elliptic
  functions to {Schwarzschild} null geodesics,'' {\em Classical and Quantum
  Gravity}, vol.~29, p.~065016, Mar. 2012.

\bibitem{munoz_orbits_2014}
G.~Muñoz, ``Orbits of massless particles in the {Schwarzschild} metric:
  {Exact} solutions,'' {\em American Journal of Physics}, vol.~82,
  pp.~564--573, June 2014.

\bibitem{kraniotis_gravitational_2014}
G.~V. Kraniotis, ``Gravitational lensing and frame dragging of light in the
  {Kerr}–{Newman} and the {Kerr}–{Newman} (anti) de {Sitter} black hole
  spacetimes,'' {\em General Relativity and Gravitation}, vol.~46, p.~1818,
  Nov. 2014.

\bibitem{de_falco_approximate_2016}
V.~De~Falco, M.~Falanga, and L.~Stella, ``Approximate analytical calculations
  of photon geodesics in the {Schwarzschild} metric,'' {\em Astronomy \&
  Astrophysics}, vol.~595, p.~A38, Nov. 2016.

\bibitem{soroushfar_detailed_2016}
S.~Soroushfar, R.~Saffari, S.~Kazempour, S.~Grunau, and J.~Kunz, ``Detailed
  study of geodesics in the {Kerr}-{Newman}-({A}){dS} spacetime and the
  rotating charged black hole spacetime in f ( {R} ) gravity,'' {\em Physical
  Review D}, vol.~94, p.~024052, July 2016.

\bibitem{barlow_asymptotically_2017}
N.~S. Barlow, S.~J. Weinstein, and J.~A. Faber, ``An asymptotically consistent
  approximant for the equatorial bending angle of light due to {Kerr} black
  holes,'' {\em Classical and Quantum Gravity}, vol.~34, p.~135017, July 2017.

\bibitem{uniyal_null_2018}
R.~Uniyal, H.~Nandan, and K.~D. Purohit, ``Null geodesics and observables
  around the {Kerr}–{Sen} black hole,'' {\em Classical and Quantum Gravity},
  vol.~35, p.~025003, Jan. 2018.

\bibitem{villanueva_null_2018}
J.~R. Villanueva, F.~Tapia, M.~Molina, and M.~Olivares, ``Null paths on a
  toroidal topological black hole in conformal {Weyl} gravity,'' {\em The
  European Physical Journal C}, vol.~78, p.~853, Oct. 2018.

\bibitem{chatterjee_analytic_2019}
A.~K. Chatterjee, K.~Flathmann, H.~Nandan, and A.~Rudra, ``Analytic solutions
  of the geodesic equation for {Reissner}-{Nordström}–(anti–)de {Sitter}
  black holes surrounded by different kinds of regular and exotic matter
  fields,'' {\em Physical Review D}, vol.~100, p.~024044, July 2019.

\bibitem{hsiao_equatorial_2020}
Y.-W. Hsiao, D.-S. Lee, and C.-Y. Lin, ``Equatorial light bending around
  {Kerr}-{Newman} black holes,'' {\em Physical Review D}, vol.~101, p.~064070,
  Mar. 2020.

\bibitem{gralla_null_2020}
S.~E. Gralla and A.~Lupsasca, ``Null geodesics of the {Kerr} exterior,'' {\em
  Physical Review D}, vol.~101, p.~044032, Feb. 2020.

\bibitem{hendi_simulation_2020}
S.~H. Hendi, A.~M. Tavakkoli, S.~Panahiyan, B.~E. Panah, and E.~Hackmann,
  ``Simulation of geodesic trajectory of charged {BTZ} black holes in massive
  gravity,'' {\em The European Physical Journal C}, vol.~80, p.~524, June 2020.

\bibitem{fathi_classical_2020}
M.~Fathi, M.~Olivares, and J.~R. Villanueva, ``Classical tests on a charged
  {Weyl} black hole: bending of light, {Shapiro} delay and {Sagnac} effect,''
  {\em The European Physical Journal C}, vol.~80, p.~51, Jan. 2020.

\bibitem{kraniotis_gravitational_2021}
G.~V. Kraniotis, ``Gravitational redshift/blueshift of light emitted by
  geodesic test particles, frame-dragging and pericentre-shift effects, in the
  {Kerr}–{Newman}–de {Sitter} and {Kerr}–{Newman} black hole
  geometries,'' {\em The European Physical Journal C}, vol.~81, p.~147, Feb.
  2021.

\bibitem{fathi_study_2022}
M.~Fathi, M.~Olivares, and J.~R. Villanueva, ``Study of null and time-like
  geodesics in the exterior of a {Schwarzschild} black hole with quintessence
  and cloud of strings,'' {\em The European Physical Journal C}, vol.~82,
  p.~629, July 2022.

\bibitem{battista_geodesic_2022}
E.~Battista and G.~Esposito, ``Geodesic motion in {Euclidean} {Schwarzschild}
  geometry,'' {\em The European Physical Journal C}, vol.~82, p.~1088, Dec.
  2022.

\bibitem{fathi_spherical_2023}
M.~Fathi, M.~Olivares, and J.~R. Villanueva, ``Spherical photon orbits around a
  rotating black hole with quintessence and cloud of strings,'' {\em The
  European Physical Journal Plus}, vol.~138, p.~7, Jan. 2023.

\bibitem{omwoyo_black_2023}
E.~Omwoyo, H.~Belich, J.~C. Fabris, and H.~Velten, ``Black hole lensing in
  {Kerr}-de {Sitter} spacetimes,'' {\em The European Physical Journal Plus},
  vol.~138, p.~1043, Nov. 2023.

\bibitem{Wang:2023}
K.~{Wang}, C.-J. {Feng}, and T.~{Wang}, ``{Image of Kerr-de Sitter black holes:
  An additional avenue for testing the cosmological constant},'' {\em arXiv
  e-prints}, p.~arXiv:2309.16944, Sept. 2023.

\bibitem{Synge:1960}
J.~Synge, {\em Relativity: The General Theory}.
\newblock No.~v. 1 in North-Holland series in physics, North-Holland Publishing
  Company, 1960.

\bibitem{bisnovatyi-kogan_gravitational_2009}
G.~S. Bisnovatyi-Kogan and O.~Y. Tsupko, ``Gravitational radiospectrometer,''
  {\em Gravitation and Cosmology}, vol.~15, pp.~20--27, Jan. 2009.

\bibitem{bisnovatyi-kogan_gravitational_2010}
G.~S. Bisnovatyi-Kogan and O.~Y. Tsupko, ``Gravitational lensing in a
  non-uniform plasma: {Gravitational} lensing in a non-uniform plasma,'' {\em
  Monthly Notices of the Royal Astronomical Society}, pp.~no--no, May 2010.

\bibitem{tsupko_gravitational_2013}
O.~Y. Tsupko and G.~S. Bisnovatyi-Kogan, ``Gravitational lensing in plasma:
  {Relativistic} images at homogeneous plasma,'' {\em Physical Review D},
  vol.~87, p.~124009, June 2013.

\bibitem{morozova_gravitational_2013}
V.~S. Morozova, B.~J. Ahmedov, and A.~A. Tursunov, ``Gravitational lensing by a
  rotating massive object in a plasma,'' {\em Astrophysics and Space Science},
  vol.~346, pp.~513--520, Aug. 2013.

\bibitem{bisnovatyi-kogan_gravitational_2015}
G.~S. Bisnovatyi-Kogan and O.~Y. Tsupko, ``Gravitational lensing in plasmic
  medium,'' {\em Plasma Physics Reports}, vol.~41, pp.~562--581, July 2015.

\bibitem{perlick_influence_2015}
V.~Perlick, O.~Y. Tsupko, and G.~S. Bisnovatyi-Kogan, ``Influence of a plasma
  on the shadow of a spherically symmetric black hole,'' {\em Physical Review
  D}, vol.~92, p.~104031, Nov. 2015.

\bibitem{atamurotov_optical_2015}
F.~Atamurotov, B.~Ahmedov, and A.~Abdujabbarov, ``Optical properties of black
  holes in the presence of a plasma: {The} shadow,'' {\em Physical Review D},
  vol.~92, p.~084005, Oct. 2015.

\bibitem{abdujabbarov_shadow_2016}
A.~Abdujabbarov, M.~Amir, B.~Ahmedov, and S.~G. Ghosh, ``Shadow of rotating
  regular black holes,'' {\em Physical Review D}, vol.~93, p.~104004, May 2016.

\bibitem{bisnovatyi-kogan_gravitational_2017}
G.~Bisnovatyi-Kogan and O.~Tsupko, ``Gravitational {Lensing} in {Presence} of
  {Plasma}: {Strong} {Lens} {Systems}, {Black} {Hole} {Lensing} and {Shadow},''
  {\em Universe}, vol.~3, p.~57, July 2017.

\bibitem{perlick_light_2017}
V.~Perlick and O.~Y. Tsupko, ``Light propagation in a plasma on {Kerr}
  spacetime: {Separation} of the {Hamilton}-{Jacobi} equation and calculation
  of the shadow,'' {\em Physical Review D}, vol.~95, p.~104003, May 2017.

\bibitem{schulze-koops_sachs_2017}
K.~Schulze-Koops, V.~Perlick, and D.~J. Schwarz, ``Sachs equations for light
  bundles in a cold plasma,'' {\em Classical and Quantum Gravity}, vol.~34,
  p.~215006, Nov. 2017.

\bibitem{abdujabbarov_shadow_2017}
A.~Abdujabbarov, B.~Toshmatov, Z.~Stuchlík, and B.~Ahmedov, ``Shadow of the
  rotating black hole with quintessential energy in the presence of plasma,''
  {\em International Journal of Modern Physics D}, vol.~26, p.~1750051, May
  2017.

\bibitem{liu_effects_2017}
C.-Q. Liu, C.-K. Ding, and J.-L. Jing, ``Effects of {Homogeneous} {Plasma} on
  {Strong} {Gravitational} {Lensing} of {Kerr} {Black} {Holes},'' {\em Chinese
  Physics Letters}, vol.~34, p.~090401, Aug. 2017.

\bibitem{haroon_shadow_2019}
S.~Haroon, M.~Jamil, K.~Jusufi, K.~Lin, and R.~B. Mann, ``Shadow and deflection
  angle of rotating black holes in perfect fluid dark matter with a
  cosmological constant,'' {\em Physical Review D}, vol.~99, p.~044015, Feb.
  2019.

\bibitem{kimpson_spatial_2019}
T.~Kimpson, K.~Wu, and S.~Zane, ``Spatial dispersion of light rays propagating
  through a plasma in {Kerr} space–time,'' {\em Monthly Notices of the Royal
  Astronomical Society}, vol.~484, pp.~2411--2419, Apr. 2019.

\bibitem{babar_optical_2020}
G.~Z. Babar, A.~Z. Babar, and F.~Atamurotov, ``Optical properties of
  {Kerr}–{Newman} spacetime in the presence of plasma,'' {\em The European
  Physical Journal C}, vol.~80, p.~761, Aug. 2020.

\bibitem{junior_spinning_2020}
H.~C. D.~L. Junior, L.~C.~B. Crispino, P.~V.~P. Cunha, and C.~A.~R. Herdeiro,
  ``Spinning black holes with a separable {Hamilton}–{Jacobi} equation from a
  modified {Newman}–{Janis} algorithm,'' {\em The European Physical Journal
  C}, vol.~80, p.~1036, Nov. 2020.

\bibitem{Badia:2021}
J.~Bad\'\i{}a and E.~F. Eiroa, ``{Shadow of axisymmetric, stationary and
  asymptotically flat black holes in the presence of plasma},'' 6 2021.

\bibitem{fathi_analytical_2021}
M.~Fathi, M.~Olivares, and J.~R. Villanueva, ``Analytical study of light ray
  trajectories in {Kerr} spacetime in the presence of an inhomogeneous
  anisotropic plasma,'' {\em The European Physical Journal C}, vol.~81, p.~987,
  Nov. 2021.

\bibitem{KUMAR2021100881}
R.~Kumar, B.~P. Singh, M.~S. Ali, and S.~G. Ghosh, ``Shadows of black hole
  surrounded by anisotropic fluid in rastall theory,'' {\em Physics of the Dark
  Universe}, vol.~34, p.~100881, 2021.

\bibitem{majernik2006rastalls}
V.~{Majernik} and L.~{Richterek}, ``{Rastall's gravity equations and Mach's
  Principle},'' {\em arXiv e-prints}, pp.~gr--qc/0610070, Oct. 2006.

\bibitem{Visser_2020}
M.~Visser, ``The kiselev black hole is neither perfect fluid, nor is it
  quintessence,'' {\em Classical and Quantum Gravity}, vol.~37, p.~045001, jan
  2020.

\bibitem{Azreg:2014}
M.~Azreg-A\"{\i}nou, ``Generating rotating regular black hole solutions without
  complexification,'' {\em Phys. Rev. D}, vol.~90, p.~064041, Sep 2014.

\bibitem{azreg-ainou_static_2014}
M.~Azreg-Aïnou, ``From static to rotating to conformal static solutions:
  rotating imperfect fluid wormholes with(out) electric or magnetic field,''
  {\em The European Physical Journal C}, vol.~74, p.~2865, May 2014.

\bibitem{perlick_calculating_2022}
V.~Perlick and O.~Y. Tsupko, ``Calculating black hole shadows: {Review} of
  analytical studies,'' {\em Physics Reports}, vol.~947, pp.~1--39, Feb. 2022.

\bibitem{Breuer:1980}
R.~A. {Breuer} and J.~{Ehlers}, ``{Propagation of High-Frequency
  Electromagnetic Waves Through a Magnetized Plasma in Curved Space-Time. I},''
  {\em Proceedings of the Royal Society of London Series A}, vol.~370,
  pp.~389--406, Mar. 1980.

\bibitem{Atamurotov:2015}
A.~Abdujabbarov, M.~Amir, B.~Ahmedov, and S.~G. Ghosh, ``Shadow of rotating
  regular black holes,'' {\em Phys. Rev. D}, vol.~93, p.~104004, May 2016.

\bibitem{Perlick:2017}
V.~Perlick and O.~Y. Tsupko, ``Light propagation in a plasma on kerr spacetime:
  Separation of the hamilton-jacobi equation and calculation of the shadow,''
  {\em Phys. Rev. D}, vol.~95, p.~104003, May 2017.

\bibitem{Bicak:1975}
J.~{Bicak} and P.~{Hadrava}, ``{General-relativistic radiative transfer theory
  in refractive and dispersive media.},'' {\em Astron. Astrophys.}, vol.~44,
  pp.~389--399, Nov. 1975.

\bibitem{PhysRevD.45.525}
R.~Kulsrud and A.~Loeb, ``Dynamics and gravitational interaction of waves in
  nonuniform media,'' {\em Phys. Rev. D}, vol.~45, pp.~525--531, Jan 1992.

\bibitem{Muhleman:1966}
D.~O. Muhleman and I.~D. Johnston, ``Radio propagation in the solar
  gravitational field,'' {\em Phys. Rev. Lett.}, vol.~17, pp.~455--458, Aug
  1966.

\bibitem{Perlick_ray_2000}
V.~Perlick, {\em Ray {Optics}, {Fermat}’s {Principle}, and {Applications} to
  {General} {Relatively}}, vol.~61 of {\em Lecture {Notes} in {Physics}}.
\newblock Berlin, Heidelberg: Springer Berlin Heidelberg, 2000.

\bibitem{Tsupko:2013}
O.~Y. Tsupko and G.~S. Bisnovatyi-Kogan, ``Gravitational lensing in plasma:
  Relativistic images at homogeneous plasma,'' {\em Phys. Rev. D}, vol.~87,
  p.~124009, Jun 2013.

\bibitem{Yan:2019}
H.~Yan, ``Influence of a plasma on the observational signature of a high-spin
  kerr black hole,'' {\em Phys. Rev. D}, vol.~99, p.~084050, Apr 2019.

\bibitem{huang_revisiting_2018}
Y.~Huang, Y.-P. Dong, and D.-J. Liu, ``Revisiting the shadow of a black hole in
  the presence of a plasma,'' {\em International Journal of Modern Physics D},
  vol.~27, p.~1850114, Sept. 2018.

\bibitem{Carter:1968}
B.~Carter, ``Global structure of the kerr family of gravitational fields,''
  {\em Phys. Rev.}, vol.~174, pp.~1559--1571, Oct 1968.

\bibitem{Mino:2003}
Y.~Mino, ``Perturbative approach to an orbital evolution around a supermassive
  black hole,'' {\em Physical Review D}, vol.~67, p.~084027, Apr 2003.

\bibitem{byrd_handbook_1971}
P.~F. Byrd and M.~D. Friedman, {\em Handbook of {Elliptic} {Integrals} for
  {Engineers} and {Scientists}}.
\newblock Berlin, Heidelberg: Springer Berlin Heidelberg, 1971.

\bibitem{Gralla:2019}
S.~E. Gralla, D.~E. Holz, and R.~M. Wald, ``Black hole shadows, photon rings,
  and lensing rings,'' {\em Phys. Rev. D}, vol.~100, p.~024018, Jul 2019.

\bibitem{stoghianidis_polar_1987}
E.~Stoghianidis and D.~Tsoubelis, ``Polar orbits in the {Kerr} space-time,''
  {\em General Relativity and Gravitation}, vol.~19, pp.~1235--1249, Dec. 1987.

\bibitem{cramer_using_1997}
C.~R. Cramer, ``Using the {Uncharged} {Kerr} {Black} {Hole} as a
  {Gravitational} {Mirror},'' {\em General Relativity and Gravitation},
  vol.~29, pp.~445--454, Apr. 1997.

\bibitem{Teo:2003}
E.~Teo, ``Spherical {Photon} {Orbits} {Around} a {Kerr} {Black} {Hole},'' {\em
  General Relativity and Gravitation}, vol.~35, pp.~1909--1926, Nov. 2003.

\bibitem{Johannsen:2013}
T.~Johannsen, ``{PHOTON} {RINGS} {AROUND} {KERR} {AND} {KERR}-{LIKE} {BLACK}
  {HOLES},'' {\em The Astrophysical Journal}, vol.~777, p.~170, oct 2013.

\bibitem{Grenzebach:2014}
A.~Grenzebach, V.~Perlick, and C.~L\"ammerzahl, ``Photon regions and shadows of
  kerr-newman-nut black holes with a cosmological constant,'' {\em Phys. Rev.
  D}, vol.~89, p.~124004, Jun 2014.

\bibitem{charbulak_spherical_2018}
D.~Charbulák and Z.~Stuchlík, ``Spherical photon orbits in the field of
  {Kerr} naked singularities,'' {\em The European Physical Journal C}, vol.~78,
  p.~879, Nov. 2018.

\bibitem{Johnson_universal_2020}
M.~D. Johnson, A.~Lupsasca, A.~Strominger, G.~N. Wong, S.~Hadar, D.~Kapec,
  R.~Narayan, A.~Chael, C.~F. Gammie, P.~Galison, D.~C.~M. Palumbo, S.~S.
  Doeleman, L.~Blackburn, M.~Wielgus, D.~W. Pesce, J.~R. Farah, and J.~M.
  Moran, ``Universal interferometric signatures of a black hole’s photon
  ring,'' {\em Science Advances}, vol.~6, p.~eaaz1310, Mar. 2020.

\bibitem{Himwich:2020}
E.~Himwich, M.~D. Johnson, A.~Lupsasca, and A.~Strominger, ``Universal
  polarimetric signatures of the black hole photon ring,'' {\em Phys. Rev. D},
  vol.~101, p.~084020, Apr 2020.

\bibitem{Gelles:2021}
Z.~Gelles, E.~Himwich, M.~D. Johnson, and D.~C.~M. Palumbo, ``Polarized image
  of equatorial emission in the kerr geometry,'' {\em Phys. Rev. D}, vol.~104,
  p.~044060, Aug 2021.

\bibitem{Ayzenberg:2022}
D.~Ayzenberg, ``Testing gravity with black hole shadow subrings,'' {\em
  Classical and Quantum Gravity}, vol.~39, p.~105009, may 2022.

\bibitem{Das:2022}
A.~Das, A.~Saha, and S.~Gangopadhyay, ``Study of circular geodesics and shadow
  of rotating charged black hole surrounded by perfect fluid dark matter
  immersed in plasma,'' {\em Classical and Quantum Gravity}, vol.~39,
  p.~075005, mar 2022.

\bibitem{ANJUM2023101195}
A.~Anjum, M.~Afrin, and S.~G. Ghosh, ``Investigating effects of dark matter on
  photon orbits and black hole shadows,'' {\em Physics of the Dark Universe},
  vol.~40, p.~101195, 2023.

\bibitem{Chen:2023}
Y.-X. Chen, J.-H. Huang, and H.~Jiang, ``Radii of spherical photon orbits
  around kerr-newman black holes,'' {\em Phys. Rev. D}, vol.~107, p.~044066,
  Feb 2023.

\bibitem{andaru_spherical_2023}
L.~Andaru, A.~Alam, B.~Jayawiguna, and H.~Ramadhan, ``Spherical orbits around
  {Kerr}-{Newman} and regular black holes,'' preprint, In Review, Oct. 2023.

\bibitem{Tavlayan:2020}
A.~Tavlayan and B.~Tekin, ``Exact formulas for spherical photon orbits around
  kerr black holes,'' {\em Phys. Rev. D}, vol.~102, p.~104036, Nov 2020.

\bibitem{fathi_gravitational_2021}
M.~Fathi, M.~Olivares, and J.~R. Villanueva, ``Gravitational {Rutherford}
  scattering of electrically charged particles from a charged {Weyl} black
  hole,'' {\em The European Physical Journal Plus}, vol.~136, p.~420, Apr.
  2021.

\end{thebibliography}

\end{document}